 \documentclass[12pt,preprint]{aastex}

\shorttitle{Luminosity and Redshift Dependence of the Covering Factor}
\shortauthors{Toba et al.}

\begin{document}


\title{Luminosity and Redshift Dependence of the Covering Factor of AGNs viewed with WISE and SDSS}


\author{Y.Toba 		\altaffilmark{1,2,3},
   		S.Oyabu		\altaffilmark{4},
   		H.Matsuhara	\altaffilmark{1,2},
   		M.Malkan	\altaffilmark{5},
		P.Gandhi	\altaffilmark{6},
		T.Nakagawa	\altaffilmark{2},
		N.Isobe		\altaffilmark{2}, 			
		M.Shirahata	\altaffilmark{2,7},  
		N.Oi		\altaffilmark{2},  
  		Y.Ohyama	\altaffilmark{8},
	   	S.Takita	\altaffilmark{2},
   		C.Yamauchi	\altaffilmark{9},
   		\and		
		K.Yano		\altaffilmark{2,10}, 
		}


\altaffiltext{1}{Department of Space and Astronautical Science, the Graduate University for Advanced Studies (Sokendai), 3-1-1 Yoshinodai, Chuo-ku, Sagamihara, Kanagawa 252-5210, Japan}
\email{toba@ir.isas.jaxa.jp}
 \altaffiltext{2}{Institute of Space and Astronautical Science, Japan Aerospace Exploration Agency, 3-1-1 Yoshinodai, Chuo-ku, Sagamihara, Kanagawa 252-5210, Japan}
  \altaffiltext{3}{Research Center for Space and Cosmic Evolution, Ehime University, Bunkyo-cho, Matsuyama 790-8577, Japan} 
 \altaffiltext{4}{Graduate School of Science, Nagoya University, Furo-cho, Chikusa-ku, Nagoya, Aichi 464-8602, Japan}
 \altaffiltext{5}{Department of Physics and Astronomy, University of California, Los Angeles, CA 90095-1547, USA}
 \altaffiltext{6}{Department of Physics, Durham University, Durham DH1-3LE, UK}
 \altaffiltext{7}{National Astronomical Observatory of Japan, 2-21-1 Osawa, Mitaka, Tokyo
181-8588}
 \altaffiltext{8}{Institute of Astronomy and Astrophysics, Academia Sinica, P.O. Box 23-141, Taipei 10617, Taiwan, R.O.C} 
 \altaffiltext{9}{Misato Observatory, 180 Matsugamine, Misato-cho, Kaiso-gun, Wakayama 640-1366, Japan}
 \altaffiltext{10}{Department of Physics, School of Science, The University of Tokyo, 7-3-1, Hongo, Bunkyo-ku, Tokyo 113-0033, Japan}


\begin{abstract}
In this work, we investigate the dependence of the covering factor (CF) of active
galactic nuclei (AGNs) (i) on the mid-infrared (MIR) luminosity and
(ii) on the redshift. We constructed 12- and 22-$\mu$m luminosity
functions (LFs) at 0.006 $\leq z \leq$ 0.3 using the {\it Wide-field
Infrared Survey Explorer} ({\it WISE}) data. Combining the {\it
WISE} catalog with the Sloan Digital Sky Survey (SDSS) spectroscopic
data, we selected 223,982 galaxies at 12 $\mu$m and 25,721 galaxies
at 22 $\mu$m for spectroscopic classification. We then identified
16,355 AGNs at 12 $\mu$m and 4,683 AGNs at 22 $\mu$m by their
optical emission lines and cataloged classifications in the SDSS.
Following that, we estimated the CF as the fraction of type 2 AGN in
all AGNs whose MIR emissions are dominated by the active nucleus
(not their host galaxies) based on their MIR colors.
We found that (i) the CF decreased with
increasing MIR luminosity, regardless of the choice of type 2 AGN
classification criteria, and
(ii) the CF did not change significantly with the redshift for $z \leq 0.2$.
Furthermore, we carried out various tests
to determine the influence of selection bias and confirmed similar
dependences exist even when taking these uncertainties into account.
The luminosity dependence of the CF can be explained by the receding
torus model, but the ``modified'' receding torus model gives a
slightly better fit, as suggested by Simpson.
\end{abstract}





\section{INTRODUCTION}
The popular unification scheme for active galactic nuclei (AGNs)
requires that the observed differences between type 1 and type 2
AGNs arise from the orientation \cite[e.g.,][]{Antonucci, Urry}, and
the basic premise is that all AGNs are fundamentally the same. This
scheme proposes a geometrically thick dusty torus surrounding the
AGN central engine (accretion disk and supermassive black hole),
with the torus providing anisotropic obscuration of the central
region so that sources viewed face-on are recognized as type 1 AGNs,
while those observed edge-on are type 2 AGNs. However, even if this
torus exists in all AGNs, its key parameters such as its geometry
and physical properties are still unclear.

We focus here on the geometrical covering fraction of the dust
torus, a fundamental parameter in the unification scheme. The
covering factor (CF) is defined as the fraction of the sky, as seen
from the AGN center, that is blocked by heavily obscuring material.
This corresponds to the fraction of type 2 AGNs in the entire AGN
population. Recently, some authors have claimed that the CF depends
on the luminosity and redshift. For example, \cite{Simpson+05}
examined data for 4,304 galaxies (including AGNs) from the Sloan
Digital Sky Survey (SDSS) Data Release 2 (DR2) and found that the CF
decreases with increasing [OIII] emission line luminosity, which is
believed to be isotropic. \cite{Hasinger} also reported a negative
correlation between the fraction of absorbed ($\sim$ type 2) AGNs
and the X-ray (2--10 keV) luminosity based on 1,290 AGNs selected in
the 2--10 keV band from different flux-limited surveys with very
high optical identification completeness. Furthermore,
\cite{Hasinger} found that the absorbed fraction increases
significantly with increasing redshift, saturating at a redshift of
$z$ $\sim$ 2. Recently, \cite{Toba} also confirmed the luminosity
dependence of the CF by using the {\it AKARI} mid-infrared all-sky
survey catalog \citep{Ishihara}. Some authors, however, have
questioned these dependencies by claiming that the data are affected
by various uncertainties. In particular, the observed correlations
can be explained as a selection effect, in which case they may not
necessarily have any astrophysical significance. For instance,
\cite{Dwelly} found from {\it XMM-Newton} observations of the
Chandra Deep Field-South that there is no evidence that the
absorption distribution is dependent on either the intrinsic X-ray
luminosity or the redshift. \cite{Akylas} suggested that the
apparent increase in the absorbed AGN fraction with increasing
redshift is due to a systematic overestimation of the column
densities measured in high redshift sources where the absorption
cut-off is shifted towards low energies. \cite{Lawrence+10}
attempted to carefully distinguish strict type 2 AGNs from more
lightly reddened type 1 AGNs, as well as from low-excitation
narrow-line AGNs, which were assembled from the literature. They
also showed that radio, infrared (IR), and volume-limited samples
all agree in showing that the type 2 fraction does not change with
luminosity. Therefore, it is still unclear whether the CF
intrinsically depends on the luminosity and particularly on
redshift.

To resolve this problem, it is important to conduct a statistical
analysis based on IR observations; reprocessed radiation from the
dust in the torus is re-emitted in the IR wavelength range. Mid-IR
(MIR) emission, in particular, is expected to be direct radiation
from the dust torus and uninfluenced by dust extinction. In this
paper, we estimate the CF of the dust torus using the MIR luminosity
functions (LFs) and examine the luminosity and redshift dependence
based on a statistically complete AGN sample. The LF of galaxies is
a fundamental statistical tool for describing galaxy properties,
since it should be almost entirely independent of the viewing angle.
We construct the MIR LFs using the data from the {\it Wide-field
Infrared Survey Explorer} \citep[{\it WISE}:][]{Wright}, which was
launched in 2009. {\it WISE} performed an all-sky survey with a high
sensitivity in four bands (particularly relevant to the study here
are the 12- and 22-$\mu$m bands).
While the spatial resolution of {\it WISE} is relatively poor owing
to the 40-cm diameter of the telescope, it is several orders of
magnitude better than those of the {\it Infrared Astronomical
Satellite} \citep[{\it IRAS}:][]{Neugebauer,Beichman} and {\it
AKARI} \citep{Murakami}, both of which performed previous all-sky IR
surveys.

This paper is organized as follows. Section 2 describes the
sample selection and derivation of the LFs, and the 12- and
22-$\mu$m LFs computed using the 1/$V_{\mathrm{max}}$ technique are
presented in Section 3. Our results are then compared with previous
studies. In Section 4, we consider the origin of the MIR emission.
According to an empirical method based on a {\it WISE} color--color
diagram, we extract sources that are dominated in the MIR by the
active nucleus. We then estimate the CF for those AGN-dominated MIR
objects and discuss the luminosity and redshift dependence of the CF
by analyzing the relationship between the CF and luminosity in
separate redshift bins.
This paper provides us with statistically robust results about the luminosity and redshift dependence of the CF and yields a reliable dust torus model that explains the results.
Throughout this paper, we assume a flat universe with $\Omega_k =0$, and we adopt ($\Omega_M$, $\Omega_{\Lambda}$) = (0.3, 0.7) and $H_0$ = 75 km s$^{-1}$ Mpc$^{-1}$.

\label{Section_Data_and_Analysis_WISE}
\section{DATA AND ANALYSIS}
We selected 12- and 22-$\mu$m flux-limited galaxies based on the {\it WISE} and SDSS catalogs, and these galaxies were then classified into five types according to their optical spectroscopic information in the SDSS catalog.
For spectroscopically classified galaxies, we constructed the LFs using the 1/$V_{\mathrm{max}}$ method, considering both the detection limit of the {\it WISE} and SDSS catalogs.

\subsection{Sample Selection}
\label{sample_selection}
    \begin{figure}
        \epsscale{1}
        \plotone{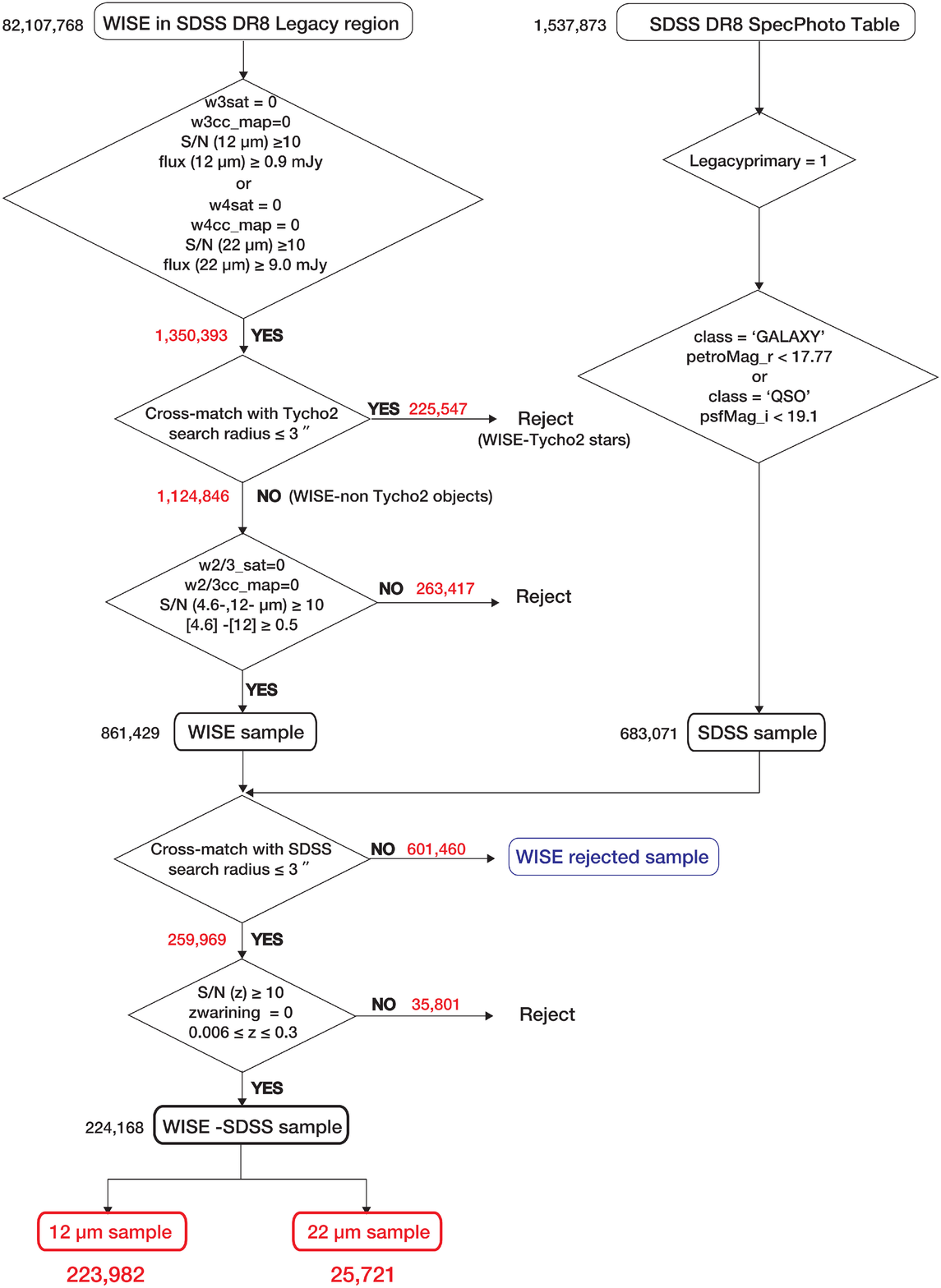}
        \caption{Flow chart of the sample selection process.}
        \label{flow_chart_sample_selection}
    \end{figure}
The {\it WISE} All-Sky Release Source Catalog provides positions and four-band (3.4-, 4.6-, 12-, and 22-$\mu$m) photometry for 563,921,584 objects.
In particular, there are 26,673,624 and 3,846,254 sources in the all-sky catalog with $\geq$10$\sigma$
detections in the 12- and 22-$\mu$m bands, respectively.
The sample used for this study was selected from {\it WISE} MIR sources with spectroscopy from the SDSS DR8 \citep{Aihara}.
In the end, we selected a 12-$\mu$m flux-limited sample of 223,982 galaxies and a 22-$\mu$m flux-limited sample of 25,721 galaxies.

\subsubsection{WISE sample}
{\it WISE} performed an all-sky survey at 3.4, 4.6, 12, and 22 $\mu$m with angular resolutions of 6.1, 6.4, 6.5, and 12.0 arcsec and a 5$\sigma$ photometric sensitivity better than 0.08, 0.11, 1, and 6 mJy (corresponding to 16.5, 15.5, 11.2, and 7.9 Vega magnitudes), respectively, in these four bands \citep{Wright}.
A flow chart of our sample selection process is shown in Figure \ref{flow_chart_sample_selection}.

We first narrowed our sample to {\it WISE} sources within the SDSS DR8 Legacy region (7966 deg$^2$).
The SDSS spectroscopic survey is performed using two multi-object fiber spectrographs on the same telescope.
Each spectroscopic fiber plug plate, referred to as a ``tile'', has a circular field-of-view with a radius of 1.49 degrees \citep{Blanton_03}, and 1794 tiles are employed in the Legacy survey.
Because the tiles are circular, there is a fraction of the sky that is covered by the overlap of tiles. The equatorial coordinates of the tile centers are contained in the {\tt sdssTileAll} table.
For the coordinates of each tile center, we searched for nearby {\it WISE} sources within a search radius of 1.49 degrees, which yielded a total of 82,107,768 {\it WISE} sources.

We then extracted 10$\sigma$-detected objects at 12 $\mu$m above 0.9
mJy ($\sim$ 11.4 mag) or at 22 $\mu$m above 9.0 mJy ($\sim$ 7.4
mag). These fluxes correspond to almost 100\% completeness flux
limits, according to the Explanatory Supplement to the {\it WISE}
All-Sky Data Release
Products\footnote{\url{http://wise2.ipac.caltech.edu/docs/release/allsky/expsup/}}.
When the sources were extracted, we also checked whether the sources
were saturated. The observed saturation levels of {\it WISE} are 1.0
($\sim$3.8 mag) and 12.0 Jy ($\sim$0.4 mag) for 12 and 22 $\mu$m,
respectively. The saturated pixel fractions listed in the catalog
are flagged as {\it w1-4sat} in each band.
We eliminated sources that had fluxes exceeding the saturation level and a high fraction
of saturated pixels (i.e., {\it WISE} sources with $w3sat \neq$ 0
for 12 $\mu$m or $w4sat \neq$ 0 for 22 $\mu$m were eliminated).
In addition, we checked sources that were contaminated or biased due to proximity to an image artifact (e.g., diffraction spikes, scattered-light halos, or optical ghosts) according to {\it w1-4cc\_map}.
A source that was unaffected by known artifacts was flagged as {\it w1-4cc\_map} = 0.
We thus eliminated sources with $w3cc\_map \neq$ 0 for 12 $\mu$m or $w4cc\_map \neq$ 0 for 22 $\mu$m.
This reduced the number of {\it WISE} samples to 1,350,393.
Note that the {\it WISE} catalog contains the Vega magnitude of each source, and
we converted these to fluxes. The zero magnitude flux densities for
12 and 22 $\mu$m are 31.674 and 8.363 Jy, respectively. Here, the
profile-fitting magnitude ({\it w1-4mpro}) was used as the magnitude
for the majority of the {\it WISE} sources. However, because the
{\it w1-4mpro} photometry is optimized for point sources and may
underestimate the true brightness of extended sources, we used the
elliptical aperture magnitude ({\it w1-4gmag}) for the extended
sources. The aperture is based on the elliptical shape reported in
the Two Micron All Sky Survey (2MASS) Extended Source Catalog (XSC).
We defined extended sources using {\it ext\_flg} in the {\it WISE}
catalog.

\begin{figure}
        \epsscale{1}
          \plottwo{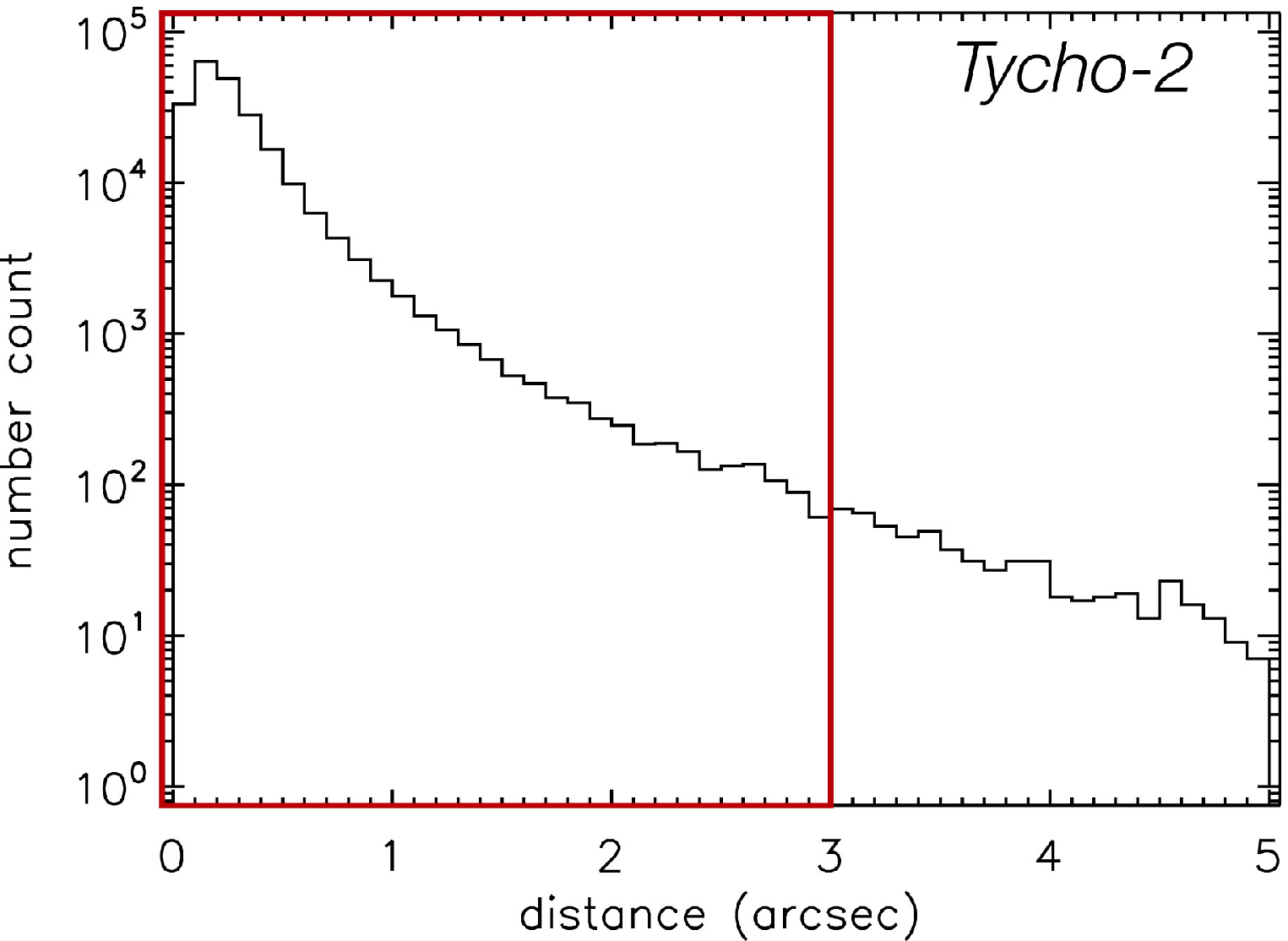}{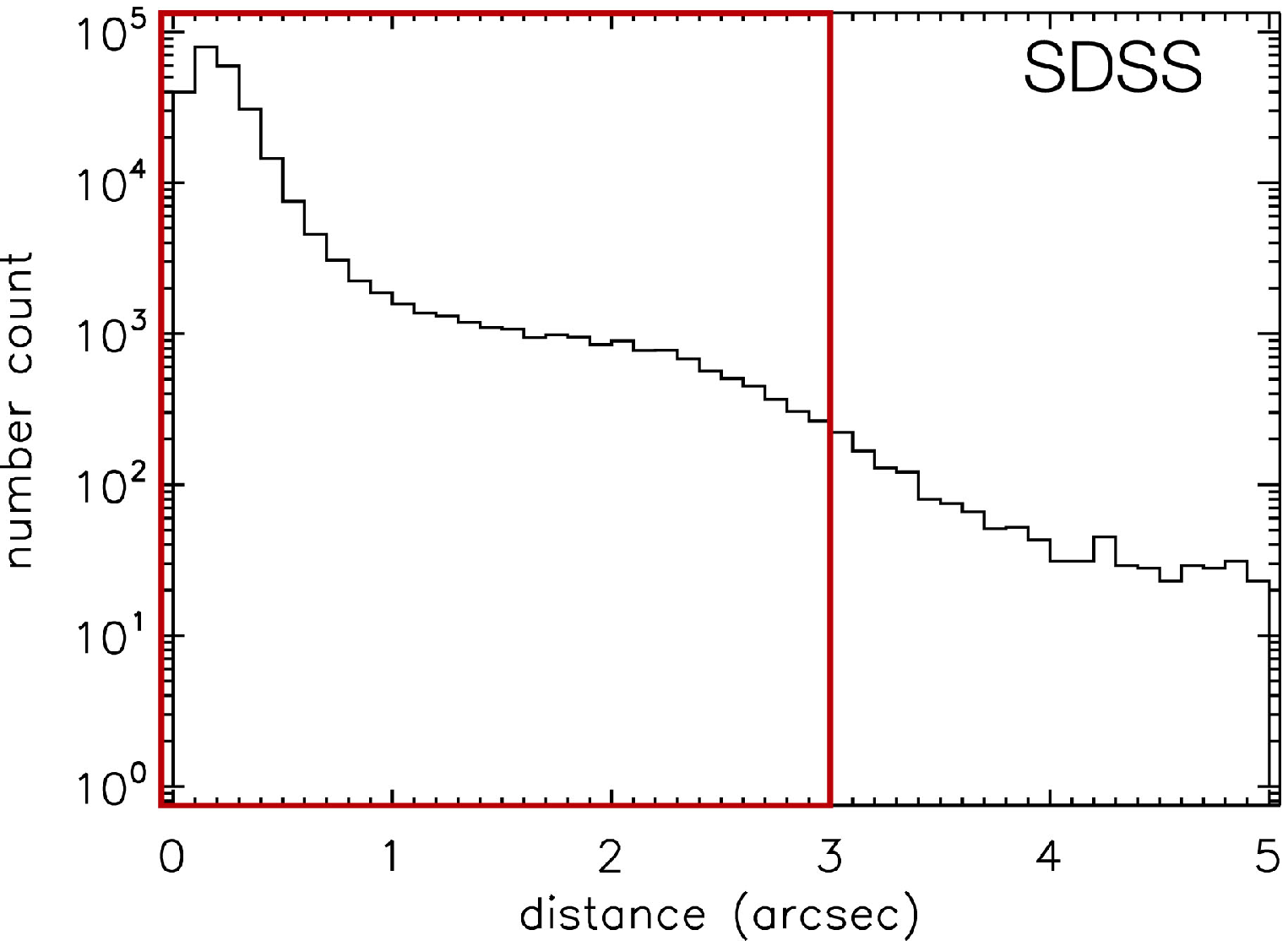}
        \caption{Histogram of the angular separation of {\it WISE} sources from the {\it Tycho-2} (left) and SDSS (right) coordinates. A search radius of 3 arcsec, as shown in red, was adopted for both sets. Cross-matching with the {\it Tycho-2} coordinates selected 225,547 objects within the search radius, while that with the SDSS coordinates selected 259,969 objects.}
        \label{cross-match_Tycho2_SDSS}
    \end{figure}

These 1,350,393 sources were then cross-identified with the {\it Tycho-2} Catalog \citep{Hog} to remove galactic bright stars.
The {\it Tycho-2} catalog contains the positions, proper motions, and two-color photometry of the 2.5 million bright stars in the sky down to the magnitude limit of the plates ($V_T \sim11.5$).
To avoid omitting high proper-motion stars, we referred to the \textit{mean position} rigorously propagated to the epoch J2000.0 by the proper motions in this catalog.
As a result, a total of 225,547 stars (hereinafter {\it WISE-Tycho 2} stars) were identified.
As shown in Figure \ref{cross-match_Tycho2_SDSS}, we adopted a 3-arcsec search radius because the star density in the SDSS spectroscopic region is at most $\sim$50 deg$^{-2}$ \citep{Hog}.
Thus, the probability of chance coincidence is less than 0.01\% (i.e., 1,350,393 $\times$ 0.0001 $\sim$ 135 sources may be misidentified), which is acceptable.

    \begin{figure}
        \epsscale{1}
        \plotone{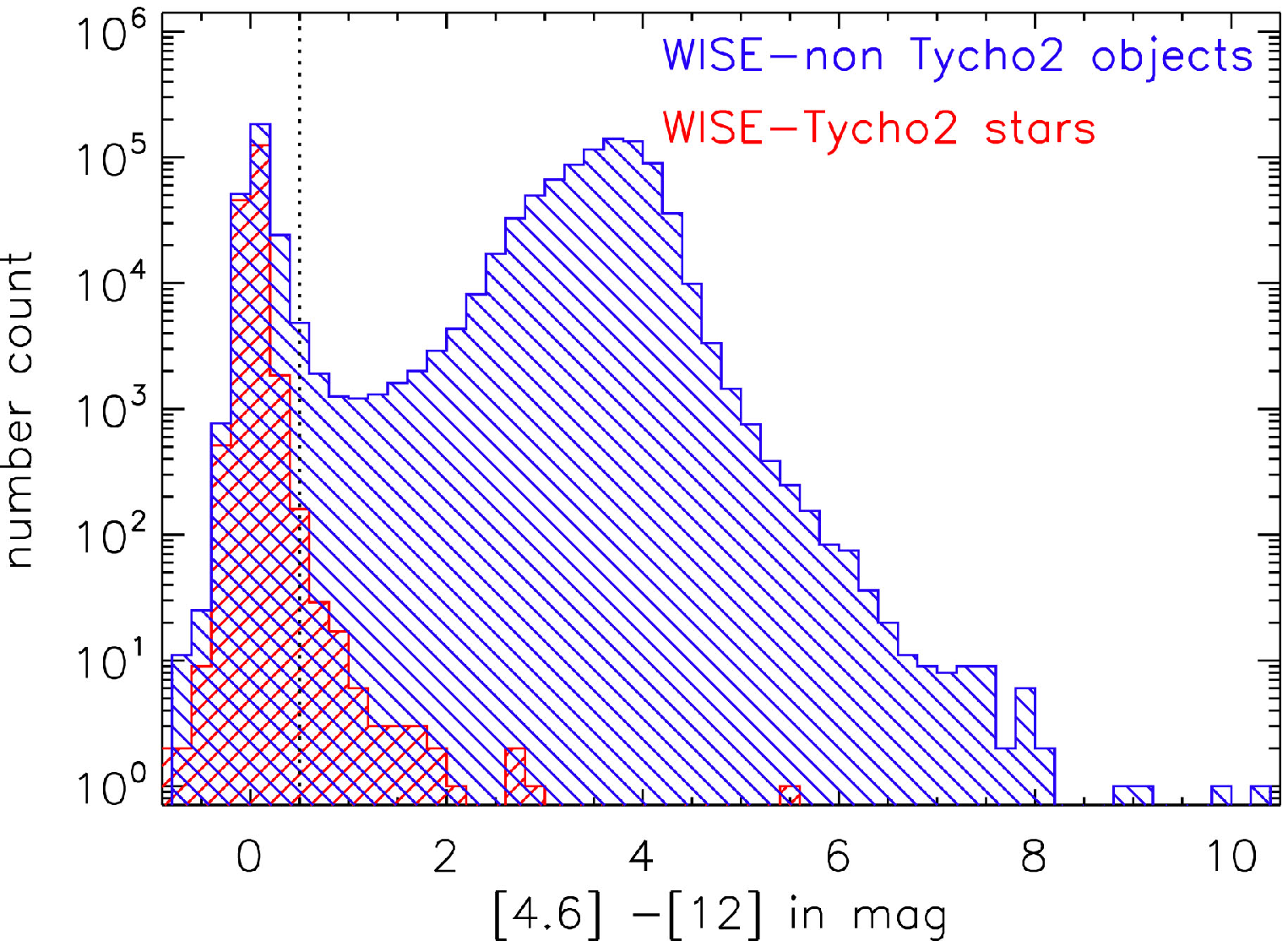}
        \caption{Color ([4.6] $-$ [12]) distribution of the {\it WISE} sources. The blue region represents {\it WISE-non Tycho 2} objects, and the red region represents {\it WISE-Tycho 2} stars. The dotted line indicates the threshold of star--galaxy separation. Objects with [4.6] $-$ [12] $\leq$ 0.5 are removed as stars.}
        \label{W2_3_distribution}
    \end{figure}

Of the 1,124,846 remaining sources (hereinafter {\it WISE-non Tycho 2} objects), we removed certain stars based on their colors.
Figure \ref{W2_3_distribution} shows a histogram of the [4.6] $-$ [12] color of the {\it WISE-non Tycho 2} objects.
Here, [4.6] and [12] represent the Vega magnitudes in the {\it WISE} 4.6- and 12-$\mu$m bands, respectively.
The zero magnitude flux density for 4.6 $\mu$m is 171.787 Jy.
To examine the color distribution of the stars, the {\it WISE-Tycho 2} stars are also plotted for comparison.
As shown in Figure \ref{W2_3_distribution}, the {\it WISE-Tycho 2} stars are located at [4.6] $-$ [12] $\sim$ 0 because the radiation from galactic stars is dominated by the Rayleigh--Jeans tail of the blackbody spectrum, thus yielding a Vega-system color near zero.
To ensure the reliability of the color value, we examined the color of objects with a signal-to-noise ratio (S/N) greater than 10, $w2/3sat = 0$, and $w2/3cc\_map =0$.
We found 263,417 objects that met the following criterion as stars:
\begin{equation}
[4.6] - [12] \leq 0.5\;.
\end{equation}
We note, however, that there are some stars in the 0.5 $<$ [4.6] $-$ [12] $<$ 1 region, but this region also may be populated with nearby elliptical galaxies with no dust.
Therefore, we adopted a robust criterion to avoid the omission of galaxies.
This left 861,429 objects in our {\it WISE} sample.

\subsubsection{SDSS sample}
\label{SDSS_sample}
The SDSS DR8 spectroscopic catalog includes the main galaxy \citep{Strauss}, luminous red galaxy (LRG)
\citep{Eisenstein}, and QSO \citep{Richards} samples.
The DR8 legacy spectroscopic survey catalog contains about 1.5 million sources and covers 7,966 deg$^2$. SDSS sources with legacyPrimary = 1 were selected from the {\tt SpecPhoto} table, created by joining spectroscopic and photometric information.
The ``legacyPrimary'' parameter is designed to choose the best available unique set of spectra of the legacy sources, and so the above criterion ensures clean spectroscopic data.
We focused on the main galaxy and QSO samples in this study; these samples are magnitude-limited objects
with \cite{Petrosian} magnitudes brighter than $r$ = 17.77 for the main galaxy sample and with point-spread function (PSF) magnitudes brighter than $i$ = 19.1 for the QSO samples at $z <$ 3.
The principal spectroscopic type (galaxy or QSO) is listed under the column headed ``CLASS'' in the {\tt SpecPhoto} table.
We thus extracted objects with (i) petroMag\_r below 17.77 mag and CLASS = ``GALAXY'' and (ii) psfMag\_i below 19.1 mag and CLASS = ``QSO''.
This process selected 683,071 SDSS objects.

\subsubsection{Cross-identification of WISE and SDSS}
We then cross-matched the {\it WISE} samples with the SDSS samples.
Using a matching radius of 3 arcsec, 259,969 {\it WISE}--SDSS
objects were selected as shown in Figure
\ref{cross-match_Tycho2_SDSS}. Adopting this search radius gives a
probability of chance coincidence of 0.05\%; \cite{Donoso}
cross-matched data from the {\it WISE} and SDSS DR7 spectroscopic
catalogs and estimated the expected false detection fraction at 3
arcsec by using random catalogs generated over the effective area.
This means that 861,429 $\times$ 0.0005 $\sim$ 431 sources may be
misidentified, which we regard as acceptable. Note that 111 {\it
WISE} sources have two SDSS counterparts because the spatial
resolution of the SDSS is better than that of {\it WISE}. These
sources are mostly close interacting systems of two members, and of
these, we chose the closest object as the optical counterpart.

\begin{figure}
        \epsscale{1}
        \plotone{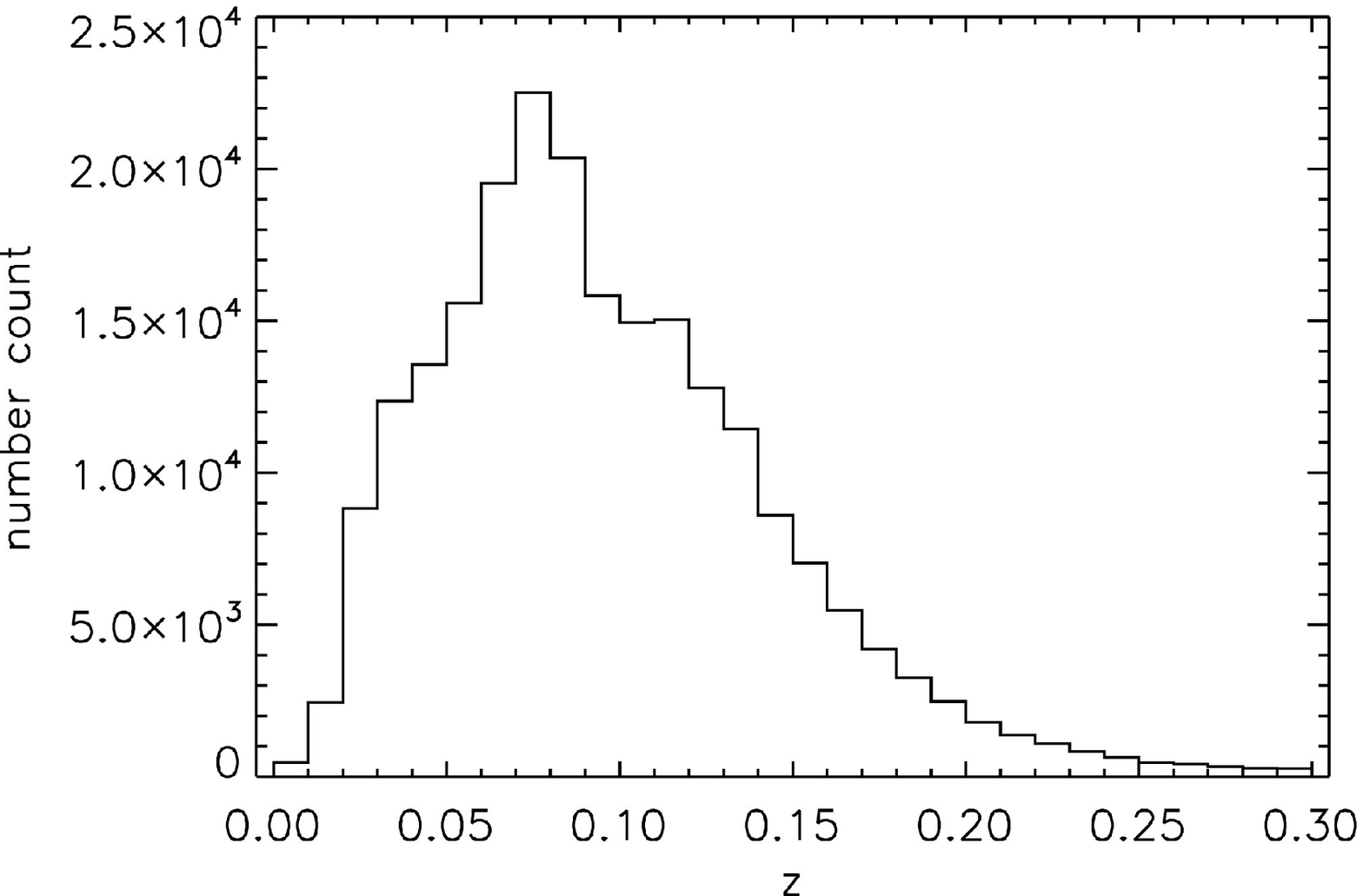}
        \caption{Redshift distribution of the 224,168 selected galaxies.}
        \label{z_dist}
    \end{figure}

>From these {\it WISE}-SDSS objects, we selected objects with {\it zWarning} = 0 and a redshift S/N greater than 10. The {\it zWarning} information is contained in the {\tt SpecPhoto} table and has flags set in suspicious cases; {\it zWarning} = 0 indicates that no problems were identified.
Among the objects with a precise estimation of the redshift, we extracted objects with 0.006 $\leq z \leq$ 0.3.
The redshift limit is applied because errors in the distance measurement are dominated by the peculiar motions of galaxies with $z \leq$ 0.006, and thus, the luminosity also has a large error.
However, for the sources at $z > 0.3$, the [NII] $\lambda \,$6583 and H$\alpha$ lines that were used for classifying the sample into several galaxy types (see Section \ref{Classification}) were shifted to around 9,000 \AA.
This wavelength almost corresponds exactly to the upper limit of the spectroscopy coverage and results in a relatively poor sensitivity. Therefore, we set the upper limit of the redshift to 0.3 to ensure a high S/N of these optical lines.
A final sample consisted of 224,168 galaxies, whose details are given in Table \ref{sample_list}.
The mean value of their redshifts is $\sim$0.1, and the redshift distribution is shown in Figure \ref{z_dist}.
Ultimately, 223,982 galaxies at 12 $\mu$m and 25,721 galaxies at 22 $\mu$m were selected through these steps.

For the selection process, we employed the ``2MASS Catalog Server Kit'' to easily construct a high-performance database server for the 2MASS Point Source Catalog (which includes 470,992,970 objects) and
several all-sky catalogs \citep{Yamauchi}.
We also used the STIL Tool Set, (STILTS\footnote{\url{http://www.star.bristol.ac.uk/~mbt/stilts/}})
which is a set of command-line tools based on the Starlink Tables Infrastructure Library \citep{Taylor}. The SDSS data were obtained from the Catalog Archive Server (CAS\footnote{\url{http://skyserver.sdss3.org/dr8/}}), a database containing catalogs of SDSS objects (both photometric and astrometric) that allows queries of their measured attributes.

\begin{deluxetable}{llrrrrcrrrrrr}
\rotate
\tabletypesize{\scriptsize}
\tablecaption{List of the 224,168 selected galaxies.\label{sample_list}}
\tablewidth{0pt}
\tablehead{
\colhead{objname} & \colhead{RA} & \colhead{DEC} & \colhead{f$_{12}$} & \colhead{f$_{22}$} & \colhead{redshift} & \colhead{type} & \colhead{w3sat} & \colhead{w4sat} & \colhead{S/N} & \colhead{S/N} & \colhead{w3cc\_map} & \colhead{w4cc\_map} \\
& \colhead{(J2000.0)} & \colhead{(J2000.0)} & \colhead{(mJy)} & \colhead{(mJy)} &  &  &  &  & (12 $\mu$m) & (22 $\mu$m) & &
}
\startdata
                          SDSS J000000.74-091320.2 &  00:00:00.73 &  -09:13:20.0 &   2.19 &   2.84 & 0.134 &         SF &0.0 & 0.0 &  14.90 &   1.10 &0.0 & 0.0 \\
                                      KUG 2357+156 &  00:00:01.98 &  +15:52:53.9 &   8.67 &  11.44 & 0.020 &         SF &0.0 & 0.0 &  27.00 &   6.40 &0.0 & 0.0 \\
                                       LCSB S0001P &  00:00:03.30 &  -10:43:15.8 &   3.37 &   5.61 & 0.083 &  Composite &0.0 & 0.0 &  33.93 &   5.68 &0.0 & 0.0 \\
                           2MASX J00000347+1411539 &  00:00:03.46 &  +14:11:53.5 &   2.09 &   3.92 & 0.115 &  Composite &0.0 & 0.0 &  13.30 &   1.90 &0.0 & 0.0 \\
                          SDSS J000004.59-105834.7 &  00:00:04.60 &  -10:58:35.0 &   1.74 &   2.79 & 0.150 &         SF &0.0 & 0.0 &  12.30 &   2.60 &0.0 & 0.0 \\
                           2MASX J00000472+0046546 &  00:00:04.74 &  +00:46:54.2 &   1.75 &   1.73 & 0.080 &        type 2 AGN &0.0 & 0.0 &  15.29 &   1.99 &0.0 & 0.0 \\
                                           ARK 591 &  00:00:07.83 &  -00:02:25.8 &   5.27 &   5.92 & 0.024 &         SF &0.0 & 0.0 &  41.76 &   6.39 &0.0 & 0.0 \\
                           2MASX J00000811+1432450 &  00:00:08.08 &  +14:32:45.9 &   4.45 &   9.06 & 0.105 &         SF &0.0 & 0.0 &  27.30 &   9.10 &0.0 & 0.0 \\
                           2MASX J00001235-1032114 &  00:00:12.34 &  -10:32:10.7 &   3.99 &   6.38 & 0.077 &         SF &0.0 & 0.0 &  24.60 &   6.80 &0.0 & 0.0 \\
                                      CGCG 382-016 &  00:00:12.78 &  +01:07:13.1 &  15.31 &  19.62 & 0.025 &         SF &0.0 & 0.0 &  67.86 &  13.92 &0.0 & 0.0 \\
                          SDSS J000013.84+003912.1 &  00:00:13.86 &  +00:39:12.0 &   1.95 &   2.31 & 0.103 &         SF &0.0 & 0.0 &  12.30 &   2.20 &0.0 & 0.0 \\
                           2MASX J00001447+1412420 &  00:00:14.44 &  +14:12:42.0 &   1.79 &   4.37 & 0.091 &      LINER &0.0 & 0.0 &  11.40 &   2.00 &0.0 & 0.0 \\
                           2MASX J00001575-0853283 &  00:00:15.78 &  -08:53:27.3 &   4.81 &   7.13 & 0.056 &  Composite &0.0 & 0.0 &  18.20 &   3.70 &0.0 & 0.0 \\
                                      KUG 2357+144 &  00:00:16.31 &  +14:43:59.8 &   4.50 &   5.73 & 0.091 &         SF &0.0 & 0.0 &  27.00 &   6.40 &0.0 & 0.0 \\
                           2MASX J00001671+1541400 &  00:00:16.75 &  +15:41:40.4 &   4.42 &   5.34 & 0.112 &         SF &0.0 & 0.0 &  27.60 &   5.50 &0.0 & 0.0 \\
                          SDSS J000018.63+154327.7 &  00:00:18.62 &  +15:43:27.9 &   1.44 &   3.92 & 0.176 &         SF &0.0 & 0.0 &  10.50 &   1.90 &0.0 & 0.0 \\
                          SDSS J000019.03-105258.9 &  00:00:19.03 &  -10:52:58.7 &   2.71 &   4.92 & 0.083 &         SF &0.0 & 0.0 &  16.90 &   4.30 &0.0 & 0.0 \\
                          SDSS J000019.89+142219.5 &  00:00:19.87 &  +14:22:19.8 &   1.81 &   2.91 & 0.094 &         SF &0.0 & 0.0 &  12.10 &   1.00 &0.0 & 0.0 \\
                          SDSS J000020.06+135001.6 &  00:00:20.04 &  +13:50:01.9 &   1.90 &   3.09 & 0.079 &         SF &0.0 & 0.0 &  13.00 &   1.10 &0.0 & 0.0 \\
                          SDSS J000020.93+001254.2 &  00:00:20.92 &  +00:12:54.4 &   2.13 &   3.98 & 0.085 &         SF &0.0 & 0.0 &  14.40 &   4.20 &0.0 & 0.0 \\
                           2MASX J00002629+0035503 &  00:00:26.31 &  +00:35:51.1 &   1.62 &   4.90 & 0.104 &        type 2 AGN &0.0 & 0.0 &  10.50 &   4.30 &0.0 & 0.0 \\
                          SDSS J000027.65+145624.6 &  00:00:27.65 &  +14:56:25.0 &   4.12 &  11.94 & 0.159 &      LINER &0.0 & 0.0 &  25.00 &  11.10 &0.0 & 0.0 \\
                           2MASX J00002809+1422530 &  00:00:28.07 &  +14:22:52.6 &   1.76 &   3.37 & 0.093 &  Composite &0.0 & 0.0 &  11.70 &   1.00 &0.0 & 0.0 \\
                          SDSS J000028.19+142509.8 &  00:00:28.19 &  +14:25:10.0 &   1.67 &   2.62 & 0.143 &         SF &0.0 & 0.0 &  11.30 &   0.70 &0.0 & 0.0 \\
                          SDSS J000030.00-103825.0 &  00:00:29.94 &  -10:38:24.7 &   2.50 &   2.89 & 0.151 &         SF &0.0 & 0.0 &  16.50 &   2.70 &0.0 & 0.0 \\
                           2MASX J00003086-0112473 &  00:00:30.89 &  -01:12:46.8 &   1.02 &   2.30 & 0.075 &         SF &0.0 & 0.0 &  10.54 &   2.74 &0.0 & 0.0 \\
                           2MASX J00003718-1102077 &  00:00:37.18 &  -11:02:07.8 &   3.94 &   9.04 & 0.151 &        type 2 AGN  &0.0 & 0.0 &  23.40 &   8.70 &0.0 & 0.0 \\
                          SDSS J000038.68+143548.1 &  00:00:38.68 &  +14:35:48.0 &   1.62 &   2.28 & 0.146 &         SF &0.0 & 0.0 &  12.30 &   2.60 &0.0 & 0.0 \\
                           2MASX J00003878+1524270 &  00:00:38.72 &  +15:24:27.3 &   1.63 &   1.88 & 0.152 &  Composite &0.0 & 0.0 &  11.80 &   0.00 &0.0 & 0.0 \\
\enddata
\tablecomments{Table \ref{sample_list} is published in its entirety in the electronic edition of the {\it Astrophysical Journal}.  A portion of the table is shown here for guidance regarding its form and content. There are 223,982 galaxies at 12 $\mu$m defined by a flux of (12 $\mu$m) $\geq$ 9.0 mJy, {\it w3sat} = 0.0,  {\it w3cc\_map} = 0.0, and an S/N of (12 $\mu$m) $\geq$ 10. There are 25,721 galaxies at 22 $\mu$m with a flux of (22 $\mu$m) $\geq$ 0.9 mJy, {\it w4sat} = 0.0, {\it w4cc\_map} = 0.0, and an S/N of (22 $\mu$m) $\geq$ 10.}
\end{deluxetable}



\subsection{Classification of Spectroscopic Galaxy Type}
\label{Classification}
    \begin{figure}
        \epsscale{0.8}
        \plotone{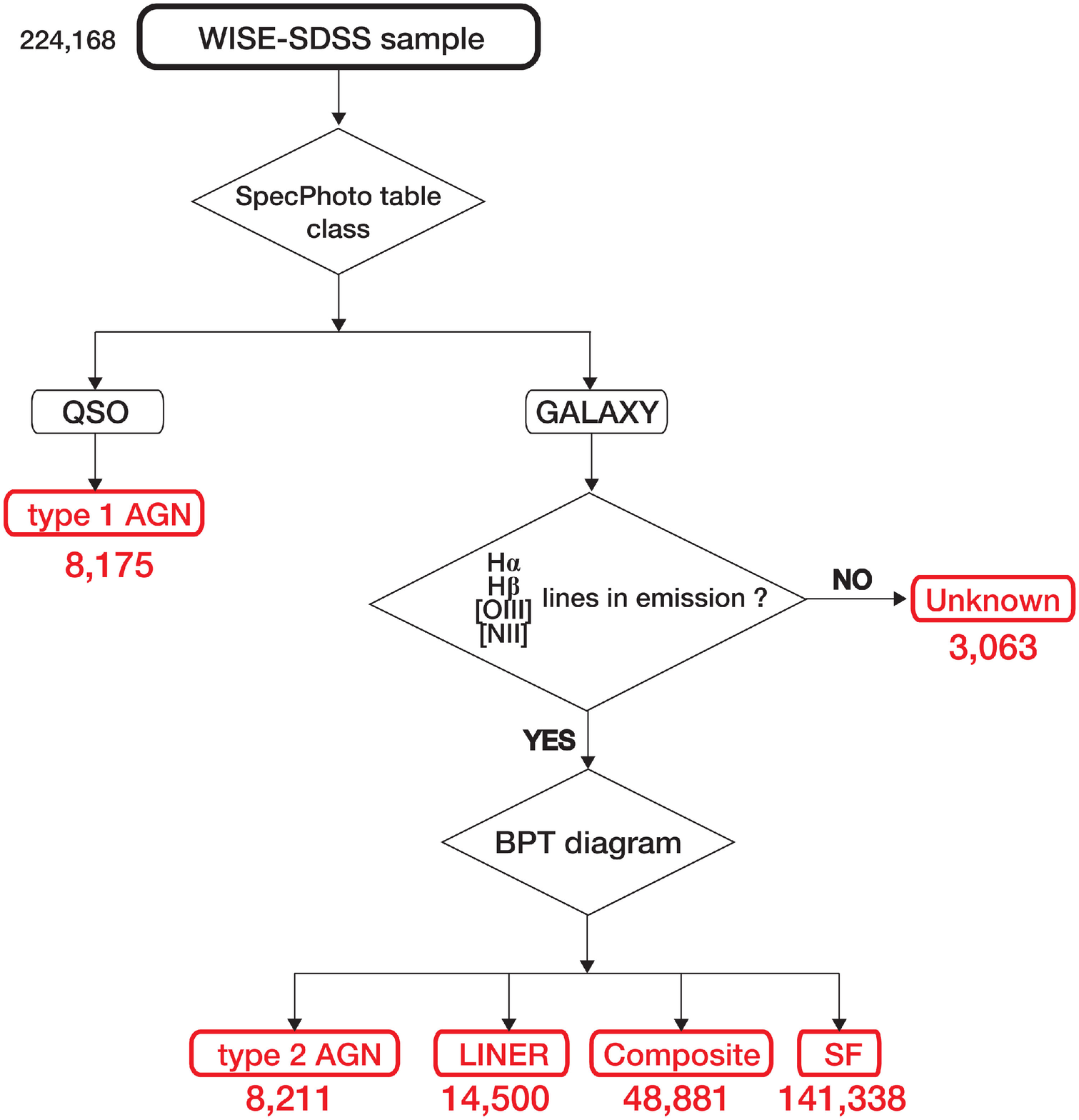}
        \caption{Outline of the type classification process.}
        \label{flow_chart_type_classification}
    \end{figure}

We spectroscopically classified the 224,168 galaxies (hereinafter
{\it WISE}-SDSS sample) into five types: type 1 AGNs including
quasars and Seyfert 1 galaxies, type 2 AGNs, low-ionization
narrow-emission-line region galaxies (LINER), galaxies that are
likely to contain both star formation and AGN activity (composite
types of galaxies, hereinafter ``Composite''), and star-forming
galaxies (SF). The classification was based on the spectroscopic
information in the {\tt SpecPhoto} table, as shown in Figure
\ref{flow_chart_type_classification}. Note that we consider Seyfert
2 galaxies (Sy2) as type 2 AGNs unless otherwise noted.

\begin{figure}
        \epsscale{0.8}
        \plotone{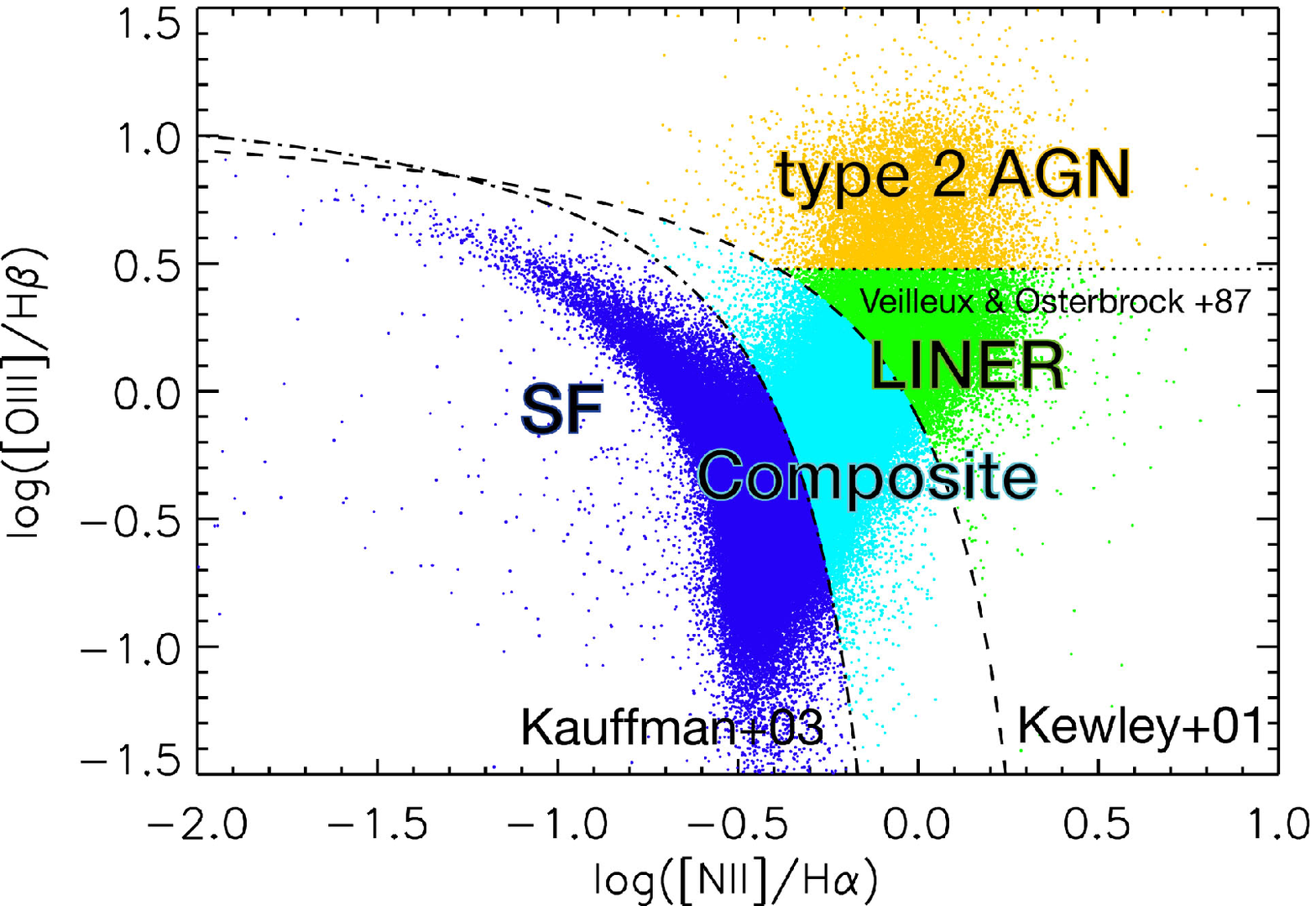}
        \caption{BPT diagram of the emission-line flux ratio [NII]/H$\alpha$ versus [OIII]/H$\beta$ for all the narrow-line galaxies for which line flux information is available. The dashed-dotted line is the criterion given by \cite{Kauffmann}, the dashed line is the criterion given by \cite{Kewley}, and the dotted line is the traditional scheme \citep[see for example,][]{Veilleux}.}
        \label{BPT}
    \end{figure}

The type 1 AGNs were identified according to the CLASS entry (CLASS = QSO) in the {\tt SpecPhoto} table. Objects for which CLASS = GALAXY were classified as type 2 AGNs, LINER, Composite, or SF by using the optical flux line ratios of [NII] $\lambda \,$6583/H$\alpha$ versus [OIII] $\lambda \,$5007/H$\beta$ \citep[BPT diagram suggested by][]{Baldwin}, as shown in Figure \ref{BPT}.
However, the BPT diagram is not able to classify a galaxies if H$\alpha$, H$\beta$, [OIII], or [NII] were not detected in the emission line.
These galaxies are classified as weak-emission-line galaxies (hereinafter, ``Unknown'').

\begin{deluxetable}{lrr}
\tablecolumns{3}
\tablewidth{0pc}
\tabletypesize{\scriptsize}
\tablecaption{Classifications of the 223,982 galaxies for the 12-$\mu$m LF and 25,721 galaxies for the 22-$\mu$m LF.\label{type_classification}}
\tablehead{
\colhead{type} & \colhead{12 $\mu$m (percentage)} & \colhead{22 $\mu$m (percentage)}
}
\startdata
			type 1 AGNs			&	  8,151 ( 3.6 \%)	&	 2,846 (11.0 \%)\\
			type 2 AGNs			&	  8,204 ( 3.7 \%) 	&	 1,837 ( 7.1 \%)\\
			LINER				&	 14,491 ( 6.5 \%) 	&	 1,477 ( 5.8 \%)\\
			Composite			&	 48,834 (21.8 \%)	&	 6,583 (25.6 \%)\\
			SF					&	141,242 (63.0 \%)	&	12,799 (49.8 \%)\\	
			Unknown				&	  3,060 ( 1.4 \%)	&	   179 ( 0.7 \%)\\	
\cline{1-3}
			All					&	223,982 (100 \%)	&	25,721 (100 \%)\\	
\enddata
\end{deluxetable}

\begin{figure}
        \epsscale{1}
         \plottwo{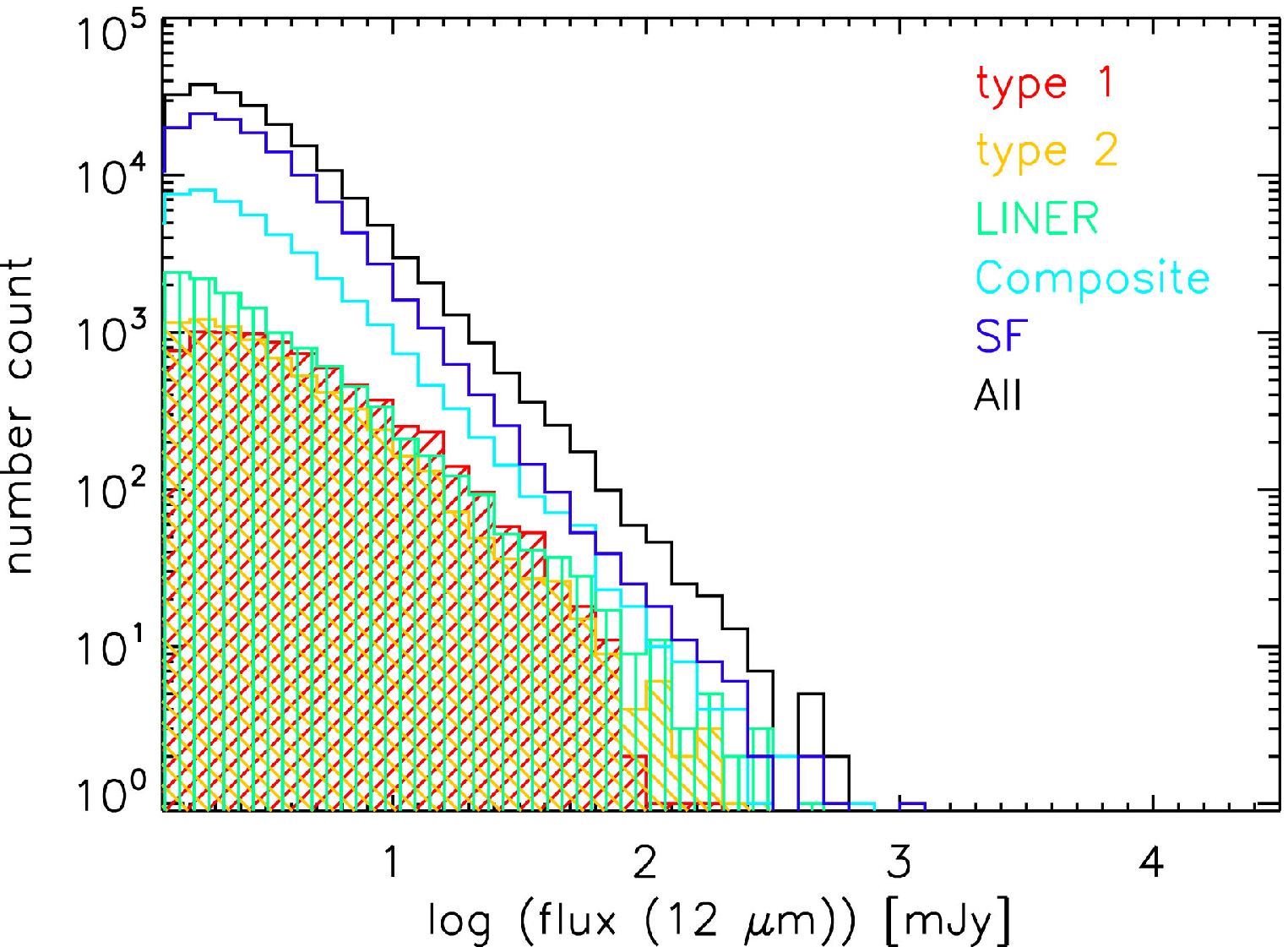}{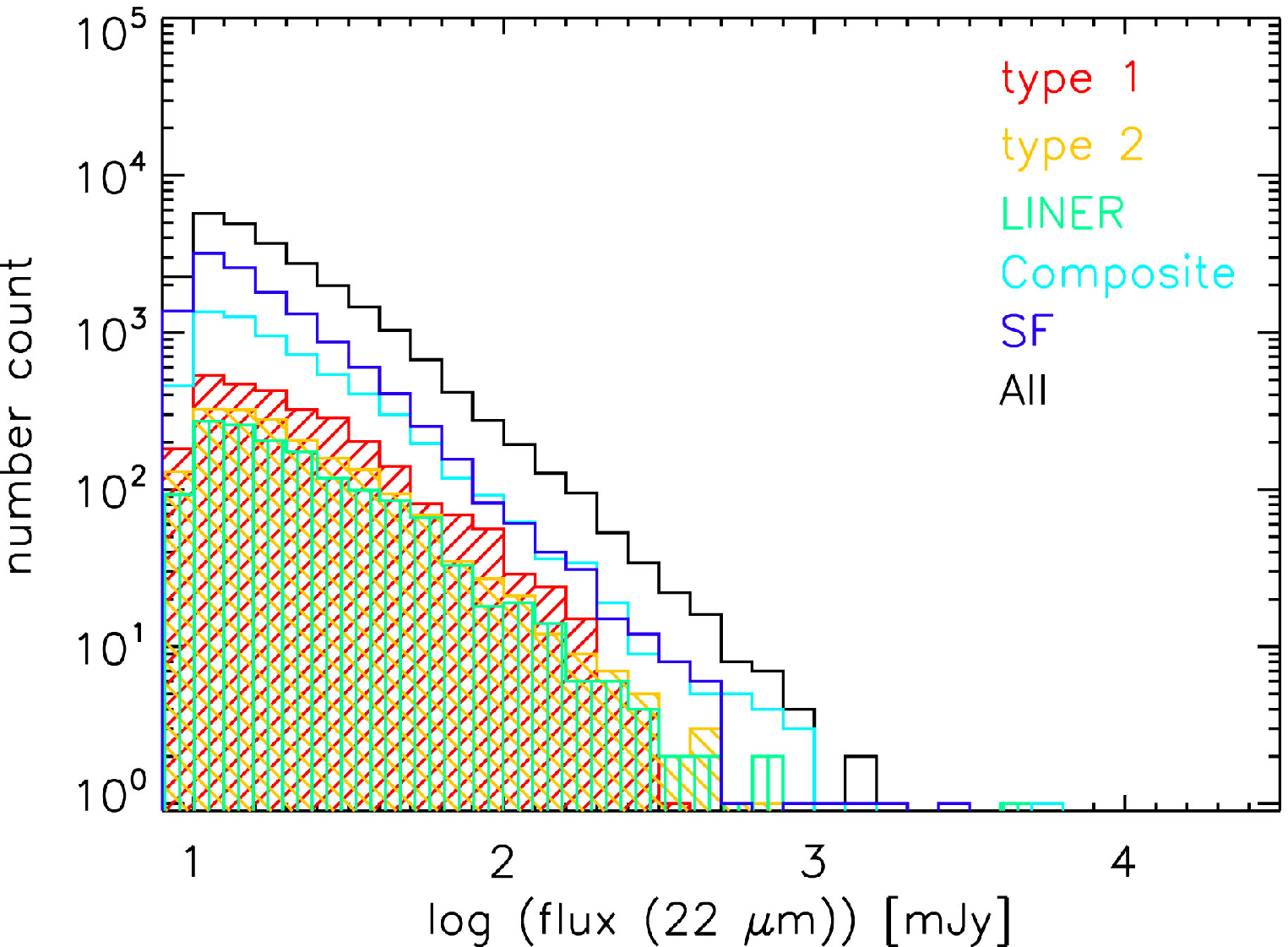}
        \caption{Flux distribution for each galaxy type at 12 $\mu$m (left) and 22 $\mu$m (right).}
        \label{flux}
    \end{figure}

\begin{figure}
        \epsscale{1}
        \plotone{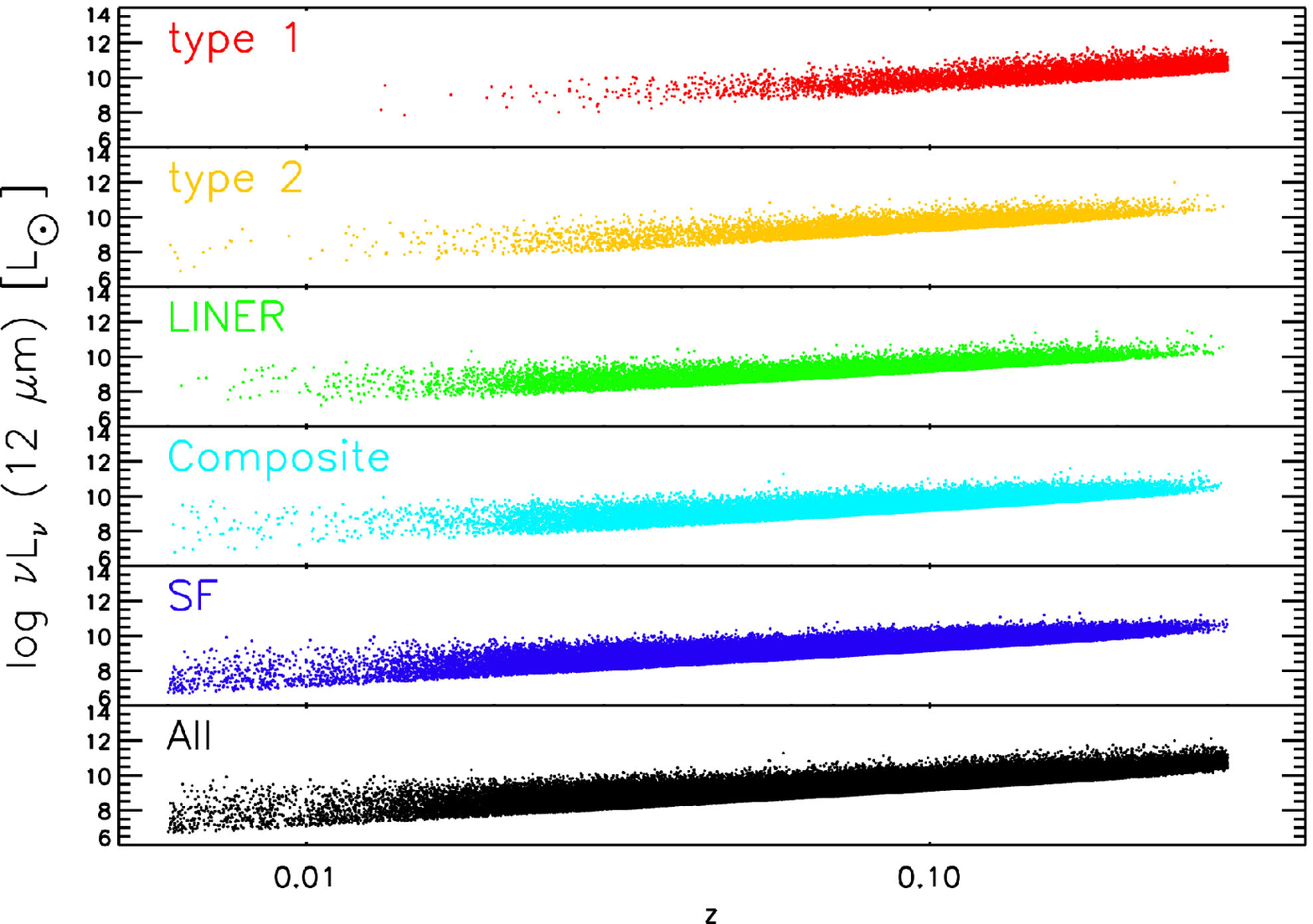}
        \caption{The 12-$\mu$m luminosities as a function of redshift for each galaxy type.}
        \label{z_L_12}
    \end{figure}

\begin{figure}
        \epsscale{1}
        \plotone{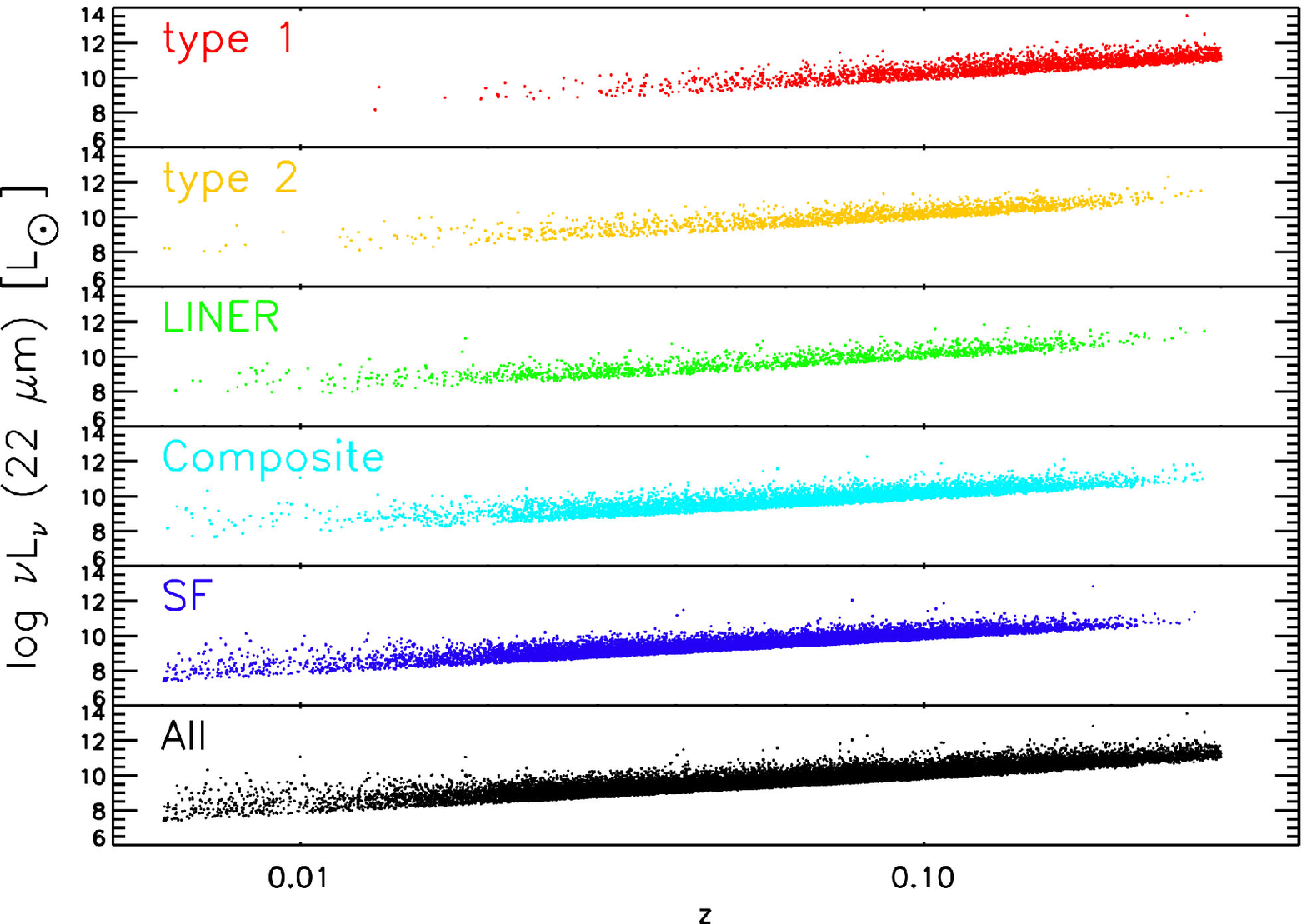}
        \caption{The 22-$\mu$m luminosities as a function of redshift for each galaxy type.}
        \label{z_L_22}
    \end{figure}

The galaxy classifications, summarized in Table
\ref{type_classification}, indicate that the 22-$\mu$m band is
especially powerful for finding AGNs. The detection rate of AGNs
(type 1 + type 2) in the 22-$\mu$m band ($\sim$18\%) is higher than
that in the 12-$\mu$m band ($\sim$7\%), since the 12-$\mu$m bandpass
includes a strong contribution from polycyclic aromatic hydrocarbon
(PAH) emission, which is unrelated to the presence of an active
nucleus.
Some authors have reported that LINERs typically have weak MIR emission \citep[e.g.,][]{Ogle}.
Thus it might be surprising that we find such a high fraction of LINERs ($\sim$6\%).  
However, our LINER classification is based on the [NII] lines rather, than the [OI] $\lambda \,$6300 or [SII] $\lambda \,$6716 and $\lambda \,$6731 lines 
others have found useful for discriminating pure LINERS \citep[e.g.,][]{Kewley_06}.
If we were to adopt the [OI]/H$\alpha$ versus [OIII]/H$\beta$ diagram, it would in fact reduce our fraction of LINERs ($\sim$ 3.5\% at 22 $\mu$m). However, as our focus in this study is on type 1 and 2 AGNs and one motivation is to check whether our findings in \cite{Toba} based on {\it AKARI} are confirmed by {\it WISE}, we employed [NII]/H$\alpha$ versus [OIII]/H$\beta$ in the same manner as \cite{Toba}.
Figure \ref{flux} presents the flux distributions at 12 and
22 $\mu$m, and the distributions of 12- and 22-$\mu$m luminosities
as a function of redshift are illustrated in Figures \ref{z_L_12}
and \ref{z_L_22}, respectively. The flux distribution does not
reveal any clear differences between each galaxy type, and we see in
the redshift distributions that type 1 and type 2 AGNs have
relatively higher redshifts.


\subsection{Derivation of Luminosity Function with 1/V$_{\mathrm{max}}$ method}
\label{Vmax_method}
Here, we derive the LFs following the 1/$V_{\mathrm{max}}$ method described by \cite{Schmidt}.
The advantage of the 1/$V_{\mathrm{max}}$ method is that it allows us to compute the LF directly from the data; no parameter dependence or model assumptions are needed.
The volume density $\phi (L)$ and its uncertainty $\sigma_{\phi (L)}$ are derived using the expressions:

\begin{equation}
\phi(L) = \sum_i^N \frac{1}{V_{max,i}}\;,
\end{equation}

\begin{equation}
\sigma_{\phi(L)} =\sqrt{\sum_i^N \frac{1}{V_{\mathrm{max},i}^2}}\;,
\end{equation}
where $V_{\mathrm{max}}$ is the maximum co-moving volume that would be enclosed at the maximum redshift at which the $i$th object could be detected.
In the context of the cosmology we adopt, $V_{\mathrm{max}}$ is
\begin{equation}
V_{\mathrm{max}}(z)  =  \frac{c}{H_0} \int_{\Omega} \int_{z_{\mathrm{min}}}^{z_{\mathrm{max}}} \frac{(1+z^{\prime})^2 D_A^2}{\sqrt{\Omega_M (1+z^{\prime})^3 + \Omega_{\Lambda}}} \mathrm{d}z^{\prime} \mathrm{d}\Omega,
\end{equation}
where $D_A$ is the angular distance for a given redshift in our adopted cosmology, $\Omega$ is the solid angle of the SDSS DR8 spectroscopic region (7966 deg$^2$ $\sim$ 2.43 str), $z_{\mathrm{min}}$ is the lower limit of the redshift bin considered, and $z_{\mathrm{max}}$ is the maximum redshift at which the object could be seen given the flux limit of the sample.
Note that when $z_{\mathrm{max}}$ is smaller than the maximum of the redshift bin considered, $V_{\mathrm{max}} = V(z_{\mathrm{max}}) - V(z_{\mathrm{min}})$.
Otherwise, $V_{\mathrm{max}}$ is equal to the volume corresponding to the bin considered, $V_{\mathrm{max}} = V_{\mathrm{bin}}$.
However, as $z_{\mathrm{max}}$ cannot be determined analytically, we calculated $z_{\mathrm{max}}$
numerically, using the following procedure.

The absolute magnitude (hereafter, we use the magnitude for descriptive purposes) of the object $M$ observed to have an apparent magnitude $m$ at a redshift $z$ is
\begin{equation}
\label{M_m} M = m - K(z) -5\log d_L(z) -25\;,
\end{equation}
where $d_L$ is the luminosity distance (measured in Mpc) for a given redshift in our adopted cosmology, and $K(z)$ is the K-correction term, which is the redshift dependence of the magnitude of any object in a given wavelength band.
When a source is artificially moved to the detection limit $m(z_{\mathrm{max}}) = m_{\mathrm{min}}$, Equation (\ref{M_m}) becomes
\begin{equation}
\label{M_m_max} M = m_{\mathrm{min}} - K(z_{\mathrm{max}}) -5 \log
d_{L_\mathrm{max}} -25\;,
\end{equation}
where $d_{L_{\mathrm{max}}}$ is the maximum luminosity distance $d_L (z_{\mathrm{max}})$.
Therefore, we numerically estimate $z_{\mathrm{max}}$ by substituting $z$ into Equations (\ref{M_m}) and (\ref{M_m_max}) step-wise in steps of $\Delta z$ (= $10^{-5}$ here) and iterating the above approach until the difference between the $M$ values obtained from Equations (\ref{M_m}) and (\ref{M_m_max}) is minimized.

Note that when we calculate $z_{\mathrm{max}}$, the difference in the detection limits of the {\it WISE} and SDSS surveys should be considered because our {\it WISE}--SDSS sample is flux- (or magnitude-) limited. For the {\it WISE} samples, the detection limit is 0.9 mJy at 12 $\mu$m and 9.0 mJy at 22 $\mu$m.
For the SDSS samples, the detection limit is 17.77 Petrosian r-band magnitude for galaxies and 19.10 PSF i-band magnitude for type 1 AGNs.
Therefore, we calculated two values of $z_{\mathrm{max}}$ for each survey considering these detection limits, and we adopted the smaller of two possible values in each case.
To compute the maximum redshift for the {\it WISE} objects, $K(z)$ in Equation (\ref{M_m}) was derived from the assumption that the spectral energy distribution (SED) of the objects in the IR region obeys a simple power law of $f(\nu) \propto \nu^{-\alpha}$, i.e.,
\begin{equation}
K_{\mathrm{WISE}} (z) = 2.5 (\alpha -1 ) \log (1 + z)\;,
\end{equation}
where the spectral index $\alpha$ is calculated using the 12- and 22-$\micron$ fluxes ($f_{12}$ and $f_{22}$, respectively) as
\begin{equation}
\label{alpha} \alpha  =  -\frac{\log \left(\frac{f_{22}}{f_{12}}
\right)}{\log \left( \frac{\nu_{22}}{\nu_{12}} \right)}\;.
\end{equation}
The frequencies at 12 and 22 $\mu$m, are $\nu_{12}$ and $\nu_{22}$, respectively. In the case of SDSS galaxies, $K(z)$ in Equation (\ref{M_m}) was computed using the K-correct (ver. 4.2) software of
\cite{Blanton}. We also assumed a power law for SDSS type 1 AGNs, with $\alpha$ calculated as
\begin{equation}
\alpha  =  -\frac{\log \left(\frac{f_r}{f_i} \right)}{\log \left(
\frac{\nu_r}{\nu_i} \right)}\;,
\end{equation}
where $f_r$ and $f_i$ are the PSF fluxes in $r$ and $i$ bands, respectively, as cataloged in the {\tt SpecPhoto} table.
The corresponding frequencies are $\nu_r$ and $\nu_i$.
Finally, for whichever maximum redshift is smaller, we adopted a maximum co-moving volume for each object $i$ of
\begin{equation}
V_{\mathrm{max},i} = \mathrm{min}
[V_{\mathrm{max},i}\,\mathrm{(WISE)},
V_{\mathrm{max},i}\,\mathrm{(SDSS)}]\;,
\end{equation}
where $V_{\mathrm{max}}$(WISE) and $V_{\mathrm{max}}$(SDSS) are obtained using $z_{\mathrm{max}}$(WISE) and
$z_{\mathrm{max}}$(SDSS), respectively.

\section{RESULTS}
We present the MIR LFs from the {\it WISE}--SDSS sample, classified
as discussed in Section \ref{Classification}, and show that the
shapes of each LF are in agreement with previous studies. We then
show each LF in different redshift bins, which indicate that AGNs
(particularly type 1 AGNs) show a certain evolution compared to
normal galaxies.

\subsection{The 12- and 22-$\micron$ Luminosity Functions}\label{LF}
\begin{figure}
       \epsscale{1}
        \plottwo{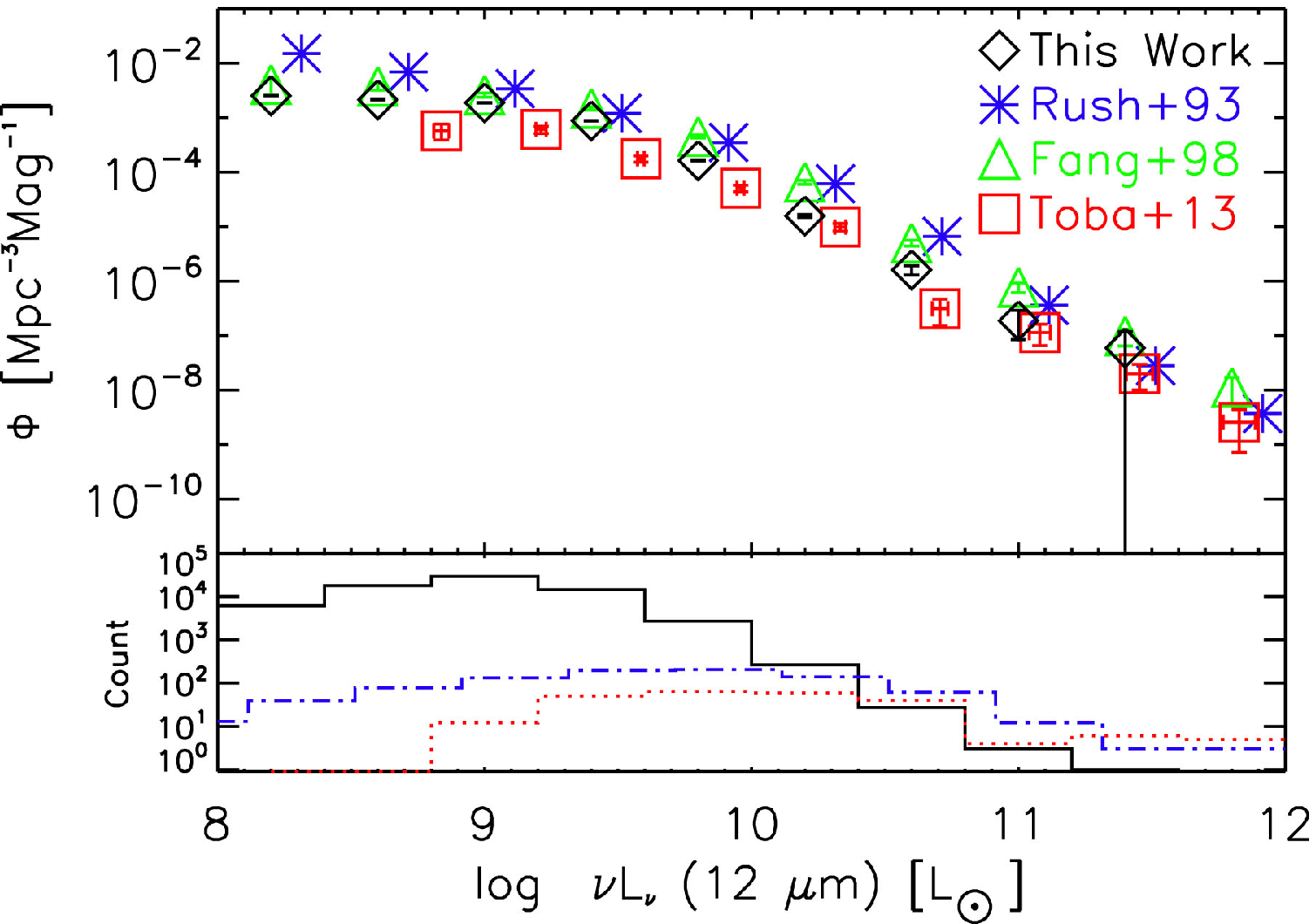}{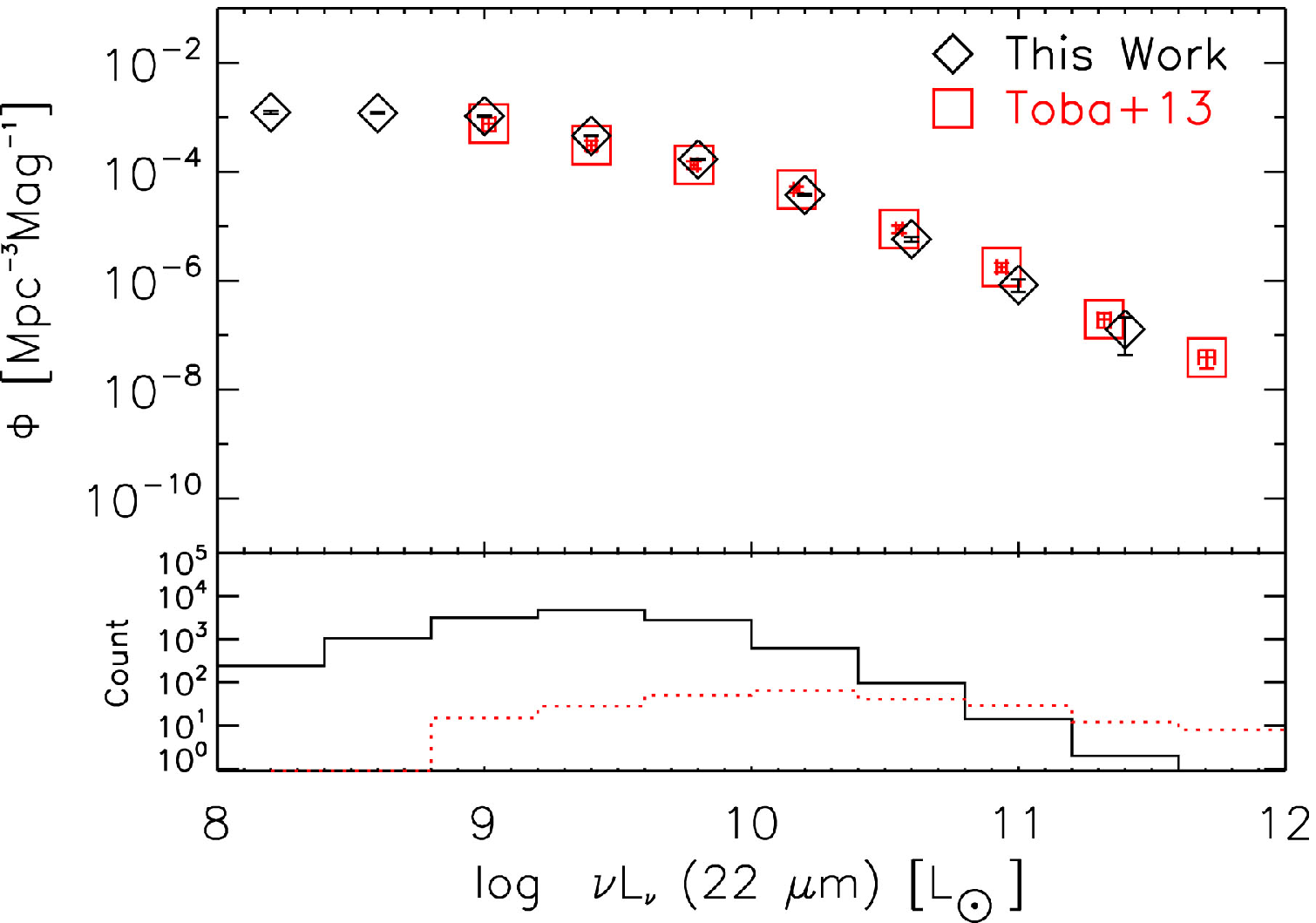}
        \caption{The 12- (left) and 22- (right) $\mu$m LFs for all galaxies for 0.006 $\leq z \leq$ 0.07. The 12-$\mu$m LFs from \cite{Rush} and \cite{Fang} and the 9- and 18-$\mu$m LFs from \cite{Toba} are also plotted for comparison. For the 9- and 18-$\mu$m LFs, we converted $\nu L_\nu (9, 18\, \micron)$ to $\nu L_\nu (12, 22\, \micron)$. The vertical error bars are calculated from the Poisson statistical uncertainty, and the horizontal error bars represent the uncertainty of the conversion to $\nu L_\nu (12, 22\, \micron)$.}
    \label{LF_ALL}
    \end{figure}

\begin{figure}
        \epsscale{1}
         \plottwo{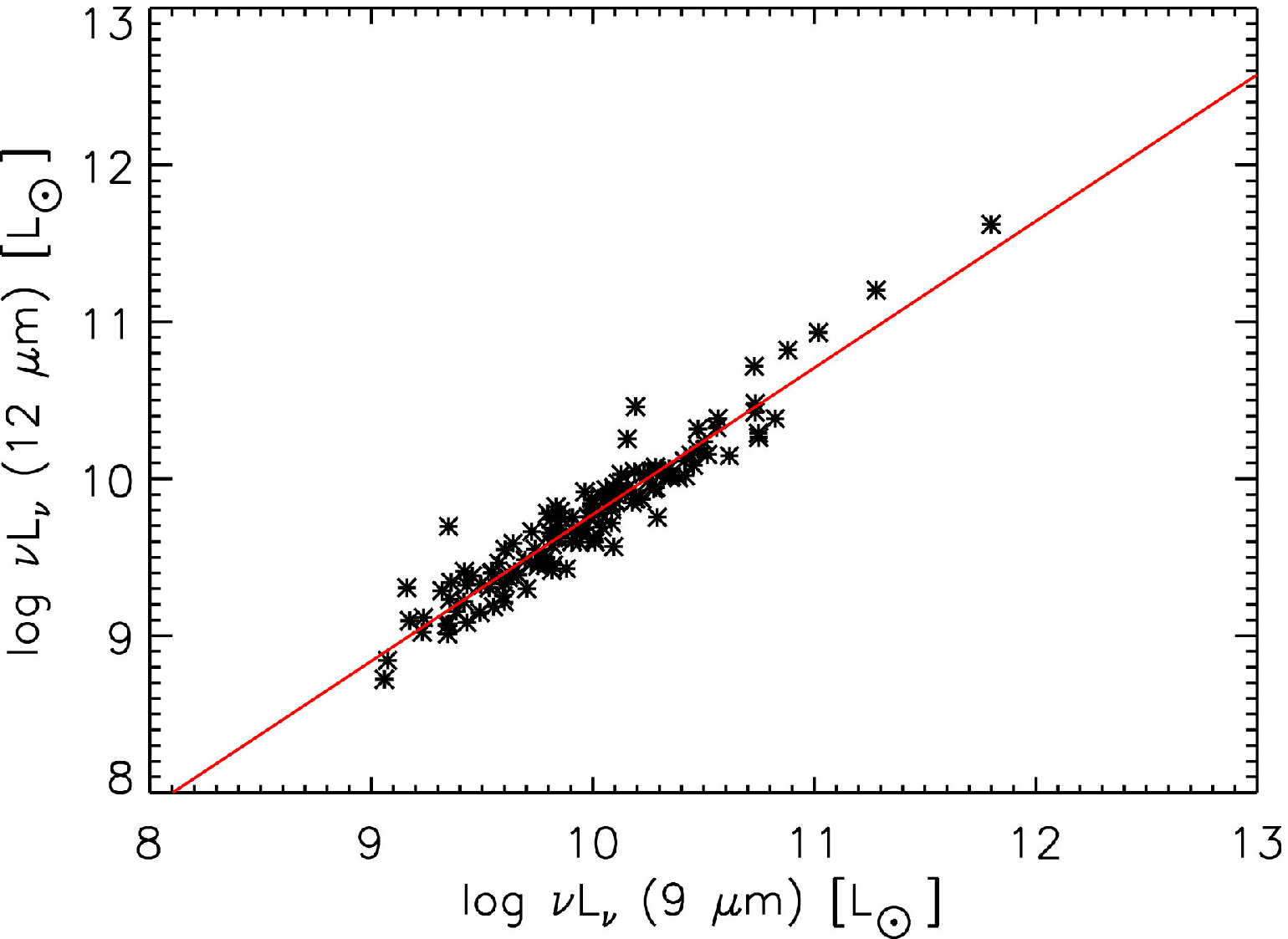}{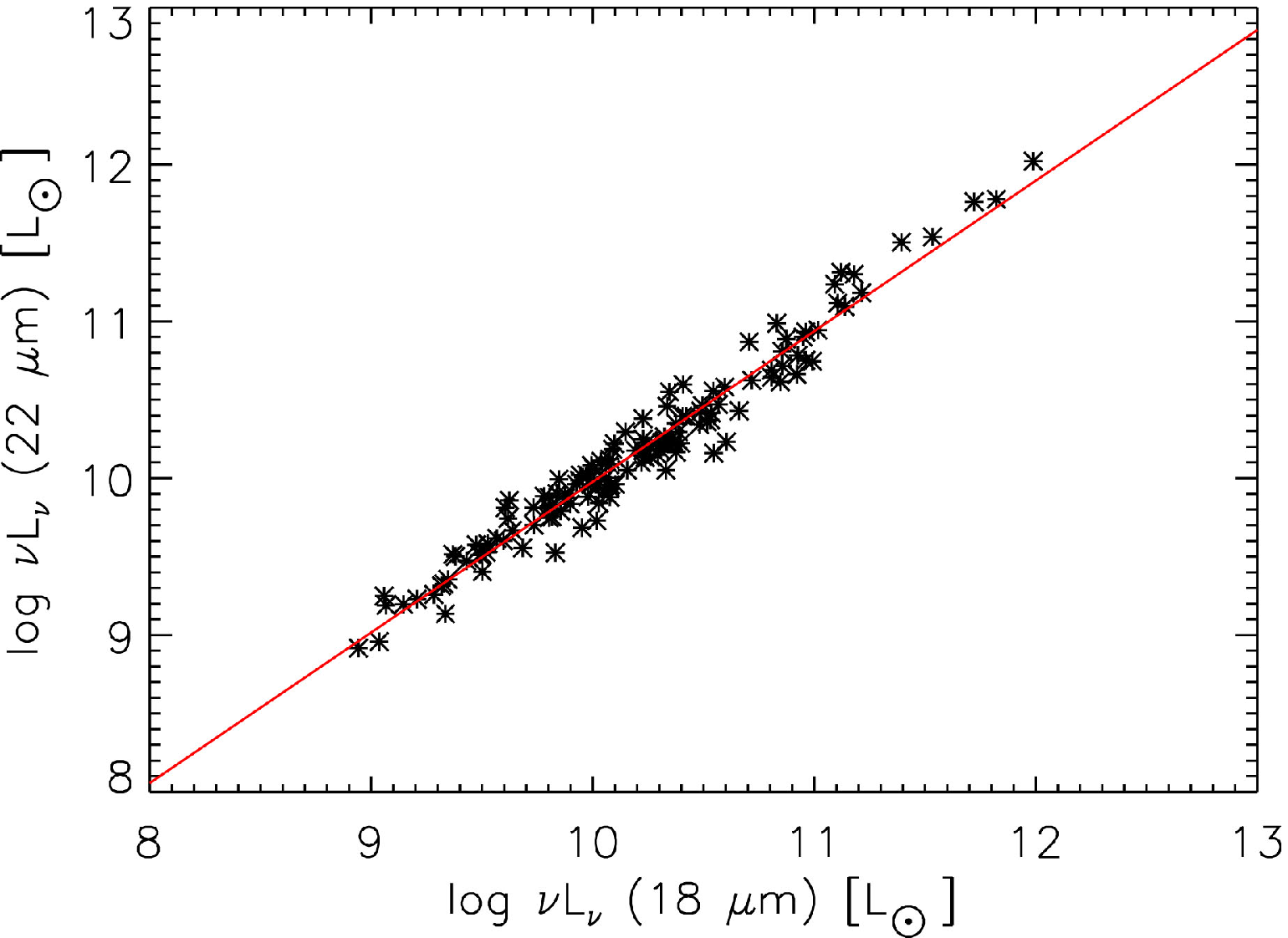}
\caption{{\it WISE} 12-$\mu$m versus {\it AKARI} 9-$\mu$m luminosities (left) and {\it WISE} 22-$\mu$m versus {\it AKARI} 18-$\mu$m luminosities (right). The red dotted line shows the best-fit linear function.}
        \label{Convert_nuLnu}
    \end{figure}

The rest-frame 12- and 22-$\mu$m LFs (i.e., the volume density of the galaxies per unit absolute magnitude range) of our {\it WISE}--SDSS galaxies at 0.006 $\leq z \leq$ 0.07, computed with the 1/$V_{\mathrm{max}}$ method, are shown in Figure \ref{LF_ALL}.
We confirm here the consistency between the LFs of {\it WISE} and {\it AKARI} \citep[e.g.,][]{Toba}.
\cite{Toba} selected 243 galaxies at 9 $\mu$m and 255 galaxies at 18 $\mu$m from the {\it AKARI} MIR all-sky survey catalog, and by combining the {\it AKARI} data with the SDSS DR7 spectroscopic data, they constructed 9- and 18-$\mu$m LFs for the first time.
To compare those LFs with ours in a similar redshift range, Figure \ref{LF_ALL} plots only local objects (0.006 $\leq z \leq$ 0.07).
Within this redshift range, the average value of the redshift ($\sim$0.04) is equal to that of \cite{Toba}.
Figure \ref{LF_ALL} also shows the {\it IRAS} 12-$\mu$m LFs \citep{Rush,Fang} for comparison.
These were derived from samples of 893 \citep{Rush} and 668 \citep{Fang} galaxies selected from the {\it IRAS} Faint Source Survey.
\cite{Fang}, in particular, corrected for the peculiar motion of the local supercluster.
Note that for the 9- and 18-$\mu$m LFs we first converted the data by cross-identifying our {\it WISE}--SDSS sample with the {\it AKARI} MIR all-sky survey catalog, selecting the 200 {\it WISE}--SDSS-{\it AKARI} sources within the 3-arcsec search radius, and calculating conversion factors by plotting $\nu L_{\nu}$(9 $\micron$) versus $\nu L_{\nu}$(12 $\micron$) and $\nu L_{\nu}$(18 $\micron$) versus $\nu L_{\nu}$(22 $\micron$), as shown in Figure \ref{Convert_nuLnu}.
We obtained the following conversion formulae:
\begin{eqnarray}
\log[\nu L_{\nu} (12 \, \micron)] & = & (0.93 \pm 0.03) \times \log[\nu L_{\nu}(9 \,\micron)] + (0.43 \pm 0.26)\;, \\
\log[\nu L_{\nu} (22 \, \micron)] & = & (0.96 \pm 0.02) \times \log[\nu L_{\nu} (18 \,\micron)] + (0.37 \pm 0.18)\;.
\end{eqnarray}
The conversion uncertainty is represented in Figure \ref{LF_ALL} as the horizontal error bars.
The shapes of the LFs obtained from previous studies \citep{Rush,Fang,Toba} are in good agreement with our derived LFs.

\begin{figure}
    \epsscale{1}
    \plottwo{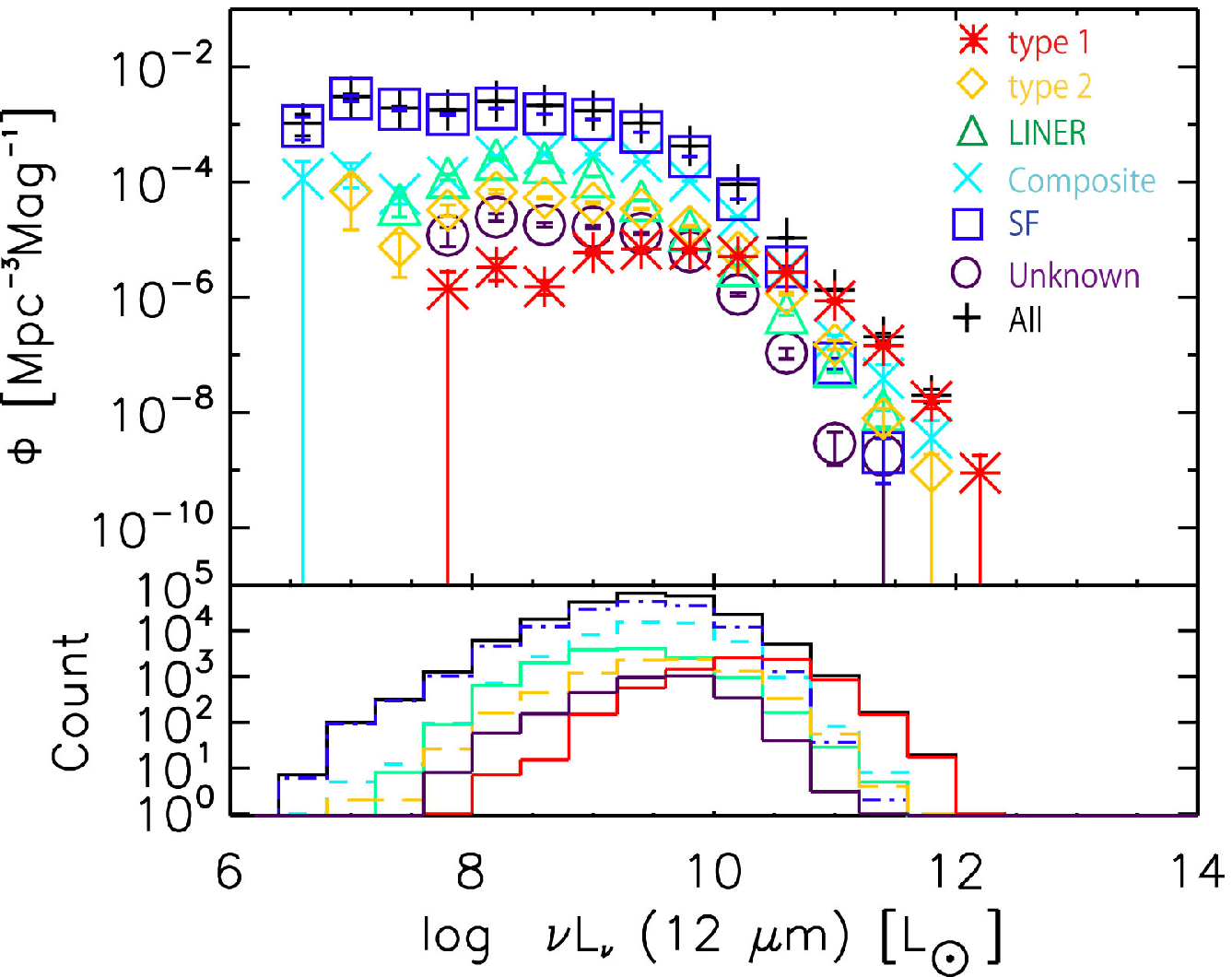}{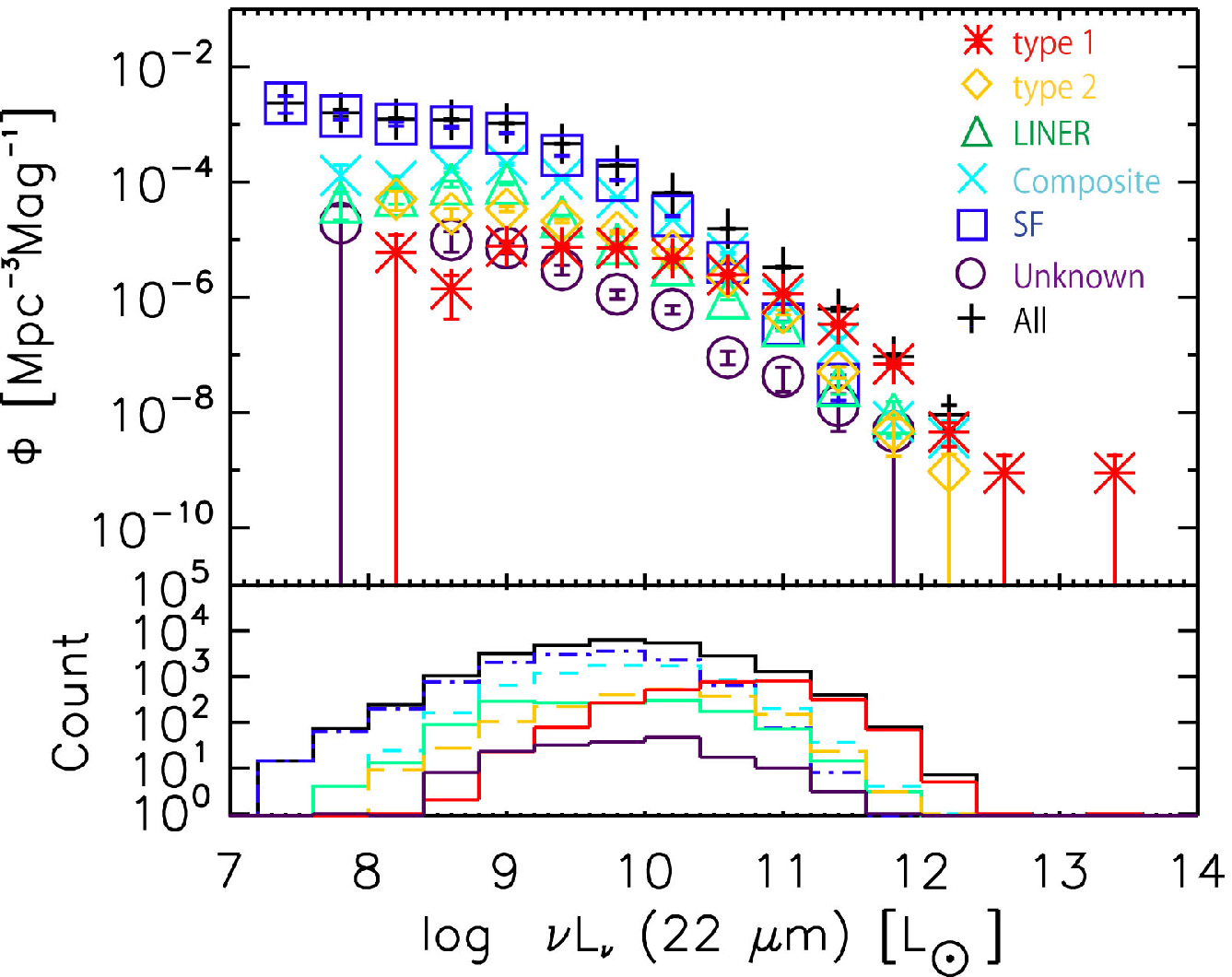}
    \caption{The 12- (left) and 22- (right) $\mu$m LFs for each galaxy type for 0.006 $\leq z \leq$ 0.3 plotted in terms of the space density as a function of luminosity. The error bars are calculated from the Poisson statistical uncertainty. The data used in these figures can be found in Tables \ref{LF_table_ALL_z}, \ref{LF_table_12_z}, and \ref{LF_table_22_z} (see Appendix \ref{LF_data}).}
    \label{LF_type}
    \end{figure}

Figure \ref{LF_type} presents the resultant LFs at 12 and 22 $\mu$m for each galaxy type for 0.006 $\leq z \leq$ 0.3.
SFs make up the majority of the objects at low luminosities, while
AGNs dominate the volume density at luminosities above
$\sim$$10^{11} L_{\odot}$. This tendency was also reported by Rush,
Malkan, \& Spinoglio (1993), who used 12-$\mu$m flux-limited samples
from the {\it IRAS} Faint Source Catalogue (FSC). Figure
\ref{LF_type} also shows that the relative number of AGNs changes
with increasing MIR luminosity; at low luminosities, type 2 AGNs
dominate the AGN population, whereas type 1 AGNs dominate at high
luminosities. \cite{Toba} also recently reported a similar trend
using {\it AKARI}. The fraction of type 2 AGNs thus changes with MIR
luminosity, which can be interpreted as a luminosity dependence of
the CF.

    \begin{figure}
    \epsscale{1}
    \plotone{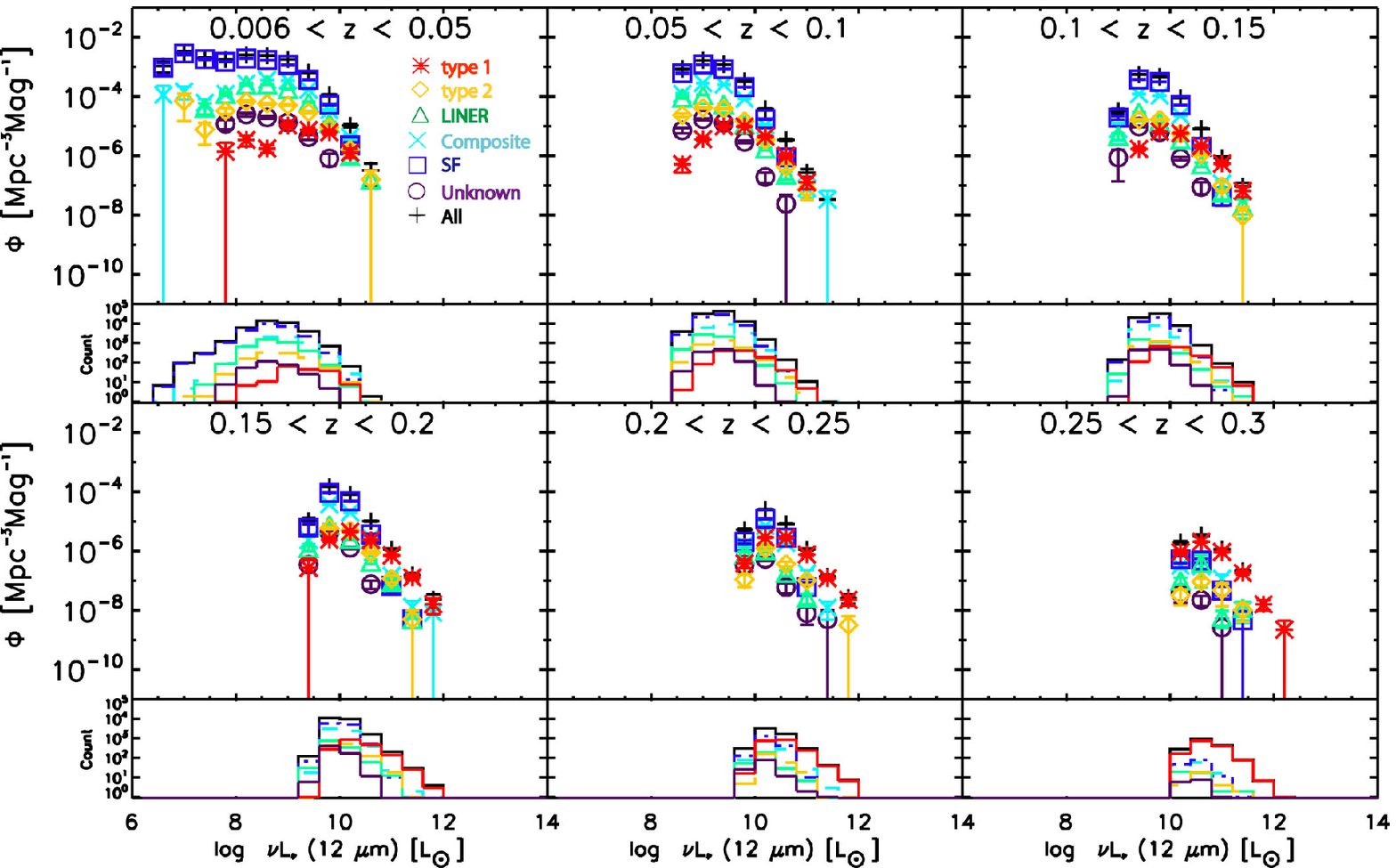}
\caption{The 12-$\mu$m LFs for each galaxy type in each redshift bin (0.006 $\leq z <$ 0.05, 0.05 $\leq z <$ 0.1, 0.1 $\leq z <$ 0.15, 0.15 $\leq z <$ 0.2, 0.2 $\leq z <$ 0.25, and 0.25 $\leq z \leq$ 0.3). The data used in this figure can be found in Tables \ref{LF_table_ALL_z} and \ref{LF_table_12_z} (see Appendix \ref{LF_data}).}
    \label{LF_type_12_z}
    \end{figure}
    \begin{figure}
   \epsscale{1}
   \plotone{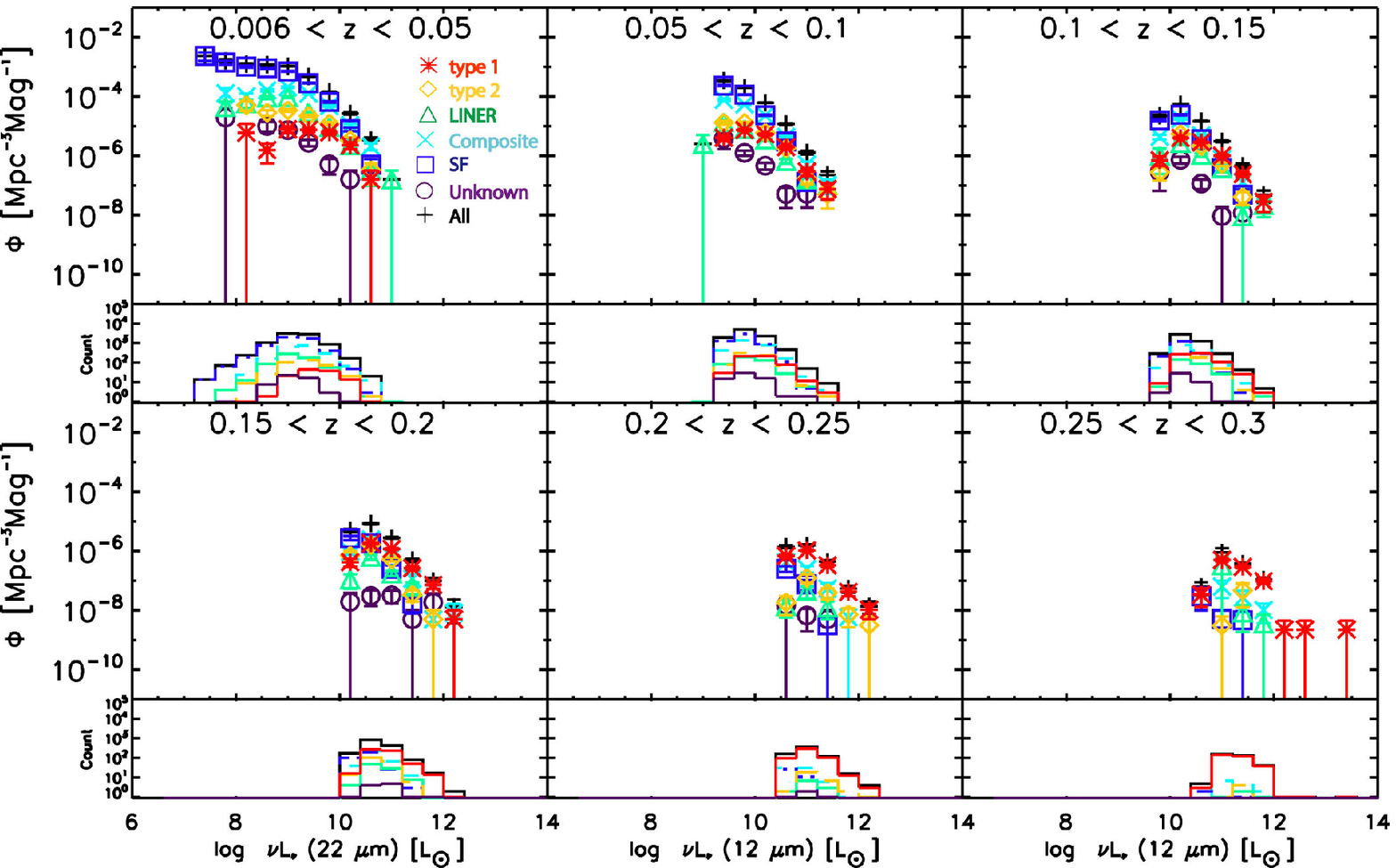}
\caption{The 22-$\mu$m LFs for each galaxy type in each redshift bin (0.006 $\leq z <$ 0.05, 0.05 $\leq z <$ 0.1, 0.1 $\leq z <$ 0.15, 0.15 $\leq z <$ 0.2, 0.2 $\leq z <$ 0.25, and 0.25 $\leq z \leq$ 0.3). The data used in this figure can be found in Tables \ref{LF_table_ALL_z} and \ref{LF_table_22_z} (see Appendix \ref{LF_data}).}
    \label{LF_type_22_z}
    \end{figure}

Figures \ref{LF_type_12_z} and \ref{LF_type_22_z} show the resultant LFs at 12 and 22 $\mu$m for each galaxy type in each of the six redshift bins (0.006 $\leq z <$ 0.05, 0.05 $\leq z <$ 0.1, 0.1 $\leq z <$ 0.15, 0.15 $\leq z <$ 0.2, 0.2 $\leq z <$ 0.25, and 0.25 $\leq z \leq$ 0.3).
The overall trends seen in Figure \ref{LF_type} are reproduced in these figures except for $z > 0.2$.
For $z > 0.2$, type 1 and type 2 AGNs dominate the volume density
over a wide range of luminosities, while for $z \leq 0.2$, their
magnitude relationship changes remarkably with increasing MIR
luminosity, as seen in Figure \ref{LF_type}. At the same time, the
overall magnitude relationship between AGNs also changes with
increasing redshift; type 2 AGNs make up the majority of the AGNs at
low redshift, while type 1 AGNs are the majority at high redshift.
This change in the fraction of type 2 AGNs with redshift can be
interpreted as a redshift dependence of the CF.

\subsection{Evolution of Luminosity Functions}\label{LF_evo}
We examined the luminosity (density) evolution of the AGN population
based on the 22-$\mu$m sample. Here, we fit the LFs for all galaxies
and AGNs using the double-power law \citep{Marshall}:
\begin{equation}
\phi(L)\mathrm{d}L = \phi^* \left\{ \left( \frac{L}{L^*}
\right)^{-\alpha} +  \left( \frac{L}{L^*} \right)^{-\beta}
\right\}^{-1} \frac{\mathrm{d}L}{L^*},
\end{equation}
where the free parameters are the characteristic luminosity $L^*$,
the normalization factor $\phi^*$, the faint-end slope $\alpha$, and
the bright-end slope $\beta$, respectively.

    \begin{figure}
        \epsscale{1}
        \plotone{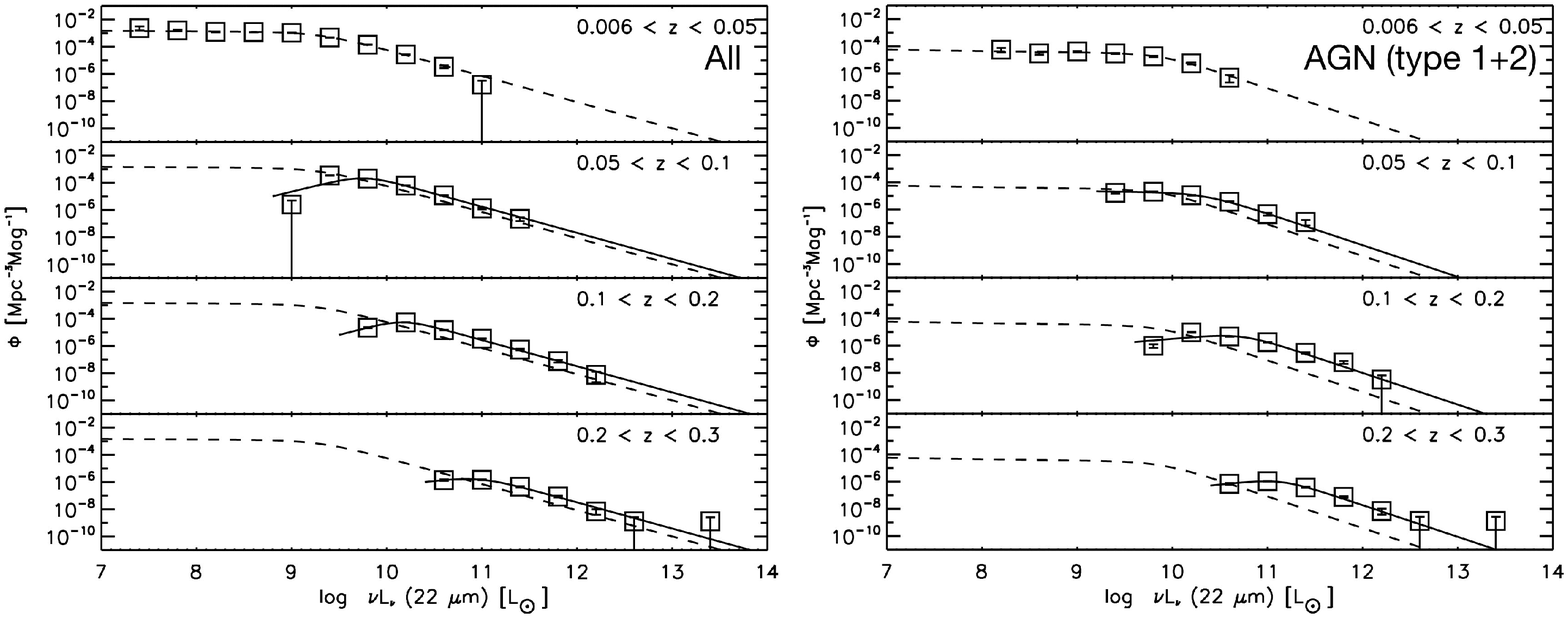}
        \caption{The 22-$\mu$m LFs for all galaxies (left) and AGNs (right)
as a function of redshift. The dashed line represents the best fit
function for 0.006 $\leq z <$ 0.05. The solid line represents the best
fit function for a fixed bright-end slope $\beta$.} \label{LF_evo_1}
    \end{figure}

\begin{figure}
        \epsscale{1}
        \plotone{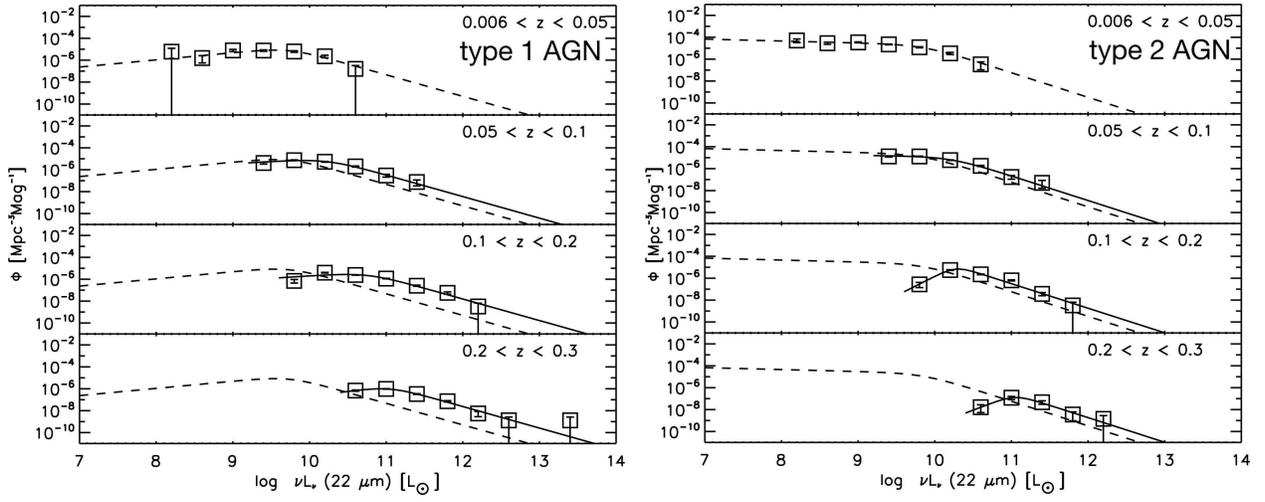}
        \caption{The 22-$\mu$m LFs for type 1 AGNs (left) and type 2 AGNs
(right) as a function of redshift. The dashed line represents the
best fit function for 0.006 $\leq z <$ 0.05. The solid line
represents the best fit function for a fixed $\beta$.}
    \label{LF_evo_2}
    \end{figure}

Figure \ref{LF_evo_1} shows the best fit for each redshift bin. Four
redshift bins (0.006 $\leq z <$ 0.05, 0.05 $\leq z <$ 0.1, 0.1 $\leq
z <$ 0.2, and 0.2 $\leq z\leq$ 0,3) are considered to keep a certain
data point (if there are less than four degrees of freedom, then we
cannot use the double-power law as a fit). The fit of the data in
the nearest redshift bin ($0.006 \leq z < 0.05$) is shown in all
panels for comparison, and to examine the evolution, fits with
$\beta$ fixed to the value of that in the nearest redshift bin
(0.006 $\leq z <$ 0.05) are also shown. 
Comparing the LF of all
galaxies with that of AGNs as a function of redshift, we see that
the LF of all galaxies does not evolve considerably with redshift,
whereas the LF of AGNs shows significant evolution. 
A comparison of the evolution of LFs for different AGN types (Figure \ref{LF_evo_2})
reveals that type 1 AGNs seem to exhibit more significant evolution
than type 2 AGNs. 
However, this difference can arise from an incompleteness of type 2 AGNs, particularly at high redshifts (z $>$ 0.2), due to the SDSS selection criterion (see also Section \ref{rejected}).

\clearpage 

\section{DISCUSSION}
In this section, we first consider the origin of the MIR emission by
using an empirical method based on a {\it WISE} color--color diagram
to extract sources that are dominated in the MIR by the active
nucleus rather than their host galaxy. The luminosity and redshift
dependence of the CF are then discussed in separate redshift bins to
disentangle the luminosity and redshift correlations. We then
consider the uncertainties in the luminosity dependence of the CF,
such as the effect of (i) the Unknown galaxies, (ii) rejected
objects in the sample selection, and (iii) optically elusive buried
AGNs. Following that, we interpret the luminosity dependence in
terms of two dust torus models (the receding torus model and the
modified receding torus model). Finally, we compare our measurements
of the CF with those of optical and hard X-ray results.

\subsection{Origin of Mid-Infrared Emission in the WISE--SDSS Sample}
\label{Xray} Before estimating the CF, we consider the origin of the
MIR emission in our AGN sample. In this study, we have assumed that
the MIR luminosity of the AGNs is dominated by emission from the
active nucleus. However, the origin of the MIR emission may not
always be an active nucleus: the emission is sometimes likely to
have a contribution from the underlying host galaxy, especially in
the low-luminosity regime.
We thus attempted to select the AGN-dominated MIR sources from the
{\it WISE}--SDSS sample by examining their MIR colors. Recently,
\cite{Mateos} suggested a highly complete and reliable MIR color
selection method for AGN candidates using the 3.4-, 4.6-, and
12-$\mu$m bands of {\it WISE}. They defined an ``AGN wedge'' based
on the {\it WISE} and wide-angle Bright Ultrahard {\it XMM-Newton}
survey (BUXS):
\begin{equation}
[3.4] - [4.6] = 0.315 \times ( [4.6] - [12] ),
\end{equation}
and
\begin{equation}
[3.4] - [4.6] =  -3.172 \times ( [4.6] - [12] ) + 7.624,
\end{equation}
where the top and bottom boundaries of the wedge are obtained by
adding y-axis ($[3.4]-[4.6]$) intercepts of +0.796 and $-$0.222,
respectively. They reported that for $L_{\mathrm{2-10 keV}} >
10^{44}$ erg s$^{-1} (\sim 10^{11} L_{\odot})$, where the AGN is
expected to dominate the MIR emission, $97.1^{+2.2}_{-4.8}$ and
$76.5^{+13.3}_{-18.4}$ percent of the BUXS type 1 and type 2 AGNs,
respectively, meet the selection criteria, i.e., a large amount of
BUXS AGNs lie in the wedge area. They also showed that compared to
other methods in the literature \citep{Jarrett,Stern}, this technique offers the highest
reliability and efficiency for detecting X-ray selected luminous AGN
populations with {\it WISE}. Therefore, we used the AGN wedge to
extract AGN-dominated {\it WISE} sources. Hereinafter, we use the
22-$\mu$m luminosity as the MIR luminosity because the 12-$\mu$m
flux is affected by PAH emission, which is unrelated to the presence
of an active nucleus as mentioned in Section \ref{Classification}
and may introduce large uncertainties.
    \begin{figure}
        \epsscale{1}
        \plotone{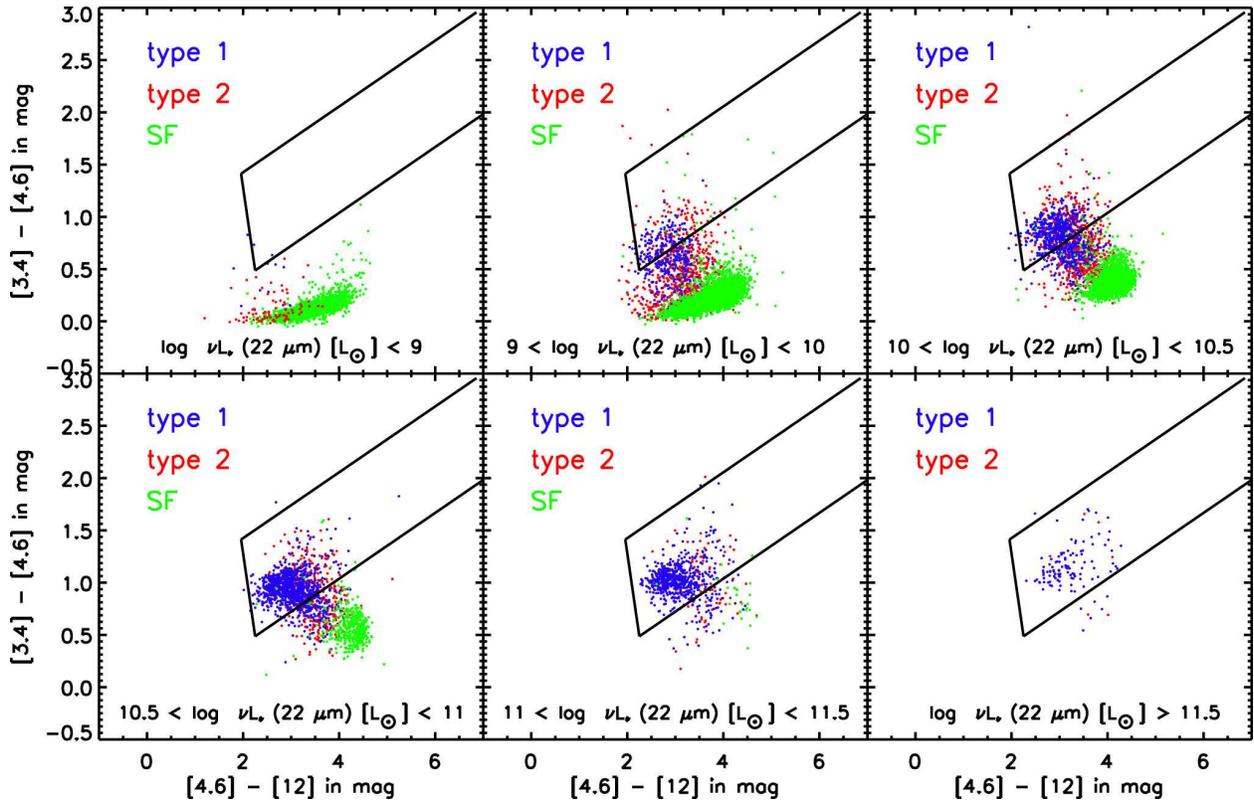}
        \caption{MIR color--color diagram for the {\it WISE}-SDSS sample at 22 $\mu$m: type 1 AGNs (blue), type 2 AGNs (red), and SFs (green) as a function of the 22-$\mu$m luminosity. The solid lines illustrate the AGN selection wedge as defined by \cite{Mateos}.}
    \label{CF_L_multi-z_AGN_22}
    \end{figure}

Figure \ref{CF_L_multi-z_AGN_22} shows the MIR colors of the {\it
WISE}--SDSS AGNs as a function of the MIR luminosity. SFs are also
plotted in the same figure for comparison. The fraction of sources
(AGNs) that are expected to be dominated by an active nucleus (i.e.,
those sources in the AGN wedge) increases with increasing MIR
luminosity. At low luminosities, the MIR emission from an AGN is
largely dominated by the large contribution from the underlying host
galaxy. At high luminosities, however, the MIR emission originates
mainly from the active nucleus. This is consistent with the result
of \cite{Mateos}.

\begin{figure}
        \epsscale{1}
        \plotone{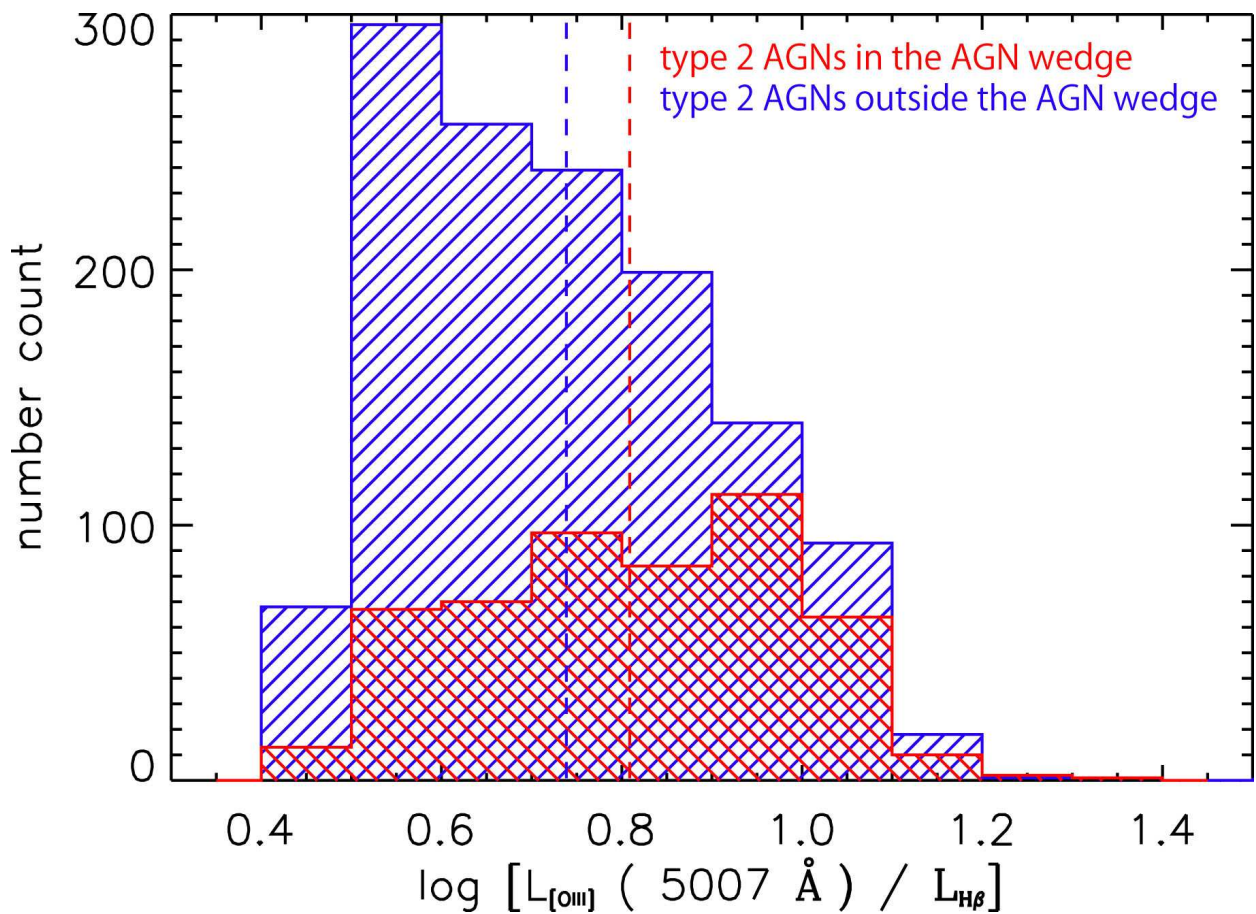}
        \caption{Distribution of the ratio of the [OIII] to H$\beta$ luminosity for type 2 AGNs in (red) and outside (blue) the AGN wedge as defined by \cite{Mateos}. The dashed lines represent the mean value of $\log [L_{\mathrm{[OIII]}}\,(5007\; {\rm\AA}) / L_{H\beta}]$ for each subsample.}
    \label{OIII_Hb}
    \end{figure}

Of the AGNs, type 2 AGNs are more affected by the star-forming
activity in their host galaxies, especially in the low-luminosity
regime, which was also reported by \cite{Mateos}. This result
indicates that there is a difference in the origin of the MIR
emission of type 1 and 2 AGNs particularly in low-luminous AGNs,
which could be interpreted as resulting from the fact that some
low-luminous (type 2) AGNs are obscured by not only a dust torus but
also their host SF. Figure \ref{OIII_Hb} shows the distribution of
the ratio of the [OIII] (5007 \AA) to H$\beta$ luminosity (believed
to be a good tracer for the strength of AGN activity) for the {\it
WISE}-SDSS type 2 AGNs in and outside the AGN wedge. As shown in
Figure \ref{OIII_Hb}, the peak of the distribution for type 2 AGNs
in the AGN wedge is relatively larger than that of the type 2 AGNs
outside the AGN wedge. To ensure the reliability of the ratio value,
we examined the ratio for objects with a S/N greater than 10 in the
[OIII] and H$\beta$ luminosities. In addition, \cite{Mateos+13}, who
adopted this technique for [OIII]-selected type 2 quasars (QSO2s)
from the SDSS, reported that the fraction of QSO2s in the AGN wedge
increases with increasing [OIII] luminosity. Therefore, we conclude
that for less powerful AGNs, the host galaxy can contribute
substantially to the MIR emission. Throughout the following
discussion, we consider the AGN-dominated objects (i.e., those in
the AGN wedge) and estimate the CF based on these objects.
We note that the [OIII] luminosity includes a contribution from HII regions, particularly in metal-poor galaxies, which means that L$_{\mathrm{[OIII}}$/L$_{\mathrm{H}\beta}$ may not always be a good tracer for AGN luminosity.
\cite{Juneau} proposed the Mass-Excitation (MEx) diagnostic to identify AGNs on the basis of their [OIII]/H$\beta$ and stellar masses.
The MEx diagnostic is a possible way to investigate the differences in the properties of samples in and outside the AGN wedge.
However, such calculations and an extended discussion of the properties based on the MEx diagram are beyond the scope of this paper.

Also, the luminosity and redshift dependences of the CF should
be discussed using the complete sample for type 1 and 2 AGNs subject
to certain criteria. Choosing objects in the WISE--SDSS sample
(i.e., extracting the objects in the AGN wedge) is equivalent to
excluding AGNs that are affected by their host galaxies.
Implementing the AGN wedge selection technique implies that the
intrinsic ratio of type 1 and type 2 AGNs is the same in and outside
the AGN wedge. This assumption is reliable because (i) the influence
of the host galaxies for each type of AGN is comparable in the
unified model, and (ii) the MIR luminosity of the nucleus for each
AGN is also comparable as reported by \cite{Gandhi}, who used
high-resolution ($\sim$0.3--0.4 arcsec) N-band filters for 12-$\mu$m
imaging data obtained with the VISIR instrument on the 8-m Very
Large Telescope.

\subsection{Luminosity and Redshift Dependence of the Covering Factor}
\label{CF_vs_L} In an effort to constrain the structure of the
hypothesized dust torus invoked by unification, we examine here the
MIR luminosity dependence of the CF. We assume that the MIR
luminosity of the AGNs is dominated by emission from the active
nucleus (see Section \ref{Xray}), and we also assume that the MIR
emission is independent of the optical classification, i.e., type 1
and type 2 AGNs should have similar continuum MIR fluxes at any
given intrinsic AGN luminosity \citep[see e.g.,][]{Horst,Gandhi}. By
integrating the LFs of the type 1 and type 2 AGNs separately, we
obtain the number density $\Phi$ for each AGN:
\begin{equation}
\Phi = \int_L \phi(L) \mathrm{d}L \sim \sum_i \phi_i(L)  \Delta L.
\end{equation}
Using these number densities, the CF and its uncertainty $\sigma_{CF}$ can be estimated as
\begin{equation}
CF = \frac{\Phi_2}{\Phi_1 + \Phi_2},
\end{equation}
and
\begin{equation}
\sigma_{CF} = CF \times \sqrt{\left( \frac{\sigma_{\Phi_{1+2}}}{\Phi_{1+2}} \right)^2 + \left( \frac{\sigma_{\Phi_2}}{\Phi_2} \right)^2},
\end{equation}
where $\Phi_1$ and $\Phi_2$ are the type 1 and type 2 AGN number densities, respectively, and $\sigma_{\Phi_1}$ and $\sigma_{\Phi_2}$ are the associated errors. Note that $\Phi_{1+2} \equiv \Phi_1 + \Phi_2$ and $\sigma_{\Phi_{1+2}} \equiv \sqrt{\sigma_{\Phi_1}^2 + \sigma_{\Phi_2}^2}$.

A point of caution here is that our flux-limited sample produces a
strong artificial correlation between the redshift and luminosity
(see Figures \ref{z_L_12} and \ref{z_L_22}). Thus, it is difficult
to decide whether it is the redshift or the luminosity that is the
more fundamental physical variable that correlates with the CF if a
sample for all luminosity and all redshift ranges is used for the
estimation. To test the intrinsic dependence of the CF on the
luminosity and redshift, we have to remove the influence of the
$z$--$L$ correlation. A simple way to do this is to analyze the CF
in separate redshift bins. \cite{Hasinger}, who also used this
diagnostic method, reported that the same trend toward a decreasing
absorption fraction (which corresponds to the CF) as a function of
X-ray luminosity, as observed in the total sample, was seen in each
of the redshift bins. We note here that the redshift interval is
optimized to keep the data points more or less equal for each
redshift bin but only for $z \leq 0.2$ because the number of objects
is limited and the SDSS survey may be incomplete, especially for
type 2 AGNs for $z > 0.2$. It should also be noted that we assume
the CF does not change within each interval, and it is under this
assumption that we test the luminosity dependence of the CF.

In addition to considering Sy2 galaxies as type 2 (obscured) AGNs,
we note that some LINERs and Composites can also show type 2
AGN-like properties. To take this into account, we also estimated
four alternative CFs that included these galaxies as type 2 AGNs:
 \begin{enumerate}
 \item Sy2s
 \item Sy2s + LINERs
 \item Sy2s + Composites
 \item Sy2s + LINERs + Composites
 \end{enumerate}

\begin{figure}
        \epsscale{1}
        \plotone{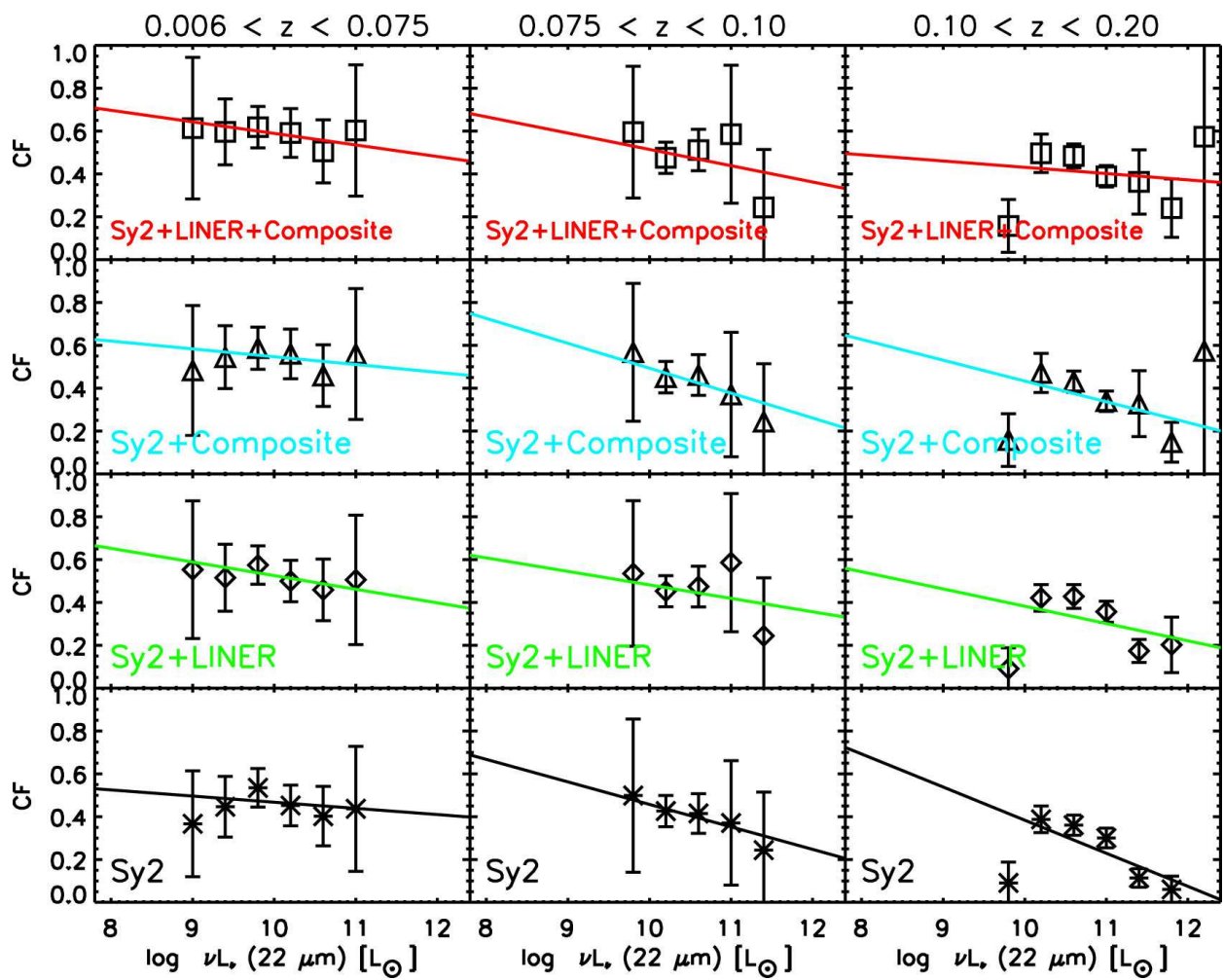}
        \caption{Variation in the CF with the 22-$\mu$m luminosity in different redshift bins for AGN-dominated MIR sources. The solid line shows the best-fit linear function determined in each redshift bin.}
    \label{CF_multi_z_22_lin_AGN}
    \end{figure}

We reiterate here that all the sources used in calculating the CF
are expected to be AGN dominant (i.e., in the AGN wedge). Figure
\ref{CF_multi_z_22_lin_AGN} shows these CFs as a function of the MIR
luminosity in different redshift bins; the CF values are also listed
in Table \ref{CF_table_22_revise}.
In the $0.006 \leq z < 0.1$ redshift range (see the left and middle panel in Figure \ref{CF_multi_z_22_lin_AGN}), the CF does not change with MIR luminosity
significantly within error, although the slope of the linear
function that we fitted shows a negative correlation. In the $0.1
\leq z \leq 0.2$ redshift bin, however, we can see that the CF does
decrease with an increase in the MIR luminosity, regardless of the
choice of type 2 AGN classification criteria.

\begin{deluxetable}{cccccccccccc}
\tablecolumns{12}
\tablewidth{0pc}
\tabletypesize{\scriptsize}
\tablecaption{CFs as a function of the 22-$\mu$m luminosity for each type 2 AGN definition.\label{CF_table_22_revise}} 
\tablehead{
\colhead{}  &  	\multicolumn{2}{c}{Sy2s} 				& \colhead{} &	
				\multicolumn{2}{c}{Sy2s + LINERs} 		& \colhead{} & 
				\multicolumn{2}{c}{Sy2s + Composites}	& \colhead{} & 
				\multicolumn{2}{c}{Sy2s + LINERs + Composites} \\
\cline{2-3}  \cline{5-6} \cline{8-9}  \cline{11-12} \\
log L 	& 	\multicolumn{1}{c}{CF} & \multicolumn{1}{c}{$\sigma_{CF}$} & \colhead{} &
		  	\multicolumn{1}{c}{CF} & \multicolumn{1}{c}{$\sigma_{CF}$} & \colhead{} &
			\multicolumn{1}{c}{CF} & \multicolumn{1}{c}{$\sigma_{CF}$} & \colhead{} &
			\multicolumn{1}{c}{CF} & \multicolumn{1}{c}{$\sigma_{CF}$} 
		 }	 
\startdata
\cutinhead{$0.006 \leq z < 0.075$}
      9.00 & 0.37 & 0.25 & & 0.55 & 0.32 & & 0.48 & 0.30 & & 0.61 & 0.33 \\
      9.40 & 0.45 & 0.14 & & 0.52 & 0.16 & & 0.54 & 0.15 & & 0.60 & 0.15 \\
      9.80 & 0.53 & 0.09 & & 0.57 & 0.09 & & 0.59 & 0.10 & & 0.62 & 0.10 \\
      10.2 & 0.45 & 0.09 & & 0.50 & 0.10 & & 0.56 & 0.12 & & 0.59 & 0.11 \\
      10.6 & 0.40 & 0.14 & & 0.46 & 0.14 & & 0.46 & 0.14 & & 0.51 & 0.15 \\
      11.0 & 0.44 & 0.29 & & 0.51 & 0.30 & & 0.56 & 0.31 & & 0.60 & 0.31 \\
\cutinhead{$0.075 \leq z < 0.1$}
      9.80 & 0.50 & 0.36 & & 0.53 & 0.34 & & 0.57 & 0.32 & & 0.60 & 0.31 \\
      10.2 & 0.43 & 0.07 & & 0.45 & 0.07 & & 0.45 & 0.07 & & 0.48 & 0.07 \\
      10.6 & 0.42 & 0.09 & & 0.47 & 0.10 & & 0.46 & 0.09 & & 0.51 & 0.10 \\
      11.0 & 0.37 & 0.29 & & 0.59 & 0.32 & & 0.37 & 0.29 & & 0.59 & 0.32 \\
      11.4 & 0.24 & 0.27 & & 0.24 & 0.27 & & 0.24 & 0.27 & & 0.24 & 0.27 \\
\cutinhead{$0.1 \leq z \leq 0.2$}
      9.80 & 0.09 & 0.10 & & 0.09 & 0.10 & & 0.16 & 0.12 & & 0.16 & 0.12 \\
      10.2 & 0.39 & 0.06 & & 0.42 & 0.06 & & 0.47 & 0.09 & & 0.50 & 0.09 \\
      10.6 & 0.36 & 0.05 & & 0.43 & 0.06 & & 0.43 & 0.05 & & 0.48 & 0.06 \\
      11.0 & 0.30 & 0.05 & & 0.36 & 0.05 & & 0.34 & 0.05 & & 0.39 & 0.05 \\
      11.4 & 0.11 & 0.04 & & 0.17 & 0.05 & & 0.33 & 0.15 & & 0.36 & 0.15 \\
      11.8 & 0.06 & 0.06 & & 0.20 & 0.13 & & 0.15 & 0.09 & & 0.24 & 0.14 \\
      12.2 &  \multicolumn{1}{c}{\nodata} &  \multicolumn{1}{c}{\nodata} & &  \multicolumn{1}{c}{\nodata} &  \multicolumn{1}{c}{\nodata} & & 0.57 & 0.70 & & 0.57 & 0.70 \\
\cutinhead{$0.006 \leq z \leq 0.15$}
      9.00 & 0.36 & 0.25 & & 0.55 & 0.33 & & 0.48 & 0.31 & & 0.62 & 0.34 \\
      9.40 & 0.44 & 0.14 & & 0.52 & 0.16 & & 0.55 & 0.15 & & 0.60 & 0.15 \\
      9.80 & 0.52 & 0.08 & & 0.56 & 0.08 & & 0.59 & 0.09 & & 0.62 & 0.09 \\
      10.2 & 0.47 & 0.05 & & 0.51 & 0.05 & & 0.55 & 0.07 & & 0.58 & 0.07 \\
      10.6 & 0.37 & 0.04 & & 0.42 & 0.04 & & 0.43 & 0.04 & & 0.48 & 0.05 \\
      11.0 & 0.28 & 0.05 & & 0.37 & 0.07 & & 0.34 & 0.06 & & 0.42 & 0.07 \\
      11.4 & 0.14 & 0.07 & & 0.16 & 0.07 & & 0.29 & 0.11 & & 0.31 & 0.11 \\
      11.8 &  \multicolumn{1}{c}{\nodata} &  \multicolumn{1}{c}{\nodata} & & 0.54 & 0.40 & &  \multicolumn{1}{c}{\nodata} &  \multicolumn{1}{c}{\nodata} & & 0.54 & 0.40 \\
\cutinhead{$0.006 \leq z \leq 0.2$}
      9.00 & 0.36 & 0.26 & & 0.55 & 0.33 & & 0.48 & 0.31 & & 0.62 & 0.34 \\
      9.40 & 0.44 & 0.14 & & 0.52 & 0.16 & & 0.55 & 0.15 & & 0.60 & 0.15 \\
      9.80 & 0.52 & 0.08 & & 0.56 & 0.08 & & 0.59 & 0.09 & & 0.62 & 0.09 \\
      10.2 & 0.47 & 0.05 & & 0.51 & 0.05 & & 0.55 & 0.07 & & 0.58 & 0.07 \\
      10.6 & 0.38 & 0.04 & & 0.44 & 0.04 & & 0.45 & 0.04 & & 0.49 & 0.04 \\
      11.0 & 0.29 & 0.04 & & 0.36 & 0.05 & & 0.33 & 0.04 & & 0.40 & 0.05 \\
      11.4 & 0.11 & 0.04 & & 0.17 & 0.05 & & 0.26 & 0.09 & & 0.30 & 0.09 \\
      11.8 & 0.06 & 0.06 & & 0.19 & 0.12 & & 0.15 & 0.09 & & 0.23 & 0.13 \\
      12.2 &  \multicolumn{1}{c}{\nodata} &  \multicolumn{1}{c}{\nodata} & &  \multicolumn{1}{c}{\nodata} &  \multicolumn{1}{c}{\nodata} & & 0.56 & 0.69 & & 0.56 & 0.69 \\
\enddata
\end{deluxetable}

This result has been reported several times, for example, by
\cite{Maiolino}, who found that the MIR spectra of 25 AGNs taken by
the infrared spectrograph (IRS) on board the {\it Spitzer Space
Telescope} showed a negative correlation between $\nu L_{6.7}/\nu
L_{5100}$ (which corresponds to the CF) and the [OIII]$\lambda$5007
line luminosity (L$_{6.7}$ and L$_{5100}$ are the continuum
luminosities at rest-frame wavelengths of 6.7 $\mu$m and 5100 \AA,
respectively). \cite{Burlon}, who constructed AGN samples in the
local universe ($z <$ 0.1) using data from the {\it Swift}-BAT
telescope and calculated the X-ray (15--55 keV) LFs of absorbed and
unabsorbed AGNs that were classified according to their absorbing
column density ($N_H$), also found a negative correlation between
the fraction of absorbed AGNs and the hard X-ray luminosity.
Recently, some studies based on {\it WISE} data have supported the
luminosity dependence of the CF. \cite{Assef} presented the
distribution of reddening in their AGN sample selected using the
{\it WISE} color in a 9-deg$^2$ NOAO Deep Wide-Field Survey
Bo\"{o}tes field. They found that the type 1 AGN (E(B - V) $<$ 0.15)
fraction is a strong function of the AGN bolometric luminosity (in
that case, the fraction of type 1 AGNs increases with bolometric
luminosity). On the basis of the {\it WISE} and SDSS data, the CF of
quasars measured by the ratio of the torus IR luminosity to the
bolometric luminosity was also found to decrease with increasing
bolometric luminosity \cite[]{Mor,Calderone,Ma,Roseboom,Gu}. More
recently, \cite{Toba} also reported a similar trend based on the
{\it AKARI} MIR data. However, these findings were from samples
containing several hundred objects. In contrast, Figure
\ref{CF_multi_z_22_lin_AGN} includes 3,000 AGNs in total. Therefore,
compared to these previous studies, our results are statistically
robust. Furthermore, the large number of AGNs allows not only for
different definitions of type 2 AGNs but also for omitting the
influence of the contribution from their host galaxies and enables
us to estimate the luminosity dependence of the CF considering only
the AGN-dominated MIR objects. The {\it WISE} results also strongly
support our previous {\it AKARI} results \citep{Toba}.

We note that the luminosity dependence of the CF we confirmed here
is slightly weaker than previous studies. This difference may be
caused by removing a large number of objects that are affected by
the contribution of their host galaxies particularly in the
low-luminosity regime. Recently, \cite{Lusso} estimated the CF by
computing the ratio of re-processed MIR emission to intrinsic
nuclear bolometric luminosity. By subtracting the contribution from
the host galaxies and correcting the reddening effect, they showed
that the obtained CF is smaller than that without any correction
especially for low luminosity, which yields a relatively weak
luminosity dependence of the CF. Our result shows a similar
tendency. Ultimately, we conclude that the CF depends on the MIR luminosity.\\

The redshift dependence of the CF based on the acknowledgment of its
luminosity dependence is derived following the diagnostic method
presented in \cite{Hasinger}. In that study, the data in each
redshift bin was also fitted with a linear function. 
\cite{Hasinger} first estimated the average value of the slope 
of the relation between the CF and luminosity in the redshift range
0.2--3.2 and then estimated the normalization value at a luminosity
of $\log (L_X)$ = 43.75 erg s$^{-1}$, in the middle of the observed
range, as a function of the redshift by keeping the slope fixed to
the average value. To quantitatively examine the dependence of the
CF on redshift, an attempt was made to correct the systematic
selection effects, and it was found that this corrected
normalization value (i.e., the CF at $\log (L_X)$ = 43.75 erg
s$^{-1}$) increased with redshift up to $z \sim 2$. This method thus
offers a simple way to examine the redshift dependence of the CF. In
the case of our flux-limited sample, however, the above analysis
would be still affected by the luminosity dependence of the CF
because our data does not cover the same redshift range given any
luminosity unlike Hasinger's sample.
This yields an ``unfair'' normalization value as a result of fitting
in each separate redshift bin, and so to evaluate the data correctly
using this method, we need to collect a complete sample that fills
the luminosity--redshift space.

We selected a complete sample in luminosity--redshift space that
enabled us to investigate the redshift dependence of the CF from
unbiased data but at the cost of reducing the number of objects in
the sample. The sample was divided into three redshift bins and four
luminosity bins. The redshift bins were $0.05 \leq z < 0.1$, $0.1
\leq z < 0.15$, and $0.15 \leq z \leq 0.2$ in a luminosity range of
$10^{10-11.6} L_{\odot}$. We note that these redshift bins are
different from those presented in Figure \ref{CF_multi_z_22_lin_AGN}
to ensure that the same luminosity range is covered in each redshift
bin. In the same manner as \cite{Hasinger}, we investigated the
dependence of the CF on redshift as follows: We first estimated the
average value of the slope in the redshift range 0.05--0.2, and the
estimated slope values for each type 2 definition are listed in
Table \ref{average_slope}, which are in good agreement with those of
Hasinger (0.25 $\pm$ 0.06) at $0.015 < z < 0.2$.
\begin{deluxetable}{lc}
\tablecolumns{2} \tablewidth{0pc} \tabletypesize{\scriptsize}
\tablecaption{Average value of the slope of the fitted linear
function in each redshift bin. \label{average_slope}} \tablehead{
\colhead{} & average value of slope
         }
\startdata
Sy2 &   $-$0.19 $\pm$ 0.06 \\
Sy2 + LINER &  $-$0.16 $\pm$ 0.06 \\
Sy2 + Composite & $-$0.18 $\pm$ 0.07\\
Sy2 + LINER + Composite & $-$0.12 $\pm$ 0.08\\
\enddata
\end{deluxetable}

\begin{figure}
        \epsscale{1}
         \plotone{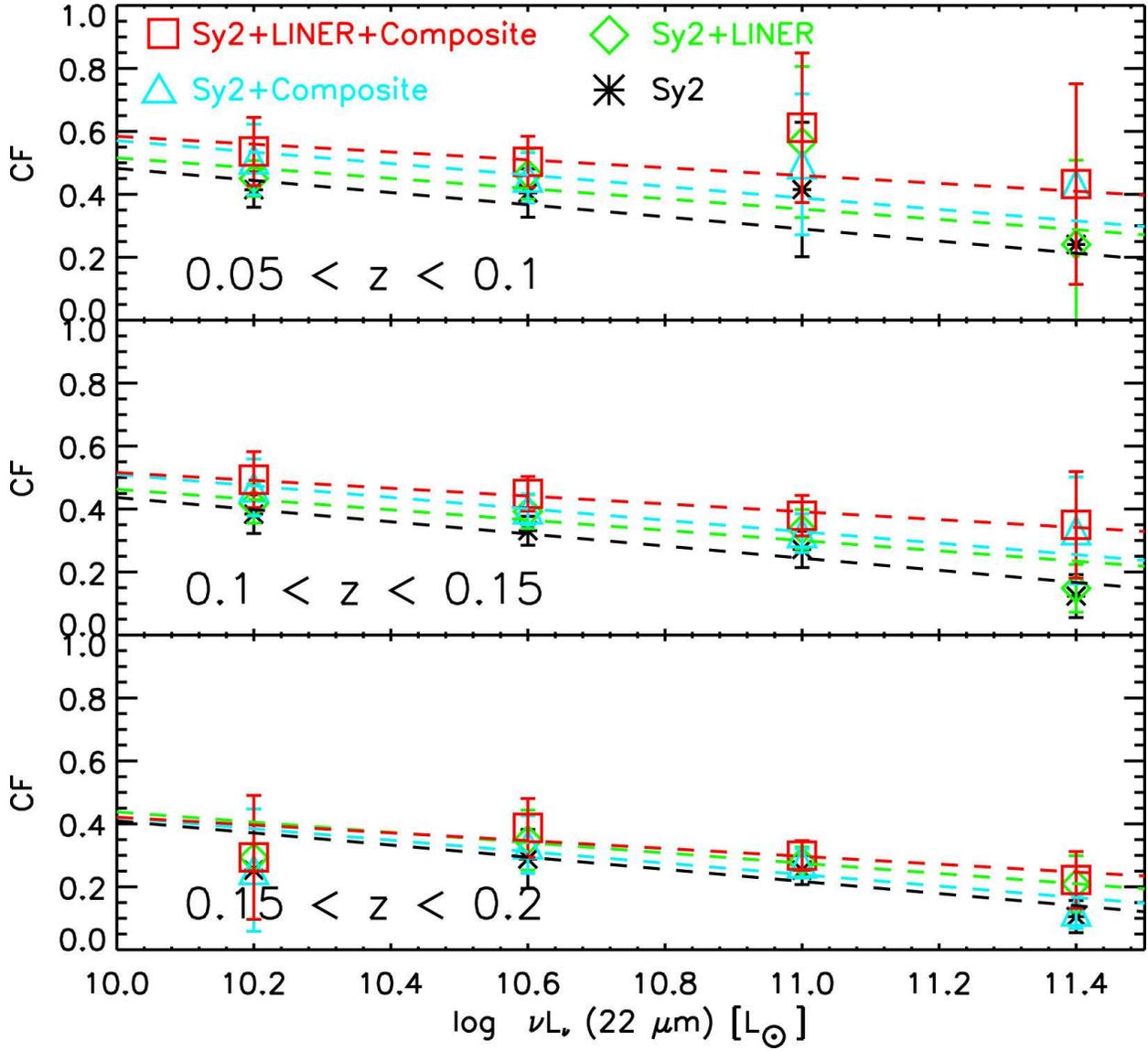}
        \caption{Variation in the CF with the 22-$\mu$m luminosity in different redshift bins. The dashed line shows the best-fit linear function with the slope fixed to the average value.}
    \label{CF_multi_z_22_lin}
    \end{figure}

We then estimated the normalization value at a luminosity of $\log[\nu
L_{\nu}$(22 $\micron$)] = 10.6 L$_{\odot}$, in the middle of the
observed range, as a function of redshift by keeping the slope fixed
to the average value (Figure \ref{CF_multi_z_22_lin}).
We found that the data are well fitted by the linear function in
each redshift bin.
    \begin{figure}
        \epsscale{1}
         \plotone{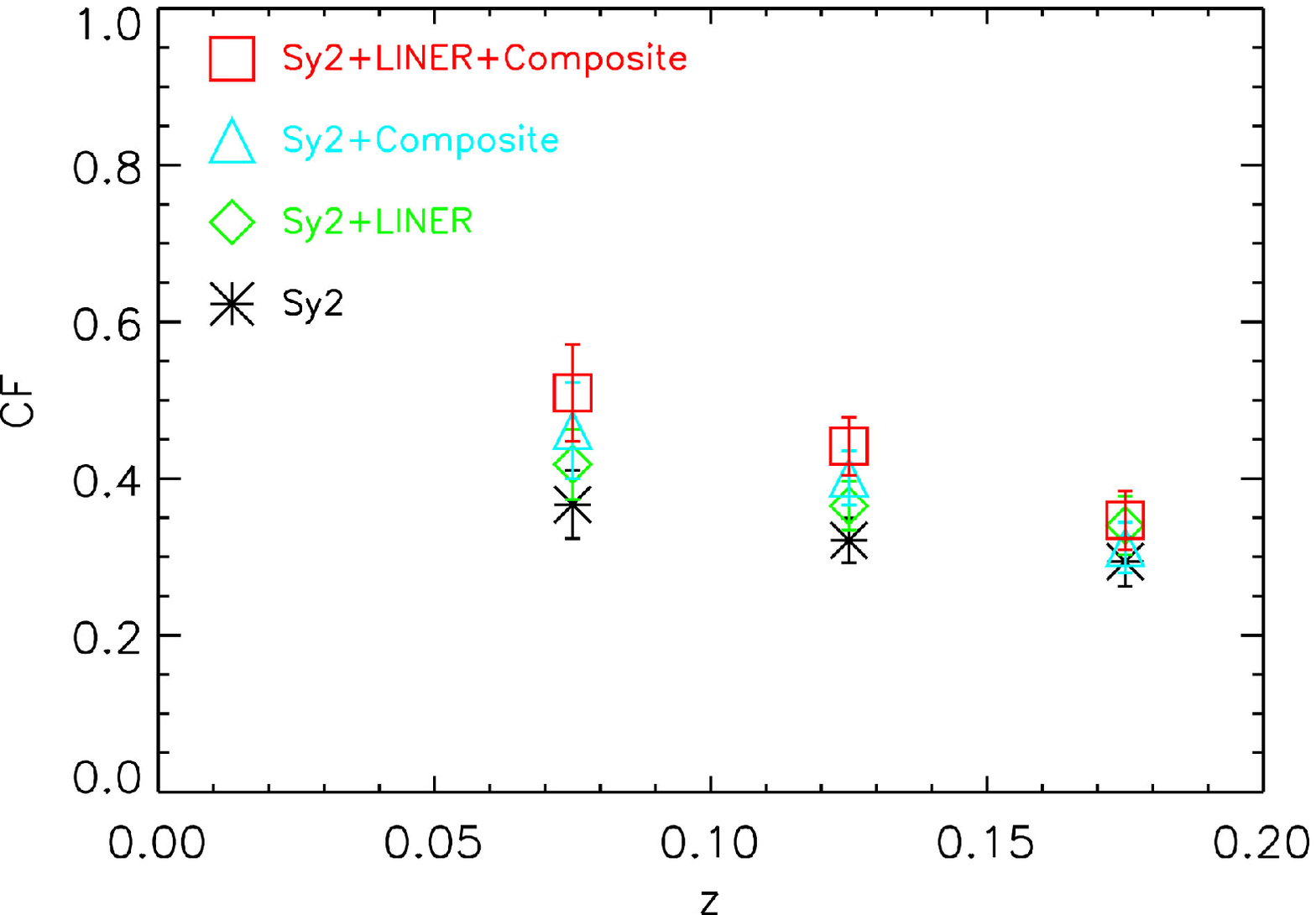}
        \caption{Dependence of the normalization of the CF at $\log[\nu L_{\nu}$(22 $\micron$)] = 10.6 L$_{\odot}$ on redshift.}
    \label{CF_multi_z_22_slope}
    \end{figure}
Figure \ref{CF_multi_z_22_slope} shows the normalization value of
the CF at $\log[\nu L_{\nu}$(22 $\micron$)] = 10.6 L$_{\odot}$ as a
function of the redshift.
We found that the CF did not change significantly with the redshift
for any type 2 AGN classification criteria. Therefore, we concluded
that the CF did not have a redshift dependence for $z \leq 0.2$.

\subsection{Uncertainties in the Luminosity Dependence}
We consider in this section the uncertainties in the luminosity
dependence of the CF. As described in Section \ref{CF_vs_L}, we
employed the AGN wedge selection technique and investigated the
luminosity dependence for selected AGN-dominated MIR objects.
However, we also consider the uncertainties without using this
technique for the benefit of users who refer to the data in Appendix
\ref{CF_without_wedge}.

\label{CF_err}
\subsubsection{Influence of the Unknown galaxies}
\label{Unknown}
A total of 3,060 and 179 galaxies were classified as Unknown galaxies in the 12- and 22-$\mu$m samples, respectively.
Even though these galaxies only constitute a small portion of the samples (1.4\% and 0.7\%, respectively), it is not clear that their influence on the CF can be ignored.
Therefore, we estimated the effect of these objects by considering the most extreme possibility, i.e., all Unknown galaxies are type 2 AGNs.
    \begin{figure}
        \epsscale{1}
         \plotone{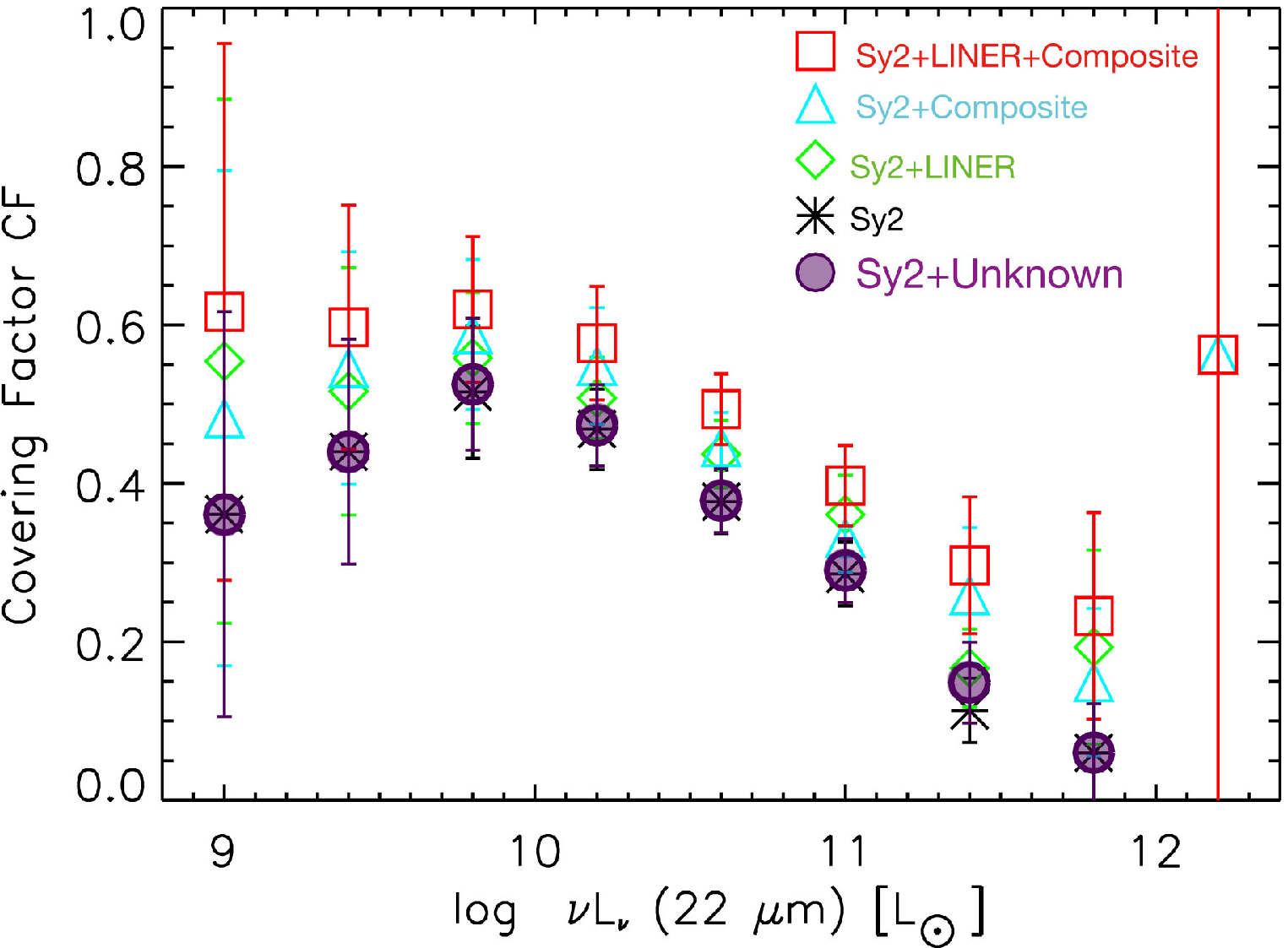}
        \caption{The CF as a function of the 22-$\micron$ luminosities including the case for which all Unknown galaxies are type 2 AGNs (purple circles) at $z \leq$ 0.2.}
    \label{CF_L_Unknown}
    \end{figure}
We should note that we investigated the influence using a sample in
all redshift ranges because the number of Unknown galaxies is too
small to estimate the CF with a high accuracy if they are divided
into separate redshift bins as in Figure
\ref{CF_multi_z_22_lin_AGN}.
We here show the result when considering the AGN wedge.
As shown in Figure \ref{CF_L_Unknown},
the CF for this case also decreases with increasing luminosity.
In the case of the all sample including the outside the AGN wedge, we also see a similar
trend (see Figure \ref{CF_L_ALL-z_12_22} in Appendix \ref{CF_without_wedge}), and so we conclude that the influence of the Unknown galaxies can be neglected.

\subsubsection{Influence of rejected objects}
\label{rejected}
When we cross-matched the {\it WISE} sample with the SDSS sample, we rejected 601,460 {\it WISE} sources that did not lie within the 3-arcsec search radius.
If these rejected objects were galaxies, then this could have an effect on our results.
We thus attempted to extract possible galaxy candidates according to the SDSS photometric information and data in the literature.

\begin{figure}
        \epsscale{0.8}
        \plotone{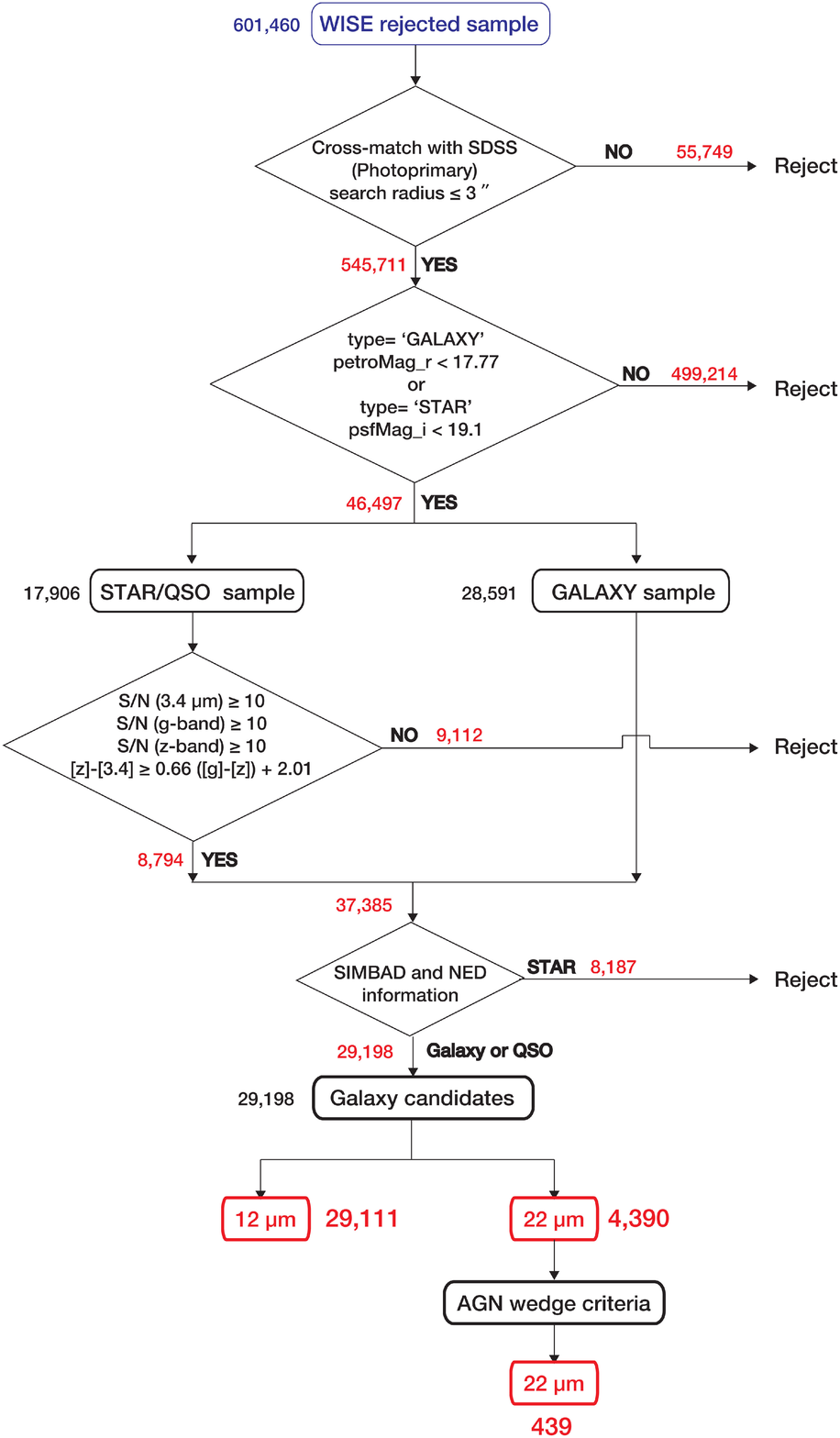}
        \caption{Flow chart for extracting the galaxy candidates from the {\it WISE}-rejected sample.}
    \label{reject_objcts}
    \end{figure}

Figure \ref{reject_objcts} shows a flow chart of the process used for the extraction.
The 601,460 {\it WISE}-rejected sources were first cross-matched with the SDSS DR8 photometric catalog, which contains 469,053,874 unique primary sources with all data cataloged in the {\tt PhotoPrimary} table on the CAS.
We again adopted a 3-arcsec search radius and selected 545,711 sources.
Of these, we extracted sources that met the SDSS spectroscopic sample selection criteria, i.e., a petroMag\_r value of less than 17.77 mag or a psfMag\_i value of less than 19.1 mag, and considered their morphologies (see Section \ref{SDSS_sample}).
The morphological information is listed under the column headed ``type'' in the {\tt PhotoPrimary} table; point-like objects are labeled as ``STAR'' (possibly including quasars) and diffuse objects are labeled as ``GALAXY'', based on the difference between the PSF and model magnitudes.
It should be noted that there is no ``QSO'' category because it is difficult to distinguish stars and quasars based on the photometry only, unlike the case for the spectroscopy-based {\tt SpecPhoto} table. Thus, an extraction of objects with (i) petroMag\_r below 17.77 mag and type = ``GALAXY'' and (ii) psfMag\_i below 19.1 mag and type = ``STAR'' yielded 46,497 sources (17,906 point-like sources and 28,591 diffuse sources).
There were 55,749 sources that did not have any SDSS photometric information, and these were rejected. While some of these objects may be galaxies, the sources are optically too faint to be detected by SDSS imaging (the exposure time per band is $\sim$60 s, and the detection limit (95\% completeness) of the r-band for point sources is 22.2 mag).
We also note that 499,214 faint sources were excluded both by this method and in the spectroscopic target selection \citep{Eisenstein, Strauss}.
We discuss the influence of these optically faint {\it WISE} sources on our result in the end of this subsection.

For the 17,906 point-like sources (hereinafter, STAR/QSO sample), we examined their color properties to remove the star objects based on the $g - z$ versus $z - [3.4]$ color--color diagram.
\cite{Wu} have also plotted this data for spectroscopically confirmed stars and quasars obtained from the SDSS DR7 and {\it WISE} catalogs and reported that most stars can be distinguished from quasars using following
criterion:
\begin{equation}
 z - [3.4] \leq 0.66 \times (g - z) + 2.01.
\end{equation}
Some high-$z$ ($z > 4$) quasars are actually lost in a field of stars when this criterion is adopted, as \cite{Wu} have mentioned.
However, as we are focusing on only low-$z$ quasars ($z \leq 0.3$), this criterion is useful.
    \begin{figure}
        \epsscale{1}
        \plotone{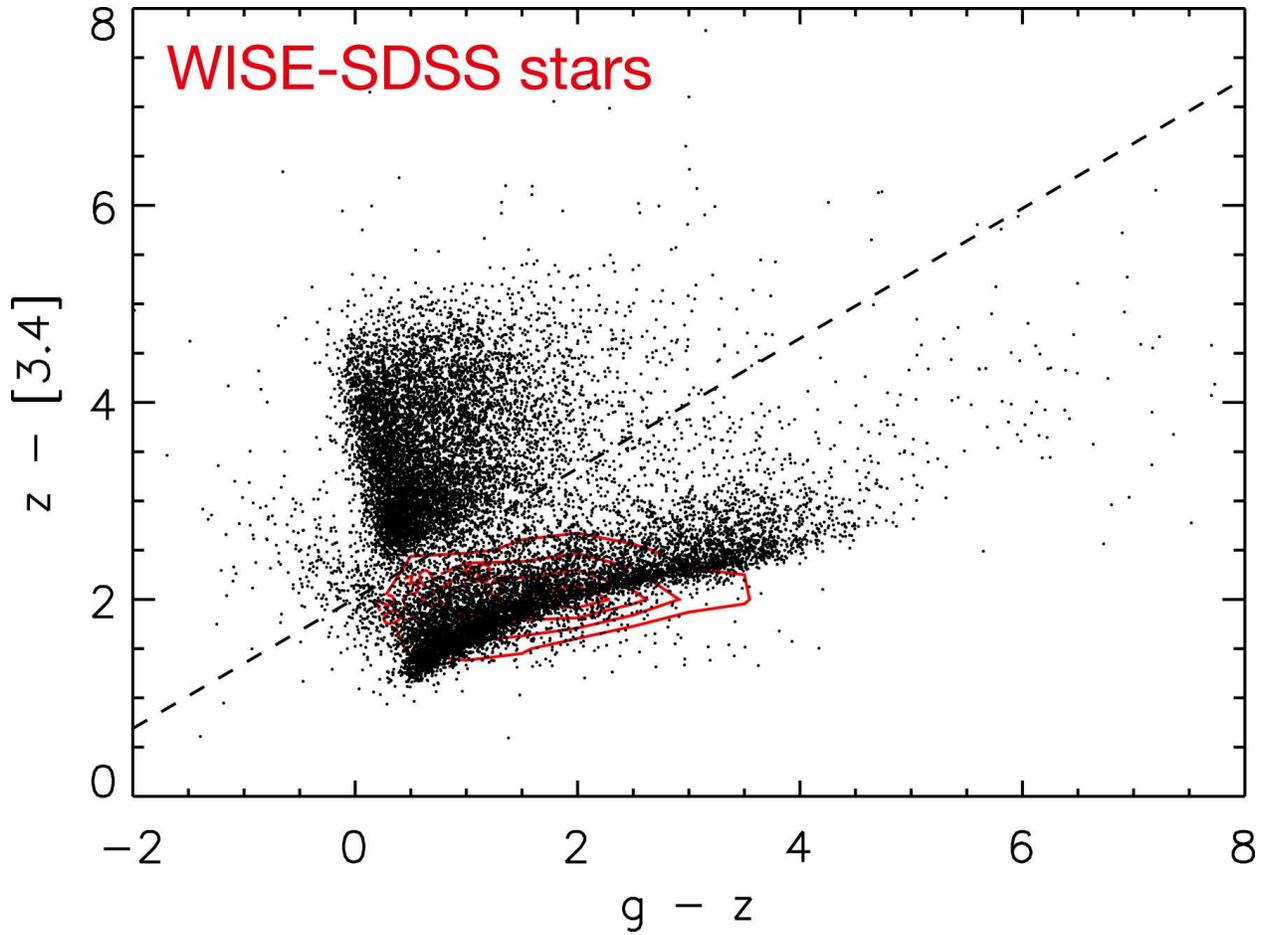}
        \caption{Distribution of the STAR/QSO sample in the $g -z$ versus $z - [3.4]$ color--color diagram. The red contours represent spectroscopically confirmed {\it WISE}--SDSS stars. The dashed line indicates the star--quasar separation criterion, $z - [3.4] = 0.66 (g - z) + 2.01$, proposed by \cite{Wu}.}
    \label{Wu_color}
    \end{figure}
The distribution of the STAR/QSO sample in the color--color diagram is shown in Figure \ref{Wu_color}.
To ensure the reliability of the color value, we examined the color for objects with a S/N of greater than 10 in the {\it g}, {\it z}, and 3.4-$\mu$m band photometry.
For comparison, spectroscopically confirmed {\it WISE}--SDSS stars are also plotted in Figure \ref{Wu_color}.
The criterion proposed by \cite{Wu} works well, and by adopting this criterion, 8,794 objects were selected as galaxy candidates.

Finally, we carefully removed stars from the 28,591 + 8,794 = 37,385
galaxy candidates by utilizing the NASA/IPAC Extragalactic Database
(NED\footnote{\url{http://ned.ipac.caltech.edu/}}) and the Set of
Identifications, Measurements, and Bibliography for Astronomical
Data (SIMBAD\footnote{\url{http://simbad.u-strasbg.fr/simbad/}})
database. The final galaxy-candidate sample consisted of 29,198
objects: 29,111 in the 12-$\mu$m sample and 4,390 in the 22-$\mu$m
sample. As the {\it WISE}--SDSS sample contained 223,982 objects in
the 12-$\mu$m sample and 25,721 objects in the 22-$\mu$m sample, the
maximum uncertainty, therefore, caused by including these new
galaxies would be 29,111/223,982 $\sim$ 0.130 (13.0\%) for the
12-$\mu$m sample and 4,390/25,721 $\sim$ 0.170 (17.0\%) for the
22-$\mu$m sample. Half of the galaxies in the {\it WISE}--SDSS
sample were classified as SF in Section \ref{Classification} (see
Table \ref{type_classification}), and we would expect half of the
galaxy candidates to be SF, even if they are all galaxies. Hence,
the influence on the estimated CF is expected to be small. In the
context of considering the AGN wedge, there are 2,922 objects in the
AGN wedge as shown in Table \ref{type_classification_AGN_wedge},
which are summarized by the type classification in the case of
considering AGN-dominated 22-$\mu$m sources. Among them, 439 objects
could be galaxies as a consequence of adopting the AGN wedge
technique for 2,922 objects. Thus the maximum uncertainty caused by
including these galaxies would be 439/2,922 $\sim$ 0.150 (15.0\%).

However, it must be noted that the above estimation is for optically bright (PetroMag\_r $<$ 17.77 for galaxies and psfMag\_i for type 1 AGNs) MIR sources.
Thus, some optically faint sources could be overlooked by our selection procedure.
We thus attempted to estimate the influence of the optically faint type 2 AGNs on our results by using deeper spectroscopic data.
The Galaxy And Mass Assembly \citep[GAMA;][]{Driver_09,Driver_11} program is a spectroscopic survey of $\sim$300,000 galaxies down to r $<$ 19.8 mag over $\sim$290 deg$^2$ using the AAOmega multi-object spectrograph on the Anglo-Australian Telescope (AAT).
Partial data obtained in the first phase of the GAMA survey has been released as Data Release 2 (DR2; Liske et al. in preparation), and this catalog provides AAT/AAOmega spectra, redshifts, and a wealth of ancillary information for 72,225 objects located in three equatorial fields (referred to as G09, G12, and G15) covering 144 deg$^2$.
The limiting Petrosian r magnitudes are 19.0 (G09 and G12) and 19.4 (G15),  two magnitudes deeper than that of the SDSS spectroscopic catalog, but the survey area is smaller than that of the SDSS.
Therefore, GAMA could be the best dataset for extracting optically faint WISE sources that were not detected by the SDSS spectroscopy.
In what follows, we extract the optically faint sources and estimate the fraction of type 2 AGNs among the {\it WISE}-rejected sample by assuming that the spatial distributions of the optically faint sources in the GAMA field are the same as those in the SDSS spectroscopic field.

We first narrowed the {\it WISE}-rejected sample to sources within the G09, G12, and G15 regions, which yielded 8,023 sources (8,013 in the 12-$\mu$m sample and 388 in the 22-$\mu$m sample).
These sources were then cross-identified with the GAMA DR2 by using a matching radius of 3 arcsec.
In this study, we used the {\tt EmLinesPhys} table, which includes the coordinates of each GAMA source and its redshift.
As a result, 4,733 sources (hereinafter WISE-nonSDSS-GAMA objects) were selected (4,732 sources in the 12-$\mu$m sample and 217 sources in the 22-$\mu$m) sample.
We then extracted type 2 AGNs, based on the BPT diagram employed in Section \ref{Classification}, with the line information obtained from the {\tt SpecLines} table \citep{Hopkins}.
In addition, we narrowed the sample down to sources with a redshift smaller than 0.2, which were adopted to evaluate the luminosity and redshift dependence of the CF.
This resulted in 163 objects being classified as type 2 AGNs at $z \leq$ 0.2 (163 sources in the 12-$\mu$m sample and 15 sources in the 22-$\mu$m sample).
In the case of the 22-$\mu$m sample, 15/217 $\sim$7\% objects were type 2 AGNs.
We note that this estimation is a lower limit because the GAMA could not detect almost 40\% of the {\it WISE}-rejected sample (hereinafter WISE-nonSDSS-nonGAMA objects), and thus some optically faint sources with a PetroMag\_r greater than 19.0 could be type 2 AGNs.
We therefore investigated the possibility that these objects exist by using NED and SIMBAD.
Among the 3,290 WISE-nonSDSS-nonGAMA objects, 2,492 and 424 objects were cross-identified with the NED and SIMBAD, respectively, by using matching radii of 3 arcsec, and we checked the existence of type 2 AGNs at 0.006 $< z <$ 0.2.
We found that there were no objects that satisfied the above criteria, although NED and SIMBAD did not have complete spectroscopic classifications for all the galaxies.
Therefore, the majority of the WISE-nonSDSS-nonGAMA objects are expected to be high-z ($>0.2$) sources, and the maximum contribution ($\sim$7\%) of type 2 AGN for the 22-$\mu$m sample mentioned above should be a reasonable estimate.
In terms of the CF (type 2 AGN fraction), this result indicates that the CF we derived in Section \ref{CF_vs_L} is an underestimation (see also Sections \ref{interpretation_CF_L} and \ref{Comparison}).

\subsubsection{Influence of optically elusive buried AGNs}
\label{Obscured}
The type classification was based on the optical spectroscopic information (see Section \ref {Classification}), but if the central engine of an AGN is enshrouded by dust covering the entire solid angle, then the bulk of the optical emission will be absorbed by the dust, and it is thus difficult to classify objects using the BPT diagram.
The presence of these ``buried'' AGNs has been reported by many authors including \cite{Oyabu}, who identified two buried AGNs based on {\it AKARI} near-IR (NIR) spectroscopic observations.
These objects do not show any AGN features in the optical spectra but do have a steep red continuum from the hot dust in the NIR spectra.

\begin{figure}
        \epsscale{1}
        \plotone{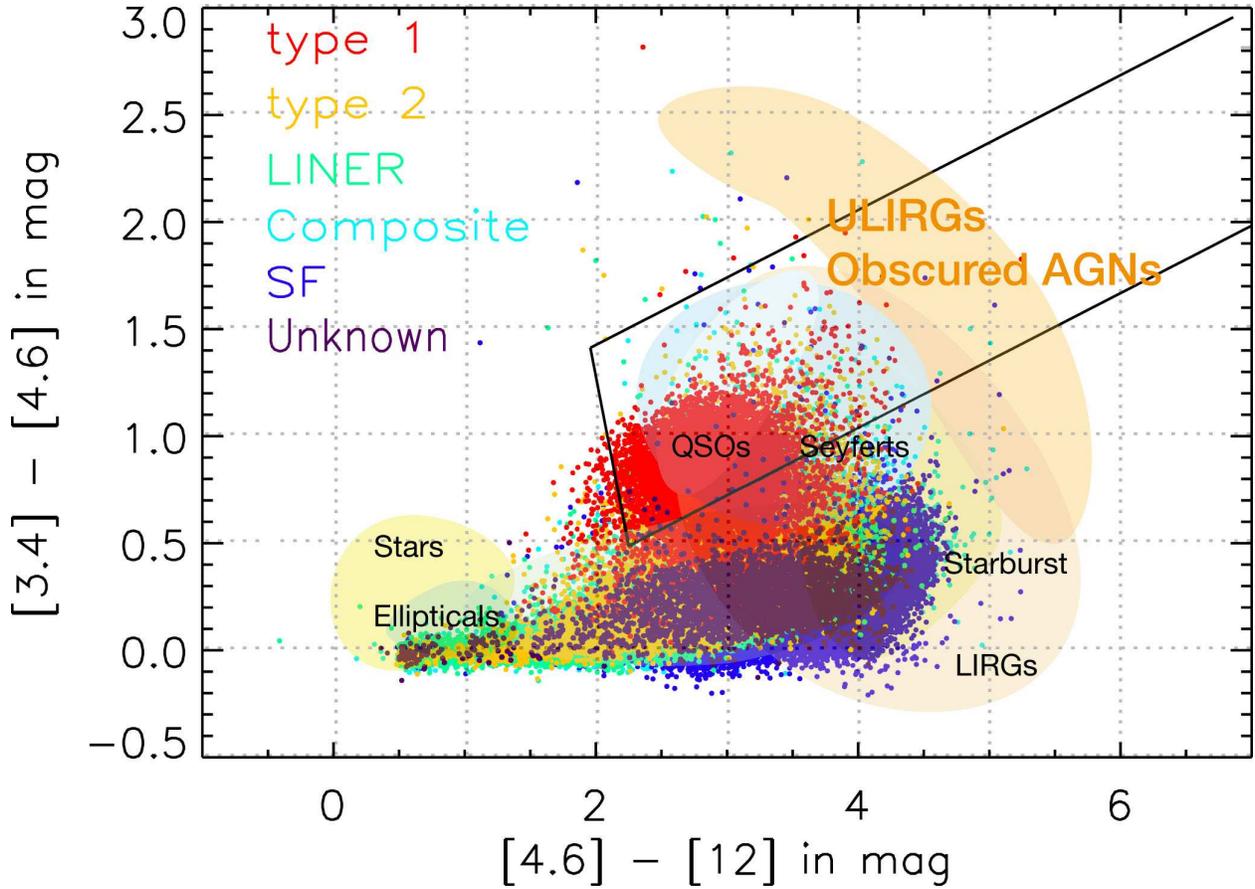}
        \caption{{\it WISE} color--color diagram of the {\it WISE}--SDSS galaxies. The shaded regions representing different galaxy types indicate areas where the photometry of redshifted sample galaxies was synthesized using simulated SEDs \citep{Wright}. The solid lines illustrate the AGN selection wedge defined from \cite{Mateos}.}
    \label{WISE_color_color}
    \end{figure}

We examined the presence of buried AGNs based on their expected {\it WISE} color, which should be very red. Figure \ref{WISE_color_color} shows the color--color diagram ($[3.4] - [4.6]$ versus $[4.6] - [12]$) for {\it WISE}--SDSS sample.
The shaded regions representing different galaxy types indicate areas where the photometry of redshifted sample galaxies was synthesized using simulated SEDs \citep{Wright}.
A few LINERs, SFs, and Composites are located in the obscured AGN
region, defined by \cite{Wright}, but the number of these obscured
AGNs is very small. In the context of considering the AGN wedge,
1.2\% of the SFs and 0.4\% of the Unknown galaxies that exist in the
AGN wedge may be candidates for buried AGNs (see Table
\ref{type_classification_AGN_wedge}). Their percentage of all
AGN-dominated objects is very small (1.6\% at most), and thus we
conclude that their influence on the CF will also be small.

\begin{deluxetable}{lr}
\tablecolumns{2}
\tablewidth{0pc}
\tabletypesize{\scriptsize}
\tablecaption{Classifications of the objects in the AGN wedge for 22-$\mu$m sample.\label{type_classification_AGN_wedge}}
\tablehead{
\colhead{type} & \colhead{number (percentage)}
}
\startdata
			type 1 AGNs			&	2,077 (71.1\%)\\
			type 2 AGNs			&	  520 (17.8\%)\\
			LINER				&	  130 (4.4\%)\\
			Composite			&	  150 (5.1\%)\\
			SF					&	   34 (1.2 \%)\\	
			Unknown				&	   11 (0.4 \%)\\	
\cline{1-2}
			All					&	 2,922 (100 \%)\\	
\enddata
\end{deluxetable}

\subsection{Interpretation of Luminosity Dependence of the Covering Factor}
\label{interpretation_CF_L}
We consider here two dust torus models that may explain the luminosity dependence of the CF.
We fit our results first to the receding torus model \citep{Lawrence+91}, which argues that an expansion of the dust sublimation radius with luminosity will push the torus to larger radii and will therefore decrease its CF.
In this model, the CF is described as a function of luminosity by \citep[e.g.,][]{Simpson+98,Simpson+05}
\begin{equation}
\label{Eq_RTM}
CF = \left( 1 + \frac{3L}{L_0} \right) ^{-0.5}\;,
\end{equation}
where $L_0$ is the luminosity at which the CF is equal to 0.5.
Here, the luminosity is based on radiation not from the dust torus but from the total output of the central engine of the AGNs.
Thus, the luminosity of concern here is not equivalent to the MIR luminosity.
However, the radiation from the central engine is thought to be strongly correlated with that from the dust torus.
For instance, \cite{Spinoglio} found that the bolometric luminosity of AGNs is proportional to the MIR luminosity, based on an examination of {\it IRAS} data.
\cite{Ichikawa} obtained MIR photometric data for a total of 128 sources in the 9-, 12-, 18-, 22-, and 25-$\mu$m bands from {\it AKARI} and {\it WISE} as well as hard X-ray (14--195 keV) data from {\it Swift} BAT. They found a good correlation between the hard X-ray and MIR luminosities over three orders of magnitude ($9 < \log \nu L_{\nu} (9,18\, \mu$m) $< 12$), which is tighter than that between the hard-X-ray luminosity and far-IR (FIR) luminosities at 90 $\mu$m.
This could indicate that the radiation from the central engine is directly connected to that from the dust torus.
Therefore, the MIR luminosity should be a good tracer of the bolometric luminosity from the central engine.

\begin{figure}
        \epsscale{1}
        \plotone{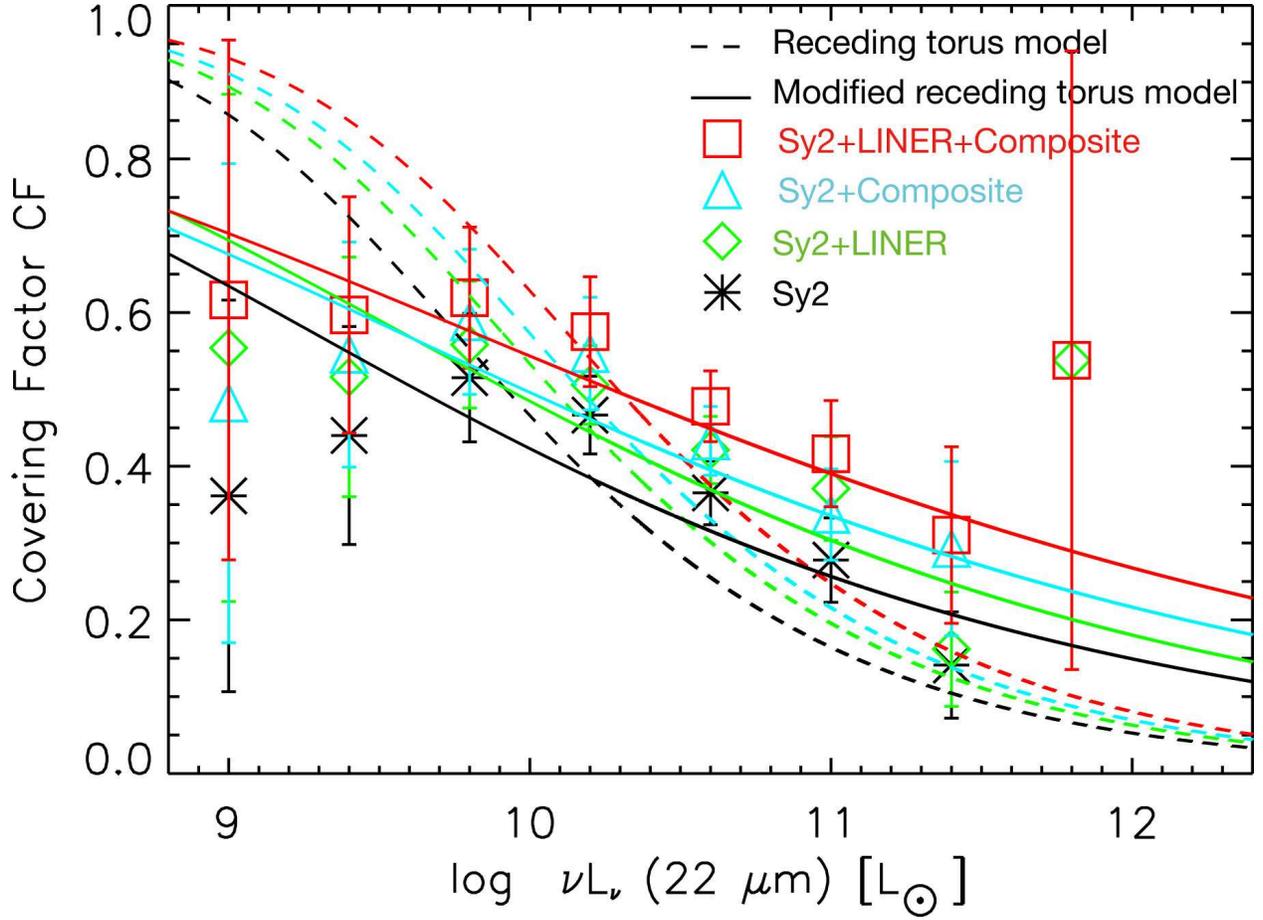}
        \caption{The CF as a function of the 22-$\mu$m luminosity for AGN-dominated MIR sources at $z \leq 0.15$. The dashed line shows the best-fit curve determined with the receding torus model. The solid line shows the best-fit curve determined with the modified receding torus model.}
    \label{CF_L_RTM_AGN}
    \end{figure}

It should be noted that we restrict the sample here to those objects at $z \leq 0.15$ to omit as much as possible the effects of optically faint WISE sources (see Section \ref{rejected}).
The relationship between the CF and MIR luminosity, derived in
Section \ref{CF_vs_L}, is compared in Figure \ref{CF_L_RTM_AGN} with
that expected from the receding torus model. Here, $L_0$ is a
free-parameter; its best-fit value and reduced chi-square value
($\chi^2 /\nu$) are listed in Table \ref{CF_RTM_AGN_para_22}. Figure
\ref{CF_L_RTM_AGN} demonstrates that the receding torus model
provides a good model for our data.

However, the receding torus model does not provide a unique explanation of the luminosity dependence; the assumption that the height of the torus is constant regardless of the luminosity is rather strict, and the value of the reduced chi-square for the CF fit is relatively large.
We therefore also considered the modified receding torus model, which was proposed by Simpson (2005) and supported by \cite{Ricci} and \cite{Lusso}.
In this model, the height of the torus ($h$) also depends on the luminosity of the AGNs,
\begin{equation}
 \label{Eq_RTM_H}
 CF = \left[ 1 + 3\left(\frac{L}{L_0} \right)^{1-2\xi} \right] ^{-0.5},
\end{equation}
where there are now two free-parameters: $L_0$ and $\xi$. The
best-fit values and $\chi^2 /\nu$ for this model are also listed in
Table \ref{CF_RTM_AGN_para_22}, and the results are also shown in
Figure \ref{CF_L_RTM_AGN}. We found that the modified receding torus
model appears to provide a better fit to the data. For this model,
$\xi$ takes positive values ($\sim$0.1--0.3; cf., Table
\ref{CF_RTM_AGN_para_22}.), which are consistent with those reported
by \cite{Simpson+05}.

The luminosity dependence of the height of the torus ($h\propto
L^{0.2-0.3}$) can be interpreted in the framework of the
radiation-limited clumpy torus model. This model was originally
suggested by \cite{Honig}, who investigated the influence of the dust
distribution on the Eddington limit of the torus and concluded that
the torus was a clumpy torus comprised of self-gravitating,
optically thick dust clouds. Clouds at small radii from the central
black hole are directly exposed to the AGN radiation pressure and
forced out to larger distances, while distant clouds are shielded
from the AGN radiation by the clouds at small radii. Both effects
determine the size of the torus.
This model gives the luminosity dependence of the height as
$h\propto L^{0.25}$, which is in good agreement with our
measurements.

\begin{deluxetable}{lcccccccc}
\tablecolumns{9}
\tablewidth{0pc}
\tabletypesize{\scriptsize}
\tablecaption{Fitting parameters of the receding torus model and modified receding torus model for the 22-$\mu$m sample in the AGN wedge.\label{CF_RTM_AGN_para_22}}
\tablehead{
\colhead{}  &  	\multicolumn{3}{c}{Receding Tours Mode} & \colhead{} &	
				\multicolumn{4}{c}{Modified Receding Torus Mode} 	 \\
\cline{2-4}  \cline{6-9} \\
\colhead{} 	& 	\multicolumn{1}{c}{$L_0$} & \multicolumn{1}{c}{$\chi^2/\nu$} & \multicolumn{1}{c}{$\chi_{\nu}^2$} & 
\colhead{} 	& 	\multicolumn{1}{c}{$L_0$} & \multicolumn{1}{c}{$\xi$} & \multicolumn{1}{c}{$\chi^2/\nu$} &  \multicolumn{1}{c}{$\chi_{\nu}^2$}
		 }	 		 
\startdata 
Sy2s & (8.33 $\pm$ 1.41)$\times 10^{9}$ & 16.34/6 & 2.72 & & (4.20 $\pm$ 2.08)$\times 10^{9}$ & (0.25 $\pm$ 0.07) & 5.26/5 & 1.05 \\
Sy2s + LINERs & (1.19 $\pm$ 0.20)$\times 10^{10}$ & 15.40/7 & 2.20 & & (8.44 $\pm$ 3.17)$\times 10^{9}$ & (0.26 $\pm$ 0.07) & 4.55/6 & 0.76 \\
Sy2s + Composites & (1.47 $\pm$ 0.27)$\times 10^{10}$ & 12.74/6 & 2.12 & & (9.51 $\pm$ 0.45)$\times 10^{9}$ & (0.29 $\pm$ 0.07) & 1.72/5 & 0.34 \\
Sy2s + LINERs + Composites & (1.96 $\pm$ 0.36)$\times 10^{10}$ & 13.52/7 & 1.93 & & (1.87 $\pm$ 0.83)$\times 10^{10}$ & (0.32 $\pm$ 0.07) & 1.09/6 & 0.18 \\
\enddata
\end{deluxetable}

\subsection{Comparison to Optical and Hard X-ray Results}
\label{Comparison}
Finally, we compare our measurement of the CF based on the MIR data
with that based on optical and hard X-ray data. Any comparison to
results obtained from a different data set should be undertaken
carefully because the luminosity and redshift ranges may be
different. In particular, as there is no consensus on the redshift
dependence of the CF at $z > 0.2$, if the CF depends strongly on
redshift as reported by previous studies \citep[e.g.,][]{La
Franca,Hasinger}, then this would affect any comparisons. Hence, we
compared our results with those obtained from \cite{Simpson+05}
and \cite{Hasinger}, who presented the relationship between the CF
and the [OIII] luminosity at $z < 0.3$ \citep{Simpson+05} and the
hard X-ray (2--10 keV) luminosity \citep{Hasinger} at $z <
0.2$\footnote{\cite{Hasinger} also treats the high redshift data ($z
< 5.2$), but we only use the data in the $0.015 < z < 0.2$ redshift
bin for the comparison. See Figure 8 in \cite{Hasinger}.}.
It should be noted that here we also restrict the sample to those objects at $z \leq 0.15$ to omit as much as possible the effects of optically faint WISE sources(see Section \ref{rejected}).

We also compared the results with those obtained from
higher energy X-ray band data, because the 2--10-keV-band-based
surveys could fail to detect heavily obscured luminous
AGN, as has been reported by, e.g., \cite{Mateos}. We referred to
two papers: \cite{Beckmann_09}, who analyzed data for 199 AGNs
supposedly detected by {\it INTEGRAL} above 20 keV and reported a
negative correlation between the fraction of absorbed/type 2 AGNs
and the hard X-ray (20--100 keV) luminosity; and \cite{Burlon}, who
also reported the same correlation on the basis of 15--55-keV data.
The sample data in both papers have been examined, and thus we
estimated the CF using the data at $z \leq$ 0.15.
We assume
here that the dependence of the CF on redshift is very weak or
almost constant even at $z \leq 0.3$, which enables us to
compare directly the luminosity dependence without considering the effect of
the redshift dependence.

\begin{figure}
        \epsscale{1}
         \plotone{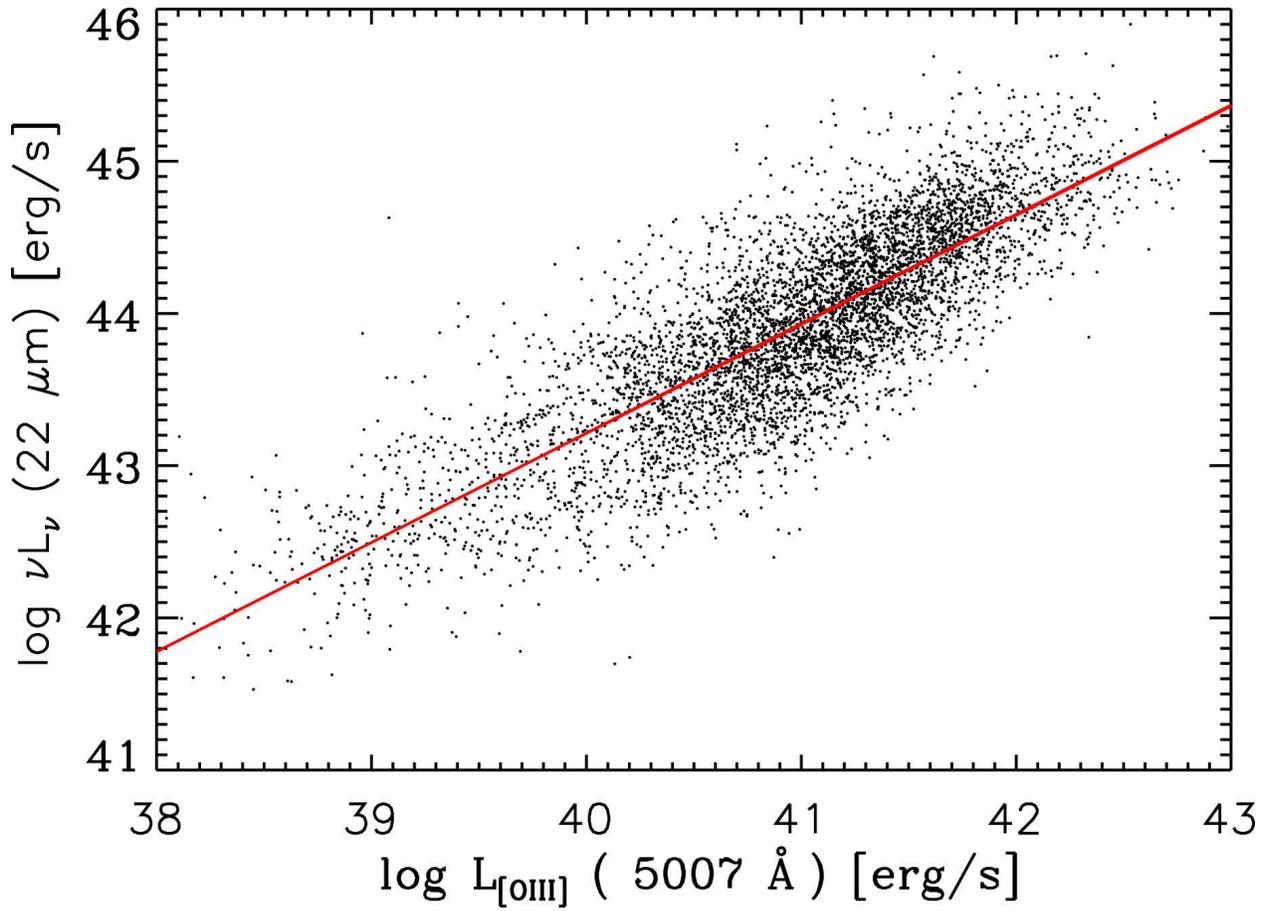}
        \caption{Plot of $\log L_{\mathrm{[OIII]}}$ versus $\nu L_{\nu}$(22 $\micron$). The solid red line shows the best-fit linear function.}
        \label{L_22_OIII}
    \end{figure}

For the comparison, we converted the [OIII] luminosity
(L$_{\mathrm{[OIII]}}$ and hard X-ray luminosity (L$_X$) to the
22-$\mu$m luminosity (L$_{\mathrm{MIR}}$) using the following
conversion formulae:
\begin{eqnarray}
\log \left(\frac{L_{\mathrm{MIR}}}{10^{43}} \right) & = & (2.36 \pm 0.01) + (0.76 \pm 0.01) \log \left(\frac{L_{\mathrm{[OIII]}}}{10^{43}}\right), \\
\label{L_MIR_HX}
\log \left(\frac{L_{\mathrm{MIR}}}{10^{43}} \right) & = & (0.19 \pm 0.05) + (1.11 \pm 0.07) \log \left(\frac{L_\mathrm{X}}{10^{43}}\right), \\
\log \left(\frac{L_{\mathrm{MIR}}}{10^{43}} \right) & = & (0.27 \pm 0.05) + (0.89 \pm 0.04) \log \left(\frac{L_\mathrm{X\,(14-195\,keV)}}{10^{43}}\right),
\end{eqnarray}
where the luminosities are normalized to $10^{43}$ erg s$^{-1}$. For
the [OIII] luminosity, we calculated the conversion factors by
plotting $\log L_{\mathrm{[OIII]}}$ versus $\log [\nu L_{\nu}$(22
$\micron$)], as shown in Figure \ref{L_22_OIII}. To ensure the
accuracy of the conversion, high-SN ($>$10) objects (type 1AGNs,
type 2 AGNs, LINERs, and Composites) are plotted. For the hard X-ray
luminosity, we used Equation (5) in \cite{Gandhi}
and Equation (2) when considering the entire sample at 18 $\mu$m in
\cite{Matsuta}.
As \cite{Gandhi} derived the relationship between the 12.3-$\mu$m
and hard X-ray luminosity, the intrinsic error associated with the
conversion in this case will be somewhat different to that of
Equation (\ref{L_MIR_HX}).
Similarly, \cite{Matsuta} derived the relationship between the
18-$\mu$m and hard X-ray (14--195 keV) luminosity from {\it AKARI}
and {\it Swift}-BAT, respectively, and thus some uncertainty may
arise from the conversion.
   \begin{figure}
        \epsscale{1}
        \plotone{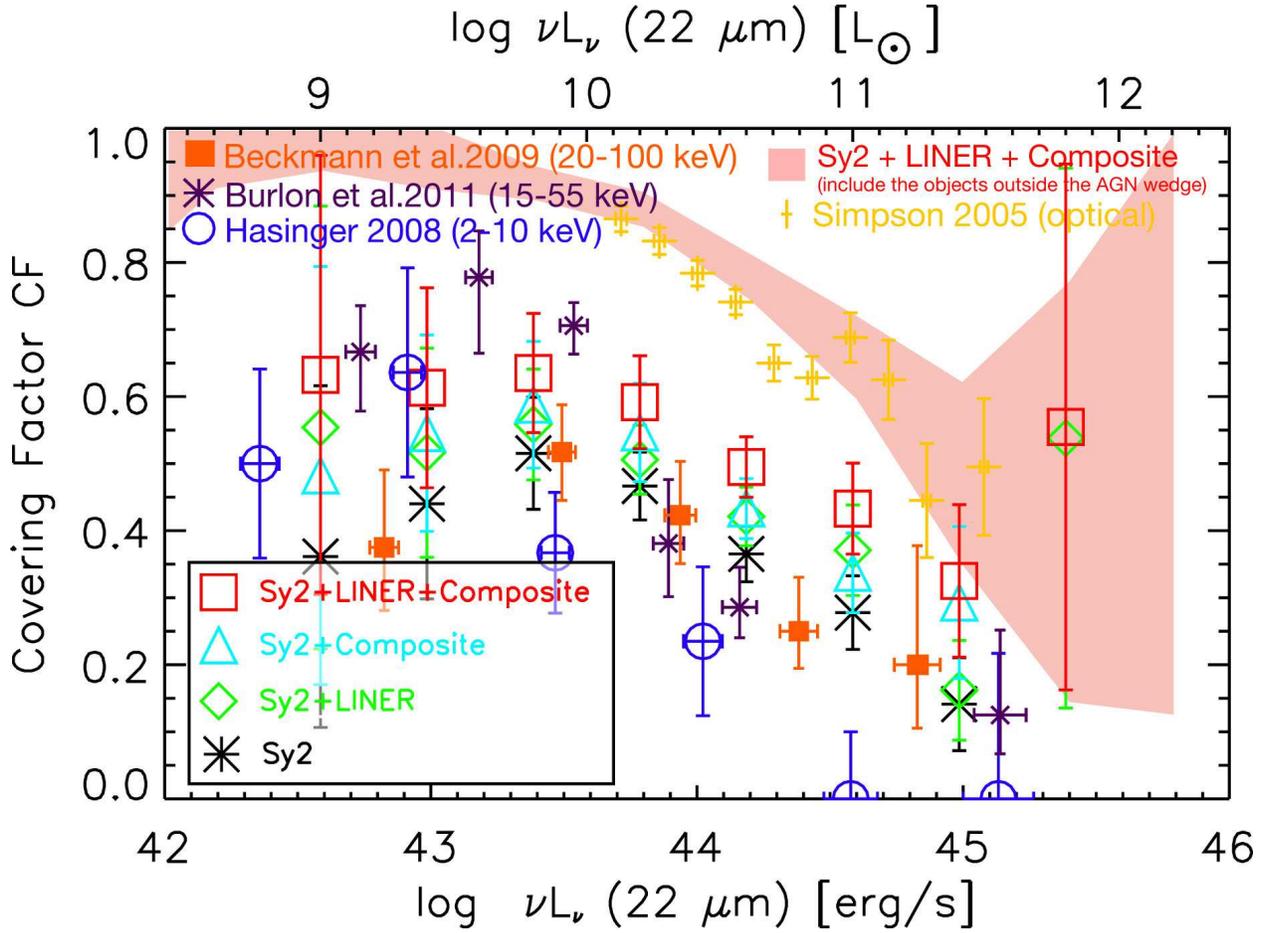}
        \caption{Comparison of our measured CF ($z \leq$ 0.15) with those of optical
\citep[yellow plus:][]{Simpson+05}, hard X-ray \citep[blue
circle:][]{Hasinger}, 15--55-keV \citep[purple asterisk:][]{Burlon}, and 20--100-keV
\citep[orange filled square:][]{Beckmann_09} studies. Errors in the
CFs estimated from \cite{Beckmann_09} and \cite{Burlon} were determined using binomial statistics \citep[see][]{Gehrels},
drawn at the 1$\sigma$ level.
The CFs including Sy2 + LINER +
Composite galaxies, which are the total sample including the outside the AGN wedge, are
also plotted (red shaded region; see Appendix
\ref{CF_without_wedge}). The conversion uncertainty, represented in
Figure, is shown as the horizontal error bars for the optical and hard
X-ray measurements.}
    \label{CF_Comparison}
    \end{figure}

Figure \ref{CF_Comparison} compares our measurements with those of
the optical \citep{Simpson+05}, hard X-ray \citep{Hasinger},
15--55-keV \citep{Burlon}, and 20--100-keV \citep{Beckmann_09}
studies. Uncertainties (1$\sigma$ level) in the CFs obtained from
\cite{Beckmann_09} and \cite{Burlon} were estimated by binomial
statistics \citep[see][]{Gehrels} in the same manner as
\cite{Burlon}.
The optically-based CF has a larger value than ours over a wide
range of luminosities, but the shape of the decrease is
similar. To examine the reasons for the differences, the CFs
obtained from objects including Sy2 + LINER + Composite galaxies,
which are the total sample including the AGN outside the wedge, are also
plotted (data are available in Table \ref{CF_table_22}). In that
case, the optical data are in good agreement with ours, which
indicates that the optical ([OIII])-based selection could be affected by
host galaxy contributions. Indeed, \cite{Caccianiga}, who investigated the
nature of all sources (35 in total) in the {\it XMM-Newton} bright
serendipitous survey, showed an optical spectrum dominated by the
light from the host galaxy with no evidence (or little evidence) for
the presence of an AGN.
In contrast, the hard X-ray based CF are consistent with ours, although our MIR-emission-based data exceed the 2--10 keV data by a substantial amount, which indicates that 2--10 keV based surveys fail to detect heavily obscured/absorbed luminous AGN as expected above.
Ultimately, our MIR selection with the AGN wedge may avoid the problems associated with optical selection
, as well as hard X-ray ($>$2 keV) based studies.
We note that the CF derived from the 22-$\mu$m sample in the AGN wedge would be an underestimate due to the lack of optically faint (PetroMag\_r $>$ 17.7) MIR sources as described in Section \ref{rejected}.
Thus, there may be a small difference between our MIR results and the optical results, although it may be difficult to fill in the gap using only optically faint type 2 AGNs.

\section{SUMMARY}
Using the {\it WISE} MIR all-sky survey, we constructed 12- and 22-$\mu$m LFs for all types of local galaxies.
Using complete optical spectroscopy of emission lines, we classified the galaxies based on the cataloged classifications in the SDSS and their emission line ratios ([NII]/H$\alpha$ and [OIII]/H$\beta$).
We classified the {\it WISE} sources into type 1 AGNs, type 2 AGNs, LINERs, Composites, and SFs.
We then calculated the number densities of the type 1 and type 2 AGNs by integrating each LF and estimated
the CF of the dust torus (the fraction of type 2 AGNs among all AGNs).
In particular, we examined the luminosity and redshift dependence of the CF for $\sim3,000$ AGN-dominated MIR sources which were extracted by examining their MIR colors.
The main results are as follows:
\begin{enumerate}
\item Less luminous AGNs in the MIR region are affected by a contribution from their host SF.
\item The CF decreases with increasing 22-$\mu$m luminosity regardless of the choice of type 2 AGN classification criteria, although this dependence is relatively weaker than previous studies.
\item The CF does not change significantly with the redshift ($z < 0.2$).
\item The luminosity dependence of the CF can be interpreted using the receding torus model. This luminosity dependence is better described by the modified receding torus model in which the height of the torus is parameterized.
\item Measurements of the CF based on optical survey data exceed our data but are in good agreement if contributions of the host galaxy (i.e., without adopting the AGN wedge selection) are not considered. In contrast, measurements of the CF based on
hard X-ray survey data are almost consistent with ours.
These trends may indicate that optical survey data is affected by
the host galaxy contribution.
\end{enumerate}
Our study has confirmed and extended previous results obtained with {\it IRAS}, {\it Spitzer}, and {\it AKARI} by constructing a much larger MIR-selected sample with {\it WISE}.
The large number of galaxies in the sample we obtained here means that the variation in the CF with the luminosity and redshift is described with a higher statistical accuracy and lower systematic errors than previous results.
We emphasize that a luminosity-dependent torus geometry destroys the simplicity of the original torus unification scheme and now requires that at least one new free function must be determined.
Our results are inconsistent with the simplest unified scheme, which expects that the CF is independent of the luminosity.
A modification of this simple zero-order unification scheme is required.
The present results with WISE have provided us with an important local benchmark for AGN studies at high redshifts.

\acknowledgments 
The authors gratefully acknowledge the anonymous referee
for a careful reading of the manuscript and very helpful
comments.
We are also deeply thankful to Drs Guenther Hasinger (Institute for Astronomy) and Chris Simpson (Liverpool John Moores University) who kindly provided data for comparison.
We also thank Drs Tadayasu Dotani (ISAS/JAXA), Yoshihiro Ueda (Kyoto University), Tohru Nagao (Ehime University), and Akihiro Doi (ISAS/JAXA) for their relevant comments. 

This publication makes use of data products from the Wide-field
Infrared Survey Explorer, which is a joint project of the University
of California, Los Angeles, and the Jet Propulsion
Laboratory/California Institute of Technology, funded by the
National Aeronautics and Space Administration. 
Funding for SDSS-III
has been provided by the Alfred P. Sloan Foundation, the
Participating Institutions, the National Science Foundation, and the
U.S. Department of Energy Office of Science. The SDSS-III web site
is \url{http://www.sdss3.org/}. SDSS-III is managed by the
Astrophysical Research Consortium for the Participating Institutions
of the SDSS-III Collaboration including the University of Arizona,
the Brazilian Participation Group, Brookhaven National Laboratory,
University of Cambridge, Carnegie Mellon University, University of
Florida, the French Participation Group, the German Participation
Group, Harvard University, the Instituto de Astrofisica de Canarias,
the Michigan State/Notre Dame/JINA Participation Group, Johns
Hopkins University, Lawrence Berkeley National Laboratory, Max
Planck Institute for Astrophysics, Max Planck Institute for
Extraterrestrial Physics, New Mexico State University, New York
University, Ohio State University, Pennsylvania State University,
University of Portsmouth, Princeton University, the Spanish
Participation Group, University of Tokyo, University of Utah,
Vanderbilt University, University of Virginia, University of
Washington, and Yale University. 
This research has made use of the
SIMBAD database, operated at CDS, Strasbourg, France. 
Also, this research
has made use of the NASA/IPAC Extragalactic Database (NED) which is
operated by the Jet Propulsion Laboratory, California Institute of
Technology, under contract with the National Aeronautics and Space
Administration.
GAMA is a joint European-Australasian project based around a spectroscopic campaign using the Anglo-Australian Telescope. The GAMA input catalogue is based on data taken from the Sloan Digital Sky Survey and the UKIRT Infrared Deep Sky Survey. Complementary imaging of the GAMA regions is being obtained by a number of independent survey programs including GALEX MIS, VST KiDS, VISTA VIKING, WISE, Herschel-ATLAS, GMRT and ASKAP providing UV to radio coverage. GAMA is funded by the STFC (UK), the ARC (Australia), the AAO, and the participating institutions. The GAMA website is \url{http://www.gama-survey.org/}.
P.G. acknowledges support from STFC grant reference
ST/J00369711.

\appendix

\section{The V/V$_{\mathrm{max}}$ Test}
Before constructing the LFs in Section \ref{Vmax_method}, we performed the standard $V/V_{\mathrm{max}}$ test \citep{Schmidt} to examine whether the spatial distribution of the sources in a sample is uniform. Here, $V$ is the volume enclosed at the redshift of an object, and $V_{\mathrm{max}}$ is the volume that would be enclosed at the maximum redshift at which the object could be detected.
If the sample is complete, the mean value of $V/V_{\mathrm{max}}$ should be 0.5.
However, the redshift evolution (i.e., the increase in the number density and/or luminosity of the sources with their redshift) is expected to increase this value.
For example, \cite{Rodighiero} reported that the bolometric IR luminosity density evolves as $(1 + z)^{3.8\pm0.4}$ in the redshift interval $0 < z < 1$ based on a combination of data from deep {\it Spitzer} surveys of the VIMOS VLT Deep Survey (VVDS-SWIRE) and GOODS fields.
In fact, the average value of $V/V_{\mathrm{max}}$ for the {\it WISE}--SDSS sample is 0.51 $\pm$ 0.001 for the 12-$\mu$m sample and 0.52 $\pm$ 0.001 for the 22-$\mu$m sample.
This means that our sample ($0.006 \leq z \leq 0.3$) could also be expected to be affected by redshift
evolution.
However, the average values for each redshift bin (we assumed that there is no evolution within a small redshift range) are close to 0.5.
Note that the redshift interval in each redshift bin is the same as that in Section \ref{LF}.
     \begin{figure}
        \plotone{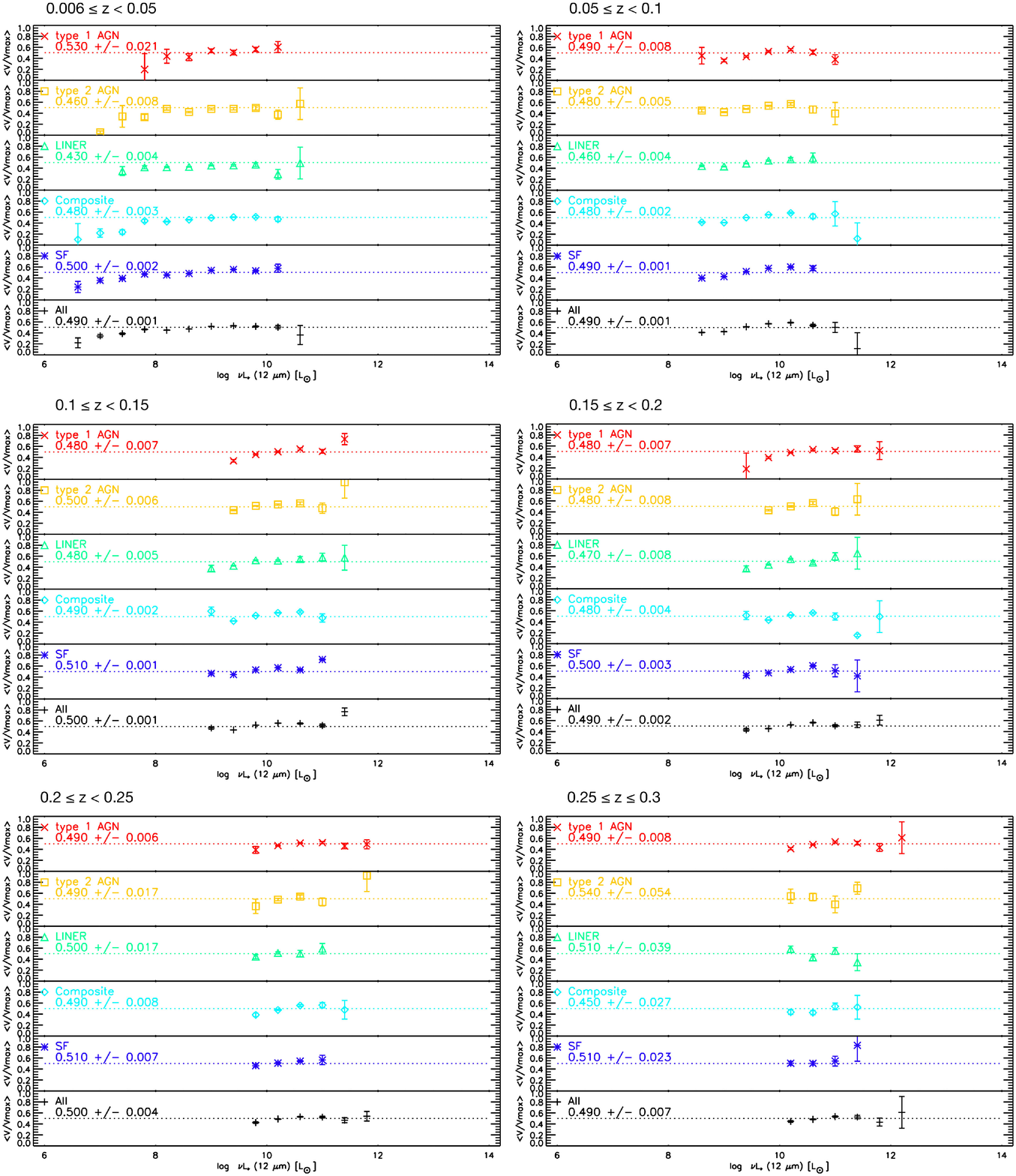}
        \caption{Average values of $V/V_{\mathrm{max}}$ for different galaxy types as a function of the 12-$\mu$m luminosity in each redshift
bin.}
    \label{V_Vmax_test_12}
    \end{figure}
     \begin{figure}
        \plotone{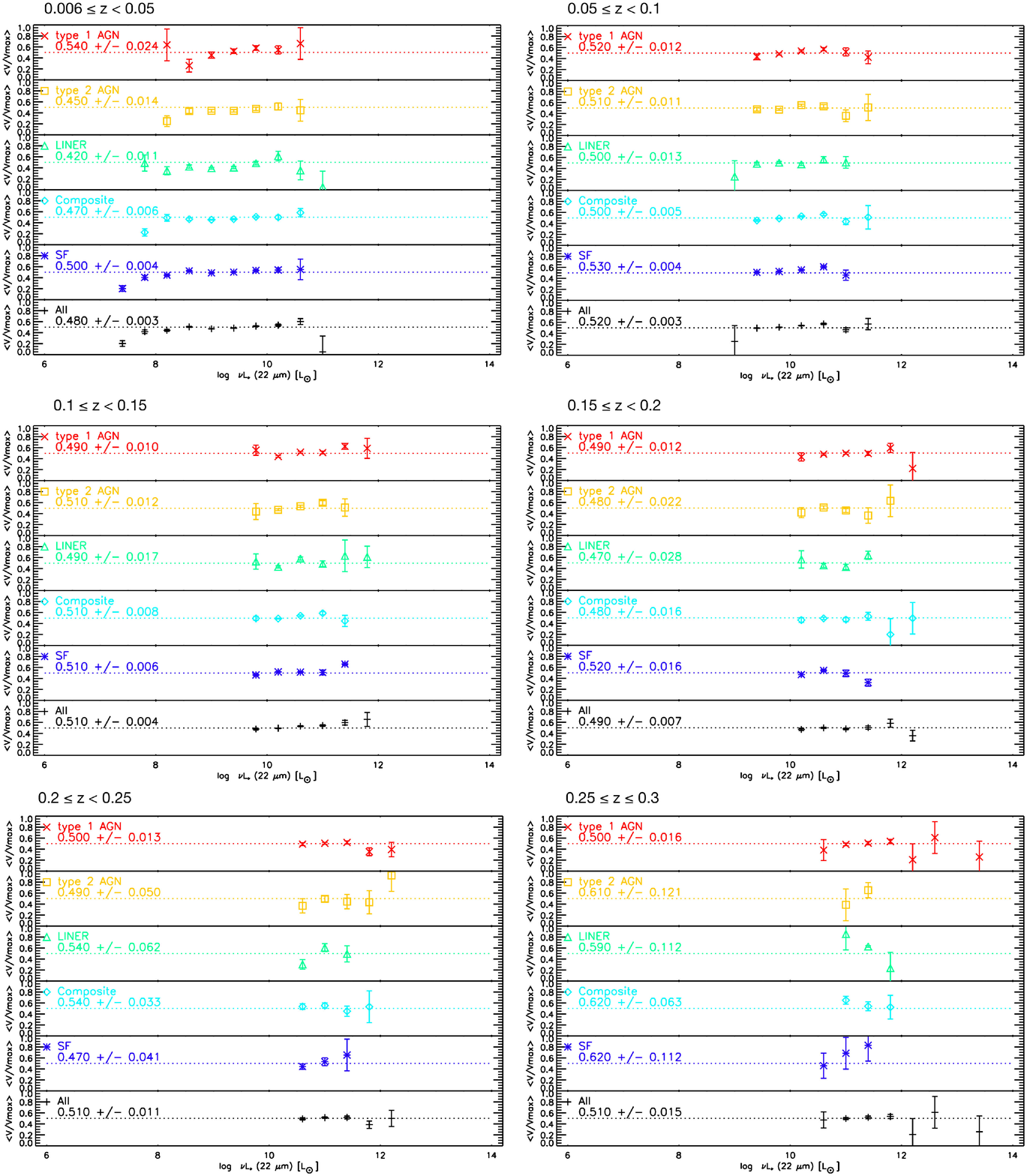}
        \caption{Average values of $V/V_{\mathrm{max}}$ for different galaxy types as a function of the 22-$\mu$m luminosity in each redshift
bin.}
    \label{V_Vmax_test_22}
    \end{figure}

Figures \ref{V_Vmax_test_12} and \ref{V_Vmax_test_22} show the average values of $V/V_{\mathrm{max}}$, $<V/V_{\mathrm{max}}>$ for each galaxy type at 12 and 22 $\mu$m in different redshift bin.
The average values are distributed around 0.5.

\section{Luminosity Functions}
\label{LF_data}
The LF data for each galaxy type in each redshift bin are summarized in Tables \ref{LF_table_ALL_z}, \ref{LF_table_12_z}, and \ref{LF_table_22_z}.
Note that the data in Table \ref{LF_table_ALL_z} is used in Figure \ref{LF_ALL}, those in Tables \ref{LF_table_12_z} and \ref{LF_table_22_z} are used in Figure \ref{LF_type}, and those in Tables \ref {LF_table_12_z} and \ref{LF_table_22_z} are used in Figures \ref{LF_type_12_z} and \ref{LF_type_22_z} (see Section \ref{LF}).

\begin{deluxetable}{rrrrrrrr}
\tablecolumns{8}
\tablewidth{0pc}
\tabletypesize{\scriptsize}
\tablecaption{LFs of all the galaxies at 12 and 22 $\mu$m for each redshift bin.\label{LF_table_ALL_z}} 
\tablehead{
\colhead{}  &  	\multicolumn{3}{c}{12 $\mu$m} & \colhead{} & \multicolumn{3}{c}{22 $\mu$m} \\	
\cline{2-4}  \cline{6-8} \\
$\log L\tablenotemark{a}$ 		& 	\multicolumn{1}{c}{$\phi$\tablenotemark{b}} & \multicolumn{1}{c}{$\sigma_{\phi}$\tablenotemark{b}} & N & \colhead{} &
  					\multicolumn{1}{c}{$\phi$} & \multicolumn{1}{c}{$\sigma_{\phi}$} & N \\
}	 
\startdata
\cutinhead{$0.006 \leq z < 0.05$}
      6.60 &  1.05$\times 10^{-3}$ &  4.18$\times 10^{-4}$ &     7 & &       \multicolumn{1}{c}{\nodata} &       \multicolumn{1}{c}{\nodata} &     0 \\
      7.00 &  3.04$\times 10^{-3}$ &  3.51$\times 10^{-4}$ &    99 & &       \multicolumn{1}{c}{\nodata} &       \multicolumn{1}{c}{\nodata} &     0 \\
      7.40 &  1.93$\times 10^{-3}$ &  1.19$\times 10^{-4}$ &   315 & &  2.34$\times 10^{-3}$ &  7.70$\times 10^{-4}$ &    14 \\
      7.80 &  1.77$\times 10^{-3}$ &  5.49$\times 10^{-5}$ &  1255 & &  1.59$\times 10^{-3}$ &  2.13$\times 10^{-4}$ &    72 \\
      8.20 &  2.52$\times 10^{-3}$ &  3.49$\times 10^{-5}$ &  6198 & &  1.23$\times 10^{-3}$ &  8.65$\times 10^{-5}$ &   242 \\
      8.60 &  2.41$\times 10^{-3}$ &  2.11$\times 10^{-5}$ & 13939 & &  1.19$\times 10^{-3}$ &  3.99$\times 10^{-5}$ &  1049 \\
      9.00 &  1.78$\times 10^{-3}$ &  1.78$\times 10^{-5}$ & 10972 & &  1.05$\times 10^{-3}$ &  2.02$\times 10^{-5}$ &  3159 \\
      9.40 &  6.43$\times 10^{-4}$ &  1.02$\times 10^{-5}$ &  4003 & &  4.72$\times 10^{-4}$ &  1.01$\times 10^{-5}$ &  2863 \\
\enddata
\tablecomments{Table \ref{LF_table_ALL_z} is published in its entirety in the 
electronic edition of the {\it Astrophysical Journal}.  A portion is 
shown here for guidance regarding its form and content.}
\tablenotetext{a}{$\mathrm{L}_{\odot}$}
\tablenotetext{b}{Mpc$^{-3}$ Mag$^{-1}$}
\end{deluxetable}

\begin{deluxetable}{rrrrrrrrrrrrrrrrrrrrrrrr}
\tablecolumns{24}
\tablewidth{0pc}
\setlength{\tabcolsep}{0.02in} 
\rotate
\tabletypesize{\scriptsize}
\tablecaption{The 12-$\micron$ LFs for each galaxy category at each redshift.\label{LF_table_12_z}}
\tablehead{
\colhead{}  &  	\multicolumn{3}{c}{type 1} & \colhead{} & \multicolumn{3}{c}{type 2} & 
\colhead{}  &	\multicolumn{3}{c}{LINER}  & \colhead{} & \multicolumn{3}{c}{Composite} & 
\colhead{}  &	\multicolumn{3}{c}{SF} 	   & \colhead{} & \multicolumn{3}{c}{Unknown} \\	
\cline{2-4}  \cline{6-8} \cline{10-12} \cline{14-16} \cline{18-20} \cline{22-24}  \\
$\log L\tablenotemark{a}$ 		& 	\multicolumn{1}{c}{$\phi$\tablenotemark{b}} & \multicolumn{1}{c}{$\sigma_{\phi}$\tablenotemark{b}} & N & \colhead{} &
					\multicolumn{1}{c}{$\phi$} & \multicolumn{1}{c}{$\sigma_{\phi}$} & N & \colhead{} &
					\multicolumn{1}{c}{$\phi$} & \multicolumn{1}{c}{$\sigma_{\phi}$} & N & \colhead{} &
					\multicolumn{1}{c}{$\phi$} & \multicolumn{1}{c}{$\sigma_{\phi}$} & N & \colhead{} &
					\multicolumn{1}{c}{$\phi$} & \multicolumn{1}{c}{$\sigma_{\phi}$} & N & \colhead{} &
  					\multicolumn{1}{c}{$\phi$} & \multicolumn{1}{c}{$\sigma_{\phi}$} & N \\
}	 
\startdata
\cutinhead{$0.006 \leq z < 0.05$}
      6.60 &        \multicolumn{1}{c}{\nodata} &        \multicolumn{1}{c}{\nodata} &     0 & &        \multicolumn{1}{c}{\nodata} &        \multicolumn{1}{c}{\nodata} &     0 & &        \multicolumn{1}{c}{\nodata} &        \multicolumn{1}{c}{\nodata} &     0 & &   1.14$\times 10^{-4}$ &   1.14$\times 10^{-4}$ &     1 & &   9.40$\times 10^{-4}$ &   4.03$\times 10^{-4}$ &     6 & &        \multicolumn{1}{c}{\nodata} &        \multicolumn{1}{c}{\nodata} &     0 \\
      7.00 &        \multicolumn{1}{c}{\nodata} &        \multicolumn{1}{c}{\nodata} &     0 & &   6.98$\times 10^{-5}$ &   5.49$\times 10^{-5}$ &     2 & &        \multicolumn{1}{c}{\nodata} &        \multicolumn{1}{c}{\nodata} &     0 & &   1.47$\times 10^{-4}$ &   6.74$\times 10^{-5}$ &     5 & &   2.82$\times 10^{-3}$ &   3.40$\times 10^{-4}$ &    92 & &        \multicolumn{1}{c}{\nodata} &        \multicolumn{1}{c}{\nodata} &     0 \\
      7.40 &        \multicolumn{1}{c}{\nodata} &        \multicolumn{1}{c}{\nodata} &     0 & &   7.72$\times 10^{-6}$ &   5.37$\times 10^{-6}$ &     2 & &   4.17$\times 10^{-5}$ &   1.69$\times 10^{-5}$ &     8 & &   6.11$\times 10^{-5}$ &   1.97$\times 10^{-5}$ &    12 & &   1.82$\times 10^{-3}$ &   1.16$\times 10^{-4}$ &   293 & &        \multicolumn{1}{c}{\nodata} &        \multicolumn{1}{c}{\nodata} &     0 \\
      7.80 &   1.39$\times 10^{-6}$ &   1.39$\times 10^{-6}$ &     1 & &   3.29$\times 10^{-5}$ &   7.14$\times 10^{-6}$ &    26 & &   1.21$\times 10^{-4}$ &   1.39$\times 10^{-5}$ &    88 & &   1.21$\times 10^{-4}$ &   1.34$\times 10^{-5}$ &    94 & &   1.48$\times 10^{-3}$ &   5.07$\times 10^{-5}$ &  1038 & &   1.20$\times 10^{-5}$ &   4.29$\times 10^{-6}$ &     8 \\
\enddata
\tablecomments{Table \ref{LF_table_12_z} is published in its entirety in the electronic edition of the {\it Astrophysical Journal}.  A portion is shown here for guidance regarding its form and content.}
\tablenotetext{a}{$\mathrm{L}_{\odot}$}
\tablenotetext{b}{Mpc$^{-3}$ Mag$^{-1}$}
\end{deluxetable}

\begin{deluxetable}{rrrrrrrrrrrrrrrrrrrrrrrr}
\tablecolumns{24}
\tablewidth{0pc}
\setlength{\tabcolsep}{0.02in} 
\rotate
\tabletypesize{\scriptsize}
\tablecaption{The 22-$\micron$ LFs for each galaxy category at each redshift.\label{LF_table_22_z}}
\tablehead{
\colhead{}  &  	\multicolumn{3}{c}{type 1} & \colhead{} & \multicolumn{3}{c}{type 2} & 
\colhead{}  &	\multicolumn{3}{c}{LINER}  & \colhead{} & \multicolumn{3}{c}{Composite} & 
\colhead{}  &	\multicolumn{3}{c}{SF} 	   & \colhead{} & \multicolumn{3}{c}{Unknown} \\	
\cline{2-4}  \cline{6-8} \cline{10-12} \cline{14-16} \cline{18-20} \cline{22-24}  \\
$\log L\tablenotemark{a}$ 		& 	\multicolumn{1}{c}{$\phi$\tablenotemark{b}} & \multicolumn{1}{c}{$\sigma_{\phi}$\tablenotemark{b}} & N & \colhead{} &
					\multicolumn{1}{c}{$\phi$} & \multicolumn{1}{c}{$\sigma_{\phi}$} & N & \colhead{} &
					\multicolumn{1}{c}{$\phi$} & \multicolumn{1}{c}{$\sigma_{\phi}$} & N & \colhead{} &
					\multicolumn{1}{c}{$\phi$} & \multicolumn{1}{c}{$\sigma_{\phi}$} & N & \colhead{} &
					\multicolumn{1}{c}{$\phi$} & \multicolumn{1}{c}{$\sigma_{\phi}$} & N & \colhead{} &
  					\multicolumn{1}{c}{$\phi$} & \multicolumn{1}{c}{$\sigma_{\phi}$} & N \\
}	 
\startdata
\cutinhead{$0.006 \leq z < 0.05$}
      7.40 &        \multicolumn{1}{c}{\nodata} &        \multicolumn{1}{c}{\nodata} &    0 & &        \multicolumn{1}{c}{\nodata} &        \multicolumn{1}{c}{\nodata} &    0 & &        \multicolumn{1}{c}{\nodata} &        \multicolumn{1}{c}{\nodata} &    0 & &        \multicolumn{1}{c}{\nodata} &        \multicolumn{1}{c}{\nodata} &    0 & &   2.34$\times 10^{-3}$ &   7.70$\times 10^{-4}$ &   14 & &        \multicolumn{1}{c}{\nodata} &        \multicolumn{1}{c}{\nodata} &    0 \\
      7.80 &        \multicolumn{1}{c}{\nodata} &        \multicolumn{1}{c}{\nodata} &    0 & &        \multicolumn{1}{c}{\nodata} &        \multicolumn{1}{c}{\nodata} &    0 & &   4.48$\times 10^{-5}$ &   2.24$\times 10^{-5}$ &    4 & &   1.29$\times 10^{-4}$ &   6.72$\times 10^{-5}$ &    4 & &   1.40$\times 10^{-3}$ &   2.00$\times 10^{-4}$ &   63 & &   1.93$\times 10^{-5}$ &   1.93$\times 10^{-5}$ &    1 \\
      8.20 &   6.02$\times 10^{-6}$ &   6.02$\times 10^{-6}$ &    1 & &   5.08$\times 10^{-5}$ &   1.82$\times 10^{-5}$ &    9 & &   5.92$\times 10^{-5}$ &   1.78$\times 10^{-5}$ &   13 & &   1.03$\times 10^{-4}$ &   2.27$\times 10^{-5}$ &   24 & &   1.01$\times 10^{-3}$ &   7.92$\times 10^{-5}$ &  195 & &        \multicolumn{1}{c}{\nodata} &        \multicolumn{1}{c}{\nodata} &    0 \\
      8.60 &   1.57$\times 10^{-6}$ &   1.01$\times 10^{-6}$ &    2 & &   2.87$\times 10^{-5}$ &   5.96$\times 10^{-6}$ &   27 & &   9.30$\times 10^{-5}$ &   1.08$\times 10^{-5}$ &   88 & &   1.62$\times 10^{-4}$ &   1.36$\times 10^{-5}$ &  163 & &   9.00$\times 10^{-4}$ &   3.53$\times 10^{-5}$ &  761 & &   1.01$\times 10^{-5}$ &   3.80$\times 10^{-6}$ &    8 \\
\enddata
\tablecomments{Table \ref{LF_table_22_z} is published in its entirety in the electronic edition of the {\it Astrophysical Journal}.  A portion is shown here for guidance regarding its form and content.}
\tablenotetext{a}{$\mathrm{L}_{\odot}$}
\tablenotetext{b}{Mpc$^{-3}$ Mag$^{-1}$}
\end{deluxetable}

\section{Luminosity and Redshift Dependences of the CF for the Sample Including the Outside the AGN wedge}
\label{CF_without_wedge} 
\begin{figure}
    \epsscale{1}
    \plottwo{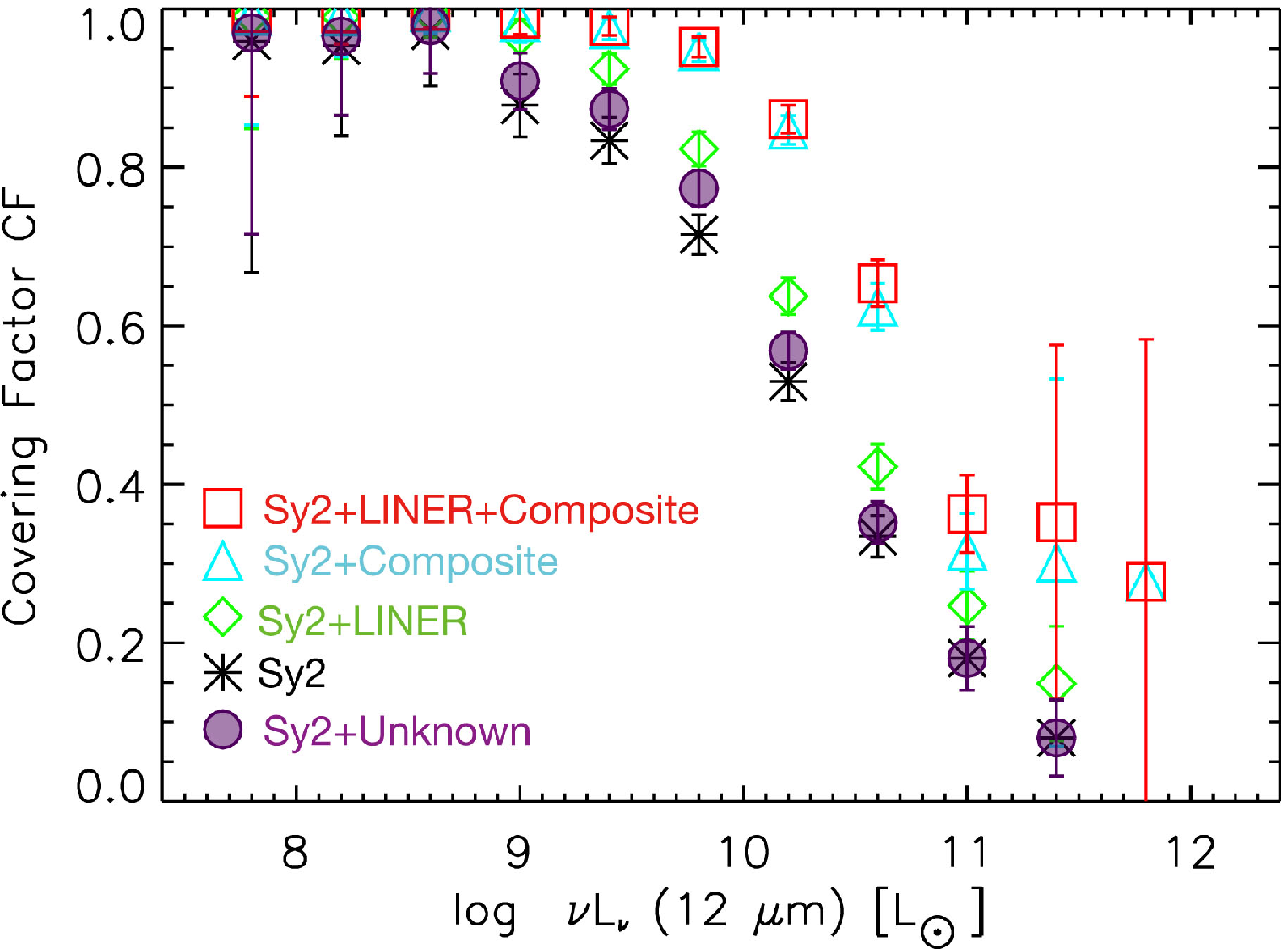}{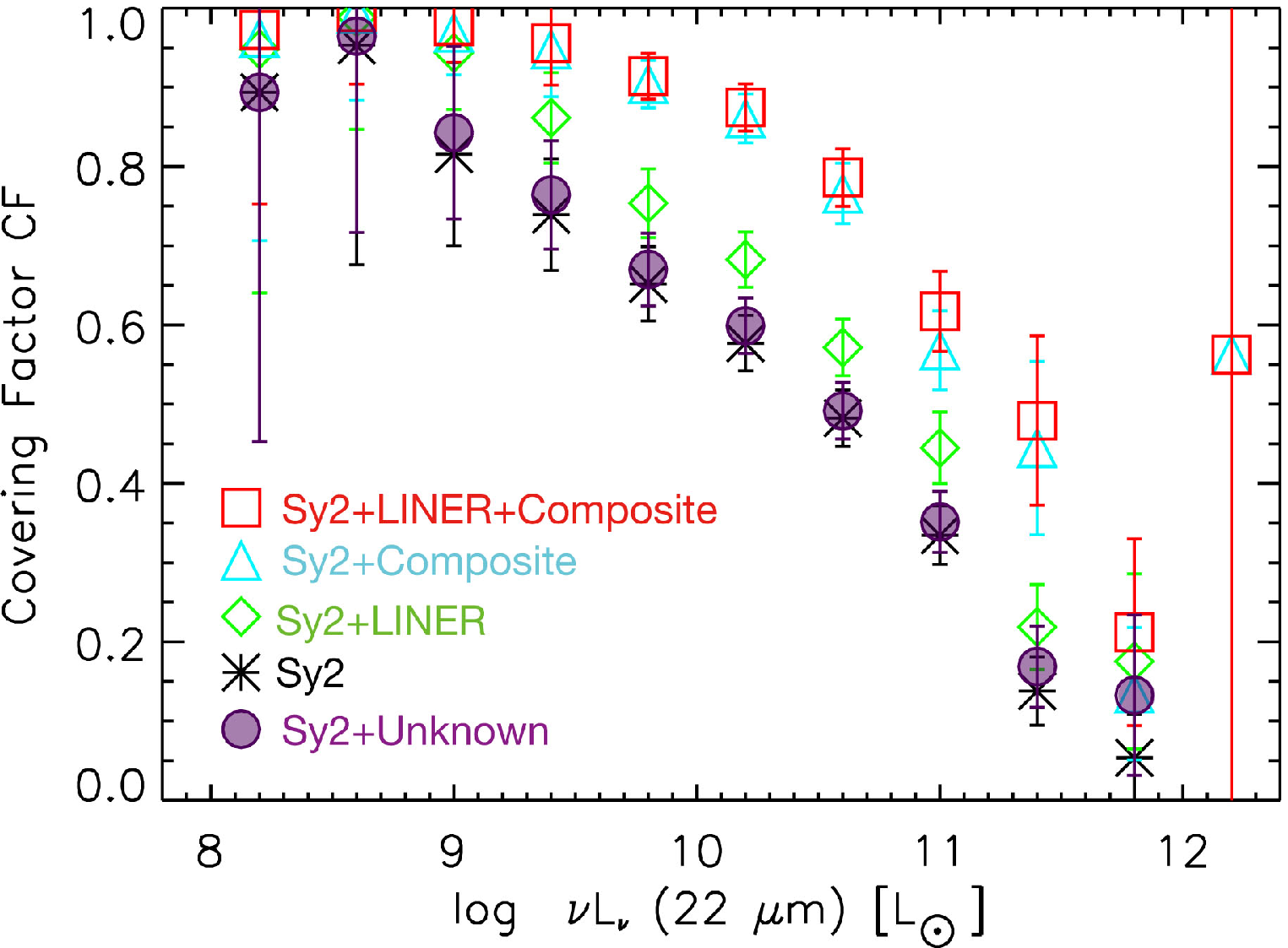}
    \caption{Variation in the CF with the 12-$\mu$m (left) and 22-$\mu$m (right) luminosities at $0.006 \leq z \leq 0.2$ without considering the AGN wedge.}
    \label{CF_L_ALL-z_12_22}
    \end{figure}

\begin{figure}
    \epsscale{1}
    \plottwo{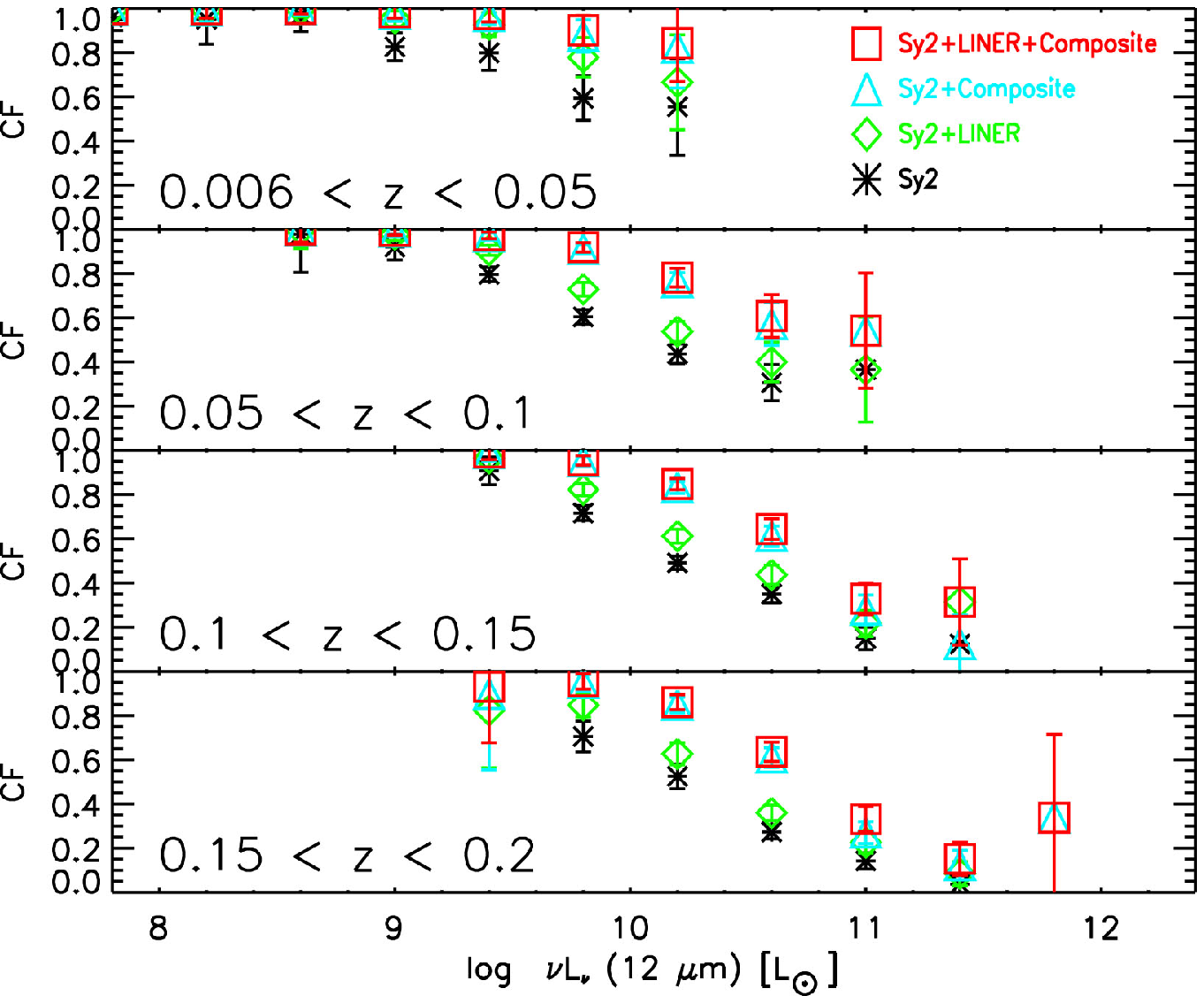}{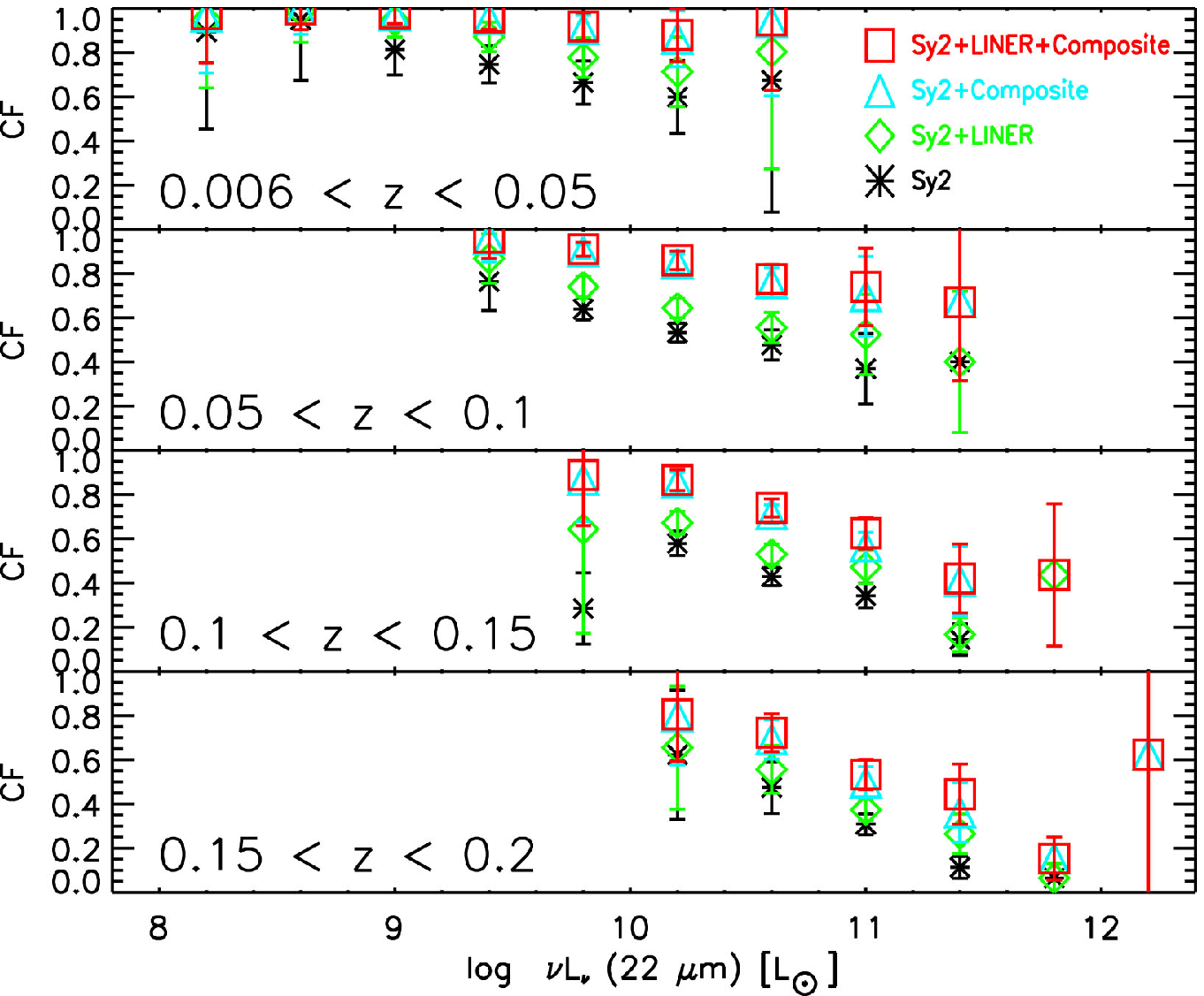}
    \caption{Variation in the CF with the 12-$\mu$m (left) and 22-$\mu$m (right) luminosities for all samples in different redshift bins without considering the AGN wedge.}
    \label{CF_L_multi-z_12_22}
    \end{figure}

In Section \ref{CF_vs_L}, we demonstrated
that the MIR emission from optically classified (especially for less
luminous) AGNs could be substantially affected by emission from
their host galaxies, and we attempted to extract the AGN-dominated
MIR objects based on their MIR colors (AGN wedge selection) to
calculate the CF. We also discussed only 22-$\mu$m-selected AGNs in
Section \ref{CF_vs_L} to omit the influence of the PAH contribution.
Nevertheless, it is worthwhile presenting the estimated CF for the sample including the outside the AGN wedge (i.e., without considering the AGN wedge) and a CF estimate from the 12-$\mu$m
sample. Hence, we show the values of the CF based on the 12- and
22-$\mu$m {\it WISE}-SDSS sample. Figures \ref{CF_L_ALL-z_12_22} and
\ref{CF_L_multi-z_12_22} represent the resultant CF at $0.006 \leq z
\leq 0.2$ and in each redshift bin, respectively. The data in these
figures are also given in Tables \ref{CF_table_12} and
\ref{CF_table_22}. As described in Section \ref{CF_vs_L}, the CF
seems to be overestimated particularly in the low-luminosity region,
compared to those in Figure \ref{CF_multi_z_22_lin_AGN}.
In addition, when comparing Figures \ref{CF_L_RTM_AGN} and \ref{CF_L_ALL-z_12_22} (right), there are many low luminosity type 2 AGNs that are missing from the ``wedge'' selected sample that are classed as AGN in the SDSS. 
In Section \ref{interpretation_CF_L}, we performed model fitting to the data in the AGN wedge and concluded that the modified receding torus model explains the data well. 
Since only AGN-dominant MIR sources were considered for Figure \ref{CF_L_RTM_AGN}, the behavior of these sources likely represents the true luminosity dependence of the CF unbiased by host emission, although some uncertainties (e.g., the influence of the optically-faint MIR sources) still remain as described in Section \ref{rejected}. 
Nevertheless, a more complete AGN sample obtained by using only the SDSS may give a more reliable picture at the low luminosity end when discussing the torus model.

\begin{deluxetable}{cccccccccccc}
\tablecolumns{12}
\tablewidth{0pc}
\tabletypesize{\scriptsize}
\tablecaption{CFs as a function of the 12-$\mu$m luminosity for each type 2 AGN definition without considering the AGN wedge.\label{CF_table_12}} 
\tablehead{
\colhead{}  &  	\multicolumn{2}{c}{Sy2s} 				& \colhead{} &	
				\multicolumn{2}{c}{Sy2s + LINERs} 		& \colhead{} & 
				\multicolumn{2}{c}{Sy2s + Composites}	& \colhead{} & 
				\multicolumn{2}{c}{Sy2s + LINERs + Composites} \\
\cline{2-3}  \cline{5-6} \cline{8-9}  \cline{11-12} \\
log L 	& 	\multicolumn{1}{c}{CF} & \multicolumn{1}{c}{$\sigma_{CF}$} & \colhead{} &
		  	\multicolumn{1}{c}{CF} & \multicolumn{1}{c}{$\sigma_{CF}$} & \colhead{} &
			\multicolumn{1}{c}{CF} & \multicolumn{1}{c}{$\sigma_{CF}$} & \colhead{} &
			\multicolumn{1}{c}{CF} & \multicolumn{1}{c}{$\sigma_{CF}$} 
		 }	 
\startdata
\cutinhead{$0.006 \leq z < 0.05$}
      7.80 & 0.96 & 0.29 & & 0.99 & 0.14 & & 0.99 & 0.14 & & 0.99 & 0.11 \\
      8.20 & 0.95 & 0.11 & & 0.99 & 0.05 & & 0.99 & 0.05 & & 0.99 & 0.04 \\
      8.60 & 0.97 & 0.07 & & 0.99 & 0.03 & & 1.00 & 0.03 & & 1.00 & 0.02 \\
      9.00 & 0.83 & 0.06 & & 0.95 & 0.04 & & 0.97 & 0.03 & & 0.98 & 0.03 \\
      9.40 & 0.80 & 0.08 & & 0.93 & 0.05 & & 0.96 & 0.04 & & 0.97 & 0.03 \\
      9.80 & 0.59 & 0.10 & & 0.78 & 0.09 & & 0.88 & 0.07 & & 0.90 & 0.07 \\
      10.2 & 0.55 & 0.22 & & 0.67 & 0.21 & & 0.82 & 0.18 & & 0.84 & 0.17 \\
\cutinhead{$0.05 \leq z < 0.1$}
      8.60 & 0.98 & 0.17 & & 1.00 & 0.08 & & 1.00 & 0.07 & & 1.00 & 0.06 \\
      9.00 & 0.92 & 0.06 & & 0.98 & 0.03 & & 0.99 & 0.02 & & 0.99 & 0.02 \\
      9.40 & 0.80 & 0.03 & & 0.91 & 0.02 & & 0.97 & 0.02 & & 0.97 & 0.01 \\
      9.80 & 0.60 & 0.03 & & 0.73 & 0.03 & & 0.91 & 0.02 & & 0.92 & 0.02 \\
      10.2 & 0.44 & 0.04 & & 0.54 & 0.05 & & 0.76 & 0.04 & & 0.78 & 0.04 \\
      10.6 & 0.31 & 0.08 & & 0.40 & 0.09 & & 0.57 & 0.10 & & 0.61 & 0.10 \\
      11.0 & 0.36 & 0.24 & & 0.36 & 0.24 & & 0.54 & 0.26 & & 0.54 & 0.26 \\
\cutinhead{$0.1 \leq z < 0.15$}
      9.40 & 0.91 & 0.06 & & 0.96 & 0.04 & & 0.99 & 0.04 & & 0.99 & 0.03 \\
      9.80 & 0.72 & 0.03 & & 0.82 & 0.03 & & 0.95 & 0.02 & & 0.95 & 0.02 \\
      10.2 & 0.49 & 0.03 & & 0.61 & 0.03 & & 0.83 & 0.03 & & 0.85 & 0.03 \\
      10.6 & 0.35 & 0.04 & & 0.44 & 0.04 & & 0.61 & 0.05 & & 0.64 & 0.05 \\
      11.0 & 0.15 & 0.05 & & 0.22 & 0.06 & & 0.28 & 0.07 & & 0.33 & 0.07 \\
      11.4 & 0.12 & 0.13 & & 0.31 & 0.19 & & 0.12 & 0.13 & & 0.31 & 0.19 \\
\cutinhead{$0.15 \leq z \leq 0.2$}
      9.40 &  \multicolumn{1}{c}{---} &  \multicolumn{1}{c}{---} & & 0.82 & 0.26 & & 0.90 & 0.34 & & 0.93 & 0.25 \\
      9.80 & 0.71 & 0.07 & & 0.85 & 0.05 & & 0.95 & 0.04 & & 0.95 & 0.03 \\
      10.2 & 0.52 & 0.06 & & 0.63 & 0.05 & & 0.85 & 0.04 & & 0.86 & 0.03 \\
      10.6 & 0.27 & 0.03 & & 0.36 & 0.03 & & 0.61 & 0.04 & & 0.64 & 0.04 \\
      11.0 & 0.14 & 0.04 & & 0.23 & 0.05 & & 0.27 & 0.05 & & 0.33 & 0.06 \\
      11.4 & 0.04 & 0.04 & & 0.08 & 0.05 & & 0.12 & 0.07 & & 0.15 & 0.07 \\
      11.8 &  \multicolumn{1}{c}{---} &  \multicolumn{1}{c}{---} & &  \multicolumn{1}{c}{---} &  \multicolumn{1}{c}{---} & & 0.34 & 0.38 & & 0.34 & 0.38 \\
\cutinhead{$0.006 \leq z \leq 0.15$}
      7.80 & 0.96 & 0.29 & & 0.99 & 0.14 & & 0.99 & 0.14 & & 0.99 & 0.11 \\
      8.20 & 0.95 & 0.11 & & 0.99 & 0.05 & & 0.99 & 0.05 & & 0.99 & 0.04 \\
      8.60 & 0.97 & 0.07 & & 0.99 & 0.03 & & 1.00 & 0.03 & & 1.00 & 0.02 \\
      9.00 & 0.88 & 0.04 & & 0.97 & 0.02 & & 0.98 & 0.03 & & 0.99 & 0.02 \\
      9.40 & 0.83 & 0.03 & & 0.92 & 0.02 & & 0.97 & 0.01 & & 0.98 & 0.01 \\
      9.80 & 0.68 & 0.02 & & 0.80 & 0.02 & & 0.94 & 0.01 & & 0.94 & 0.01 \\
      10.2 & 0.49 & 0.03 & & 0.60 & 0.03 & & 0.82 & 0.02 & & 0.84 & 0.02 \\
      10.6 & 0.34 & 0.04 & & 0.43 & 0.04 & & 0.60 & 0.04 & & 0.63 & 0.04 \\
      11.0 & 0.19 & 0.06 & & 0.24 & 0.07 & & 0.32 & 0.07 & & 0.36 & 0.07 \\
      11.4 & 0.12 & 0.13 & & 0.32 & 0.19 & & 0.43 & 0.40 & & 0.51 & 0.37 \\
\cutinhead{$0.006 \leq z \leq 0.2$}
      7.80 & 0.96 & 0.29 & & 0.99 & 0.14 & & 0.99 & 0.14 & & 0.99 & 0.11 \\
      8.20 & 0.95 & 0.11 & & 0.99 & 0.05 & & 0.99 & 0.05 & & 0.99 & 0.04 \\
      8.60 & 0.97 & 0.07 & & 0.99 & 0.03 & & 1.00 & 0.03 & & 1.00 & 0.02 \\
      9.00 & 0.88 & 0.04 & & 0.97 & 0.02 & & 0.98 & 0.03 & & 0.99 & 0.02 \\
      9.40 & 0.83 & 0.03 & & 0.92 & 0.02 & & 0.97 & 0.01 & & 0.98 & 0.01 \\
      9.80 & 0.72 & 0.02 & & 0.82 & 0.02 & & 0.95 & 0.01 & & 0.95 & 0.01 \\
      10.2 & 0.53 & 0.02 & & 0.64 & 0.02 & & 0.85 & 0.02 & & 0.86 & 0.02 \\
      10.6 & 0.33 & 0.03 & & 0.42 & 0.03 & & 0.62 & 0.03 & & 0.65 & 0.03 \\
      11.0 & 0.18 & 0.04 & & 0.25 & 0.04 & & 0.32 & 0.05 & & 0.36 & 0.05 \\
      11.4 & 0.08 & 0.05 & & 0.15 & 0.07 & & 0.30 & 0.23 & & 0.35 & 0.22 \\
      11.8 &  \multicolumn{1}{c}{---} &  \multicolumn{1}{c}{---} & &  \multicolumn{1}{c}{---} &  \multicolumn{1}{c}{---} & & 0.28 & 0.31 & & 0.28 & 0.31 \\
\enddata
\end{deluxetable}
\begin{deluxetable}{cccccccccccc}
\tablecolumns{12}
\tablewidth{0pc}
\tabletypesize{\scriptsize}
\tablecaption{CFs as a function of the 22-$\mu$m luminosity for each type 2 AGN definition without considering the AGN wedge.\label{CF_table_22}} 
\tablehead{
\colhead{}  &  	\multicolumn{2}{c}{Sy2s} 				& \colhead{} &	
				\multicolumn{2}{c}{Sy2s + LINERs} 		& \colhead{} & 
				\multicolumn{2}{c}{Sy2s + Composites}	& \colhead{} & 
				\multicolumn{2}{c}{Sy2s + LINERs + Composites} \\
\cline{2-3}  \cline{5-6} \cline{8-9}  \cline{11-12} \\
log L 	& 	\multicolumn{1}{c}{CF} & \multicolumn{1}{c}{$\sigma_{CF}$} & \colhead{} &
		  	\multicolumn{1}{c}{CF} & \multicolumn{1}{c}{$\sigma_{CF}$} & \colhead{} &
			\multicolumn{1}{c}{CF} & \multicolumn{1}{c}{$\sigma_{CF}$} & \colhead{} &
			\multicolumn{1}{c}{CF} & \multicolumn{1}{c}{$\sigma_{CF}$} 
		 }	 
\startdata
\cutinhead{$0.006 \leq z < 0.05$}
      8.20 & 0.89 & 0.44 & & 0.95 & 0.31 & & 0.96 & 0.26 & & 0.97 & 0.22 \\
      8.60 & 0.95 & 0.27 & & 0.99 & 0.14 & & 0.99 & 0.11 & & 0.99 & 0.09 \\
      9.00 & 0.81 & 0.11 & & 0.94 & 0.07 & & 0.97 & 0.05 & & 0.98 & 0.05 \\
      9.40 & 0.75 & 0.08 & & 0.87 & 0.07 & & 0.95 & 0.06 & & 0.96 & 0.05 \\
      9.80 & 0.66 & 0.10 & & 0.78 & 0.09 & & 0.91 & 0.06 & & 0.92 & 0.06 \\
      10.2 & 0.60 & 0.16 & & 0.71 & 0.16 & & 0.86 & 0.13 & & 0.88 & 0.12 \\
      10.6 & 0.68 & 0.60 & & 0.80 & 0.53 & & 0.94 & 0.33 & & 0.94 & 0.32 \\
\cutinhead{$0.05 \leq z < 0.1$}
      9.40 & 0.76 & 0.13 & & 0.87 & 0.11 & & 0.96 & 0.11 & & 0.96 & 0.10 \\
      9.80 & 0.64 & 0.05 & & 0.74 & 0.05 & & 0.90 & 0.03 & & 0.91 & 0.03 \\
      10.2 & 0.53 & 0.04 & & 0.64 & 0.04 & & 0.85 & 0.04 & & 0.86 & 0.04 \\
      10.6 & 0.48 & 0.07 & & 0.56 & 0.07 & & 0.76 & 0.07 & & 0.78 & 0.07 \\
      11.0 & 0.37 & 0.16 & & 0.52 & 0.18 & & 0.70 & 0.18 & & 0.74 & 0.18 \\
      11.4 & 0.40 & 0.32 & & 0.40 & 0.32 & & 0.67 & 0.35 & & 0.67 & 0.35 \\
\cutinhead{$0.1 \leq z < 0.15$}
      9.80 & 0.28 & 0.16 & & 0.64 & 0.47 & & 0.87 & 0.19 & & 0.89 & 0.23 \\
      10.2 & 0.58 & 0.05 & & 0.67 & 0.05 & & 0.85 & 0.05 & & 0.87 & 0.05 \\
      10.6 & 0.43 & 0.04 & & 0.53 & 0.04 & & 0.71 & 0.04 & & 0.74 & 0.04 \\
      11.0 & 0.34 & 0.06 & & 0.47 & 0.07 & & 0.56 & 0.06 & & 0.63 & 0.07 \\
      11.4 & 0.14 & 0.07 & & 0.17 & 0.08 & & 0.41 & 0.16 & & 0.42 & 0.16 \\
      11.8 &  \multicolumn{1}{c}{---} &  \multicolumn{1}{c}{---} & & 0.44 & 0.32 & &  \multicolumn{1}{c}{---} &  \multicolumn{1}{c}{---} & & 0.44 & 0.32 \\
\cutinhead{$0.15 \leq z \leq 0.2$}
      10.2 & 0.62 & 0.29 & & 0.65 & 0.28 & & 0.80 & 0.22 & & 0.81 & 0.21 \\
      10.6 & 0.47 & 0.12 & & 0.56 & 0.11 & & 0.69 & 0.09 & & 0.72 & 0.09 \\
      11.0 & 0.31 & 0.05 & & 0.37 & 0.05 & & 0.50 & 0.07 & & 0.53 & 0.07 \\
      11.4 & 0.11 & 0.05 & & 0.26 & 0.09 & & 0.36 & 0.14 & & 0.44 & 0.14 \\
      11.8 & 0.06 & 0.07 & & 0.06 & 0.07 & & 0.15 & 0.10 & & 0.15 & 0.10 \\
      12.2 &  \multicolumn{1}{c}{---} &  \multicolumn{1}{c}{---} & &  \multicolumn{1}{c}{---} &  \multicolumn{1}{c}{---} & & 0.62 & 0.77 & & 0.62 & 0.77 \\
\cutinhead{$0.006 \leq z \leq 0.15$}
      8.20 & 0.89 & 0.44 & & 0.95 & 0.31 & & 0.96 & 0.26 & & 0.97 & 0.22 \\
      8.60 & 0.95 & 0.28 & & 0.99 & 0.14 & & 0.99 & 0.11 & & 1.00 & 0.09 \\
      9.00 & 0.82 & 0.12 & & 0.94 & 0.07 & & 0.97 & 0.05 & & 0.98 & 0.05 \\
      9.40 & 0.74 & 0.07 & & 0.86 & 0.06 & & 0.95 & 0.06 & & 0.96 & 0.05 \\
      9.80 & 0.65 & 0.05 & & 0.75 & 0.04 & & 0.90 & 0.03 & & 0.91 & 0.03 \\
      10.2 & 0.57 & 0.03 & & 0.68 & 0.03 & & 0.86 & 0.03 & & 0.87 & 0.03 \\
      10.6 & 0.45 & 0.04 & & 0.54 & 0.04 & & 0.75 & 0.04 & & 0.77 & 0.04 \\
      11.0 & 0.34 & 0.05 & & 0.46 & 0.06 & & 0.59 & 0.07 & & 0.65 & 0.07 \\
      11.4 & 0.18 & 0.08 & & 0.20 & 0.08 & & 0.46 & 0.14 & & 0.47 & 0.14 \\
      11.8 &  \multicolumn{1}{c}{\nodata} &  \multicolumn{1}{c}{\nodata} & & 0.44 & 0.32 & &  \multicolumn{1}{c}{\nodata} &  \multicolumn{1}{c}{\nodata} & & 0.44 & 0.32 \\
\cutinhead{$0.006 \leq z \leq 0.2$}
      8.20 & 0.89 & 0.44 & & 0.95 & 0.31 & & 0.96 & 0.26 & & 0.97 & 0.22 \\
      8.60 & 0.95 & 0.28 & & 0.99 & 0.14 & & 0.99 & 0.11 & & 1.00 & 0.09 \\
      9.00 & 0.82 & 0.12 & & 0.94 & 0.07 & & 0.97 & 0.05 & & 0.98 & 0.05 \\
      9.40 & 0.74 & 0.07 & & 0.86 & 0.06 & & 0.95 & 0.06 & & 0.96 & 0.05 \\
      9.80 & 0.65 & 0.05 & & 0.75 & 0.04 & & 0.90 & 0.03 & & 0.91 & 0.03 \\
      10.2 & 0.58 & 0.03 & & 0.68 & 0.03 & & 0.86 & 0.03 & & 0.87 & 0.03 \\
      10.6 & 0.48 & 0.04 & & 0.57 & 0.04 & & 0.77 & 0.04 & & 0.79 & 0.04 \\
      11.0 & 0.33 & 0.04 & & 0.44 & 0.05 & & 0.57 & 0.05 & & 0.62 & 0.05 \\
      11.4 & 0.14 & 0.04 & & 0.22 & 0.05 & & 0.44 & 0.11 & & 0.48 & 0.11 \\
      11.8 & 0.05 & 0.05 & & 0.18 & 0.11 & & 0.13 & 0.08 & & 0.21 & 0.12 \\
      12.2 &  \multicolumn{1}{c}{---} &  \multicolumn{1}{c}{---} & &  \multicolumn{1}{c}{---} &  \multicolumn{1}{c}{---} & & 0.56 & 0.69 & & 0.56 & 0.69 \\
\enddata
\end{deluxetable}

\begin{figure}
    \epsscale{1}
        \plotone{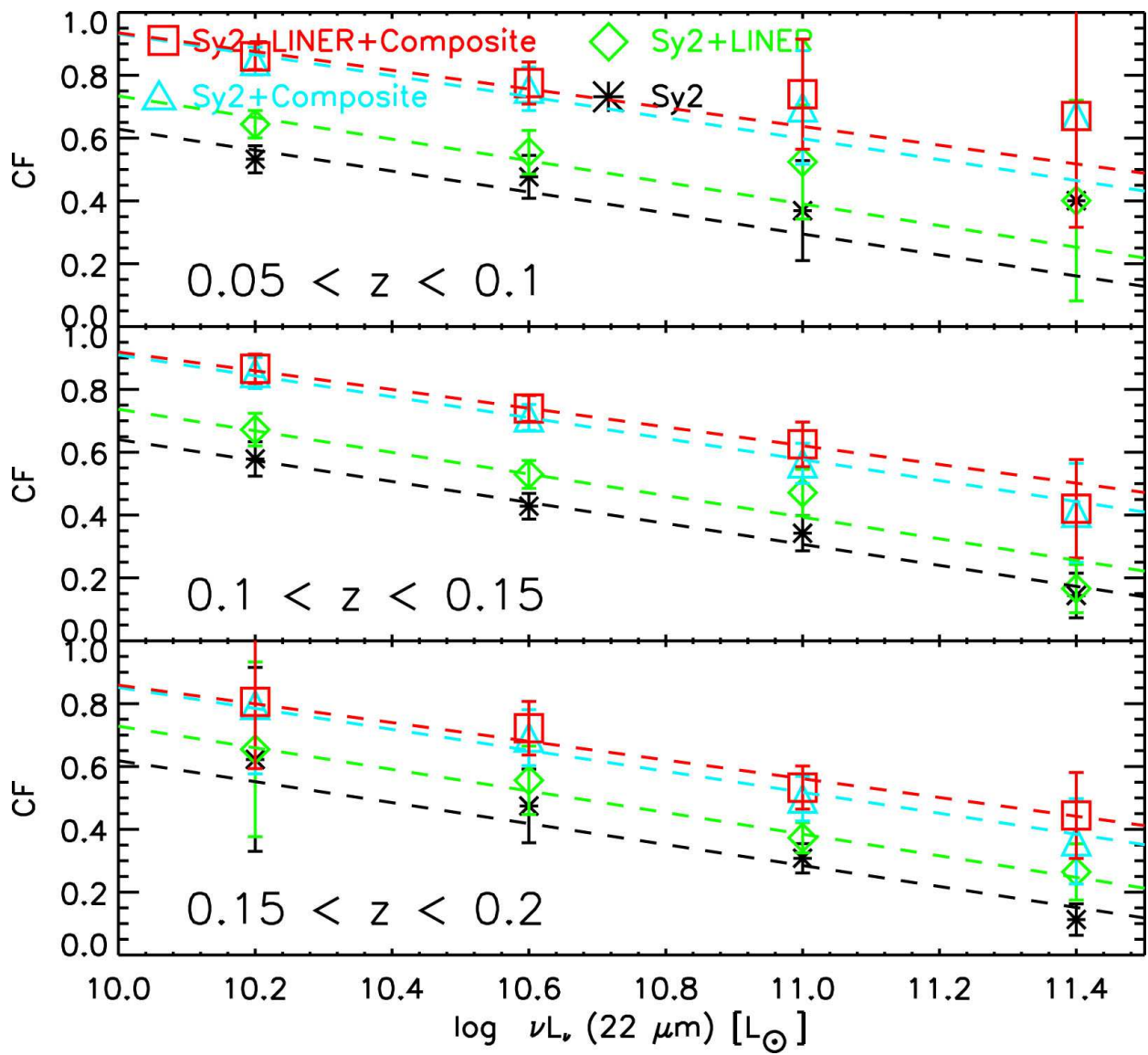}
        \caption{Variation in the CF with the 22-$\mu$m luminosity in different redshift bins without considering the AGN wedge. The dashed line shows the best-fit linear function with the slope fixed to the average value.}
    \label{CF_multi_z_22_lin_nonAGN}
    \end{figure}

\begin{figure}
        \epsscale{1}
        \plotone{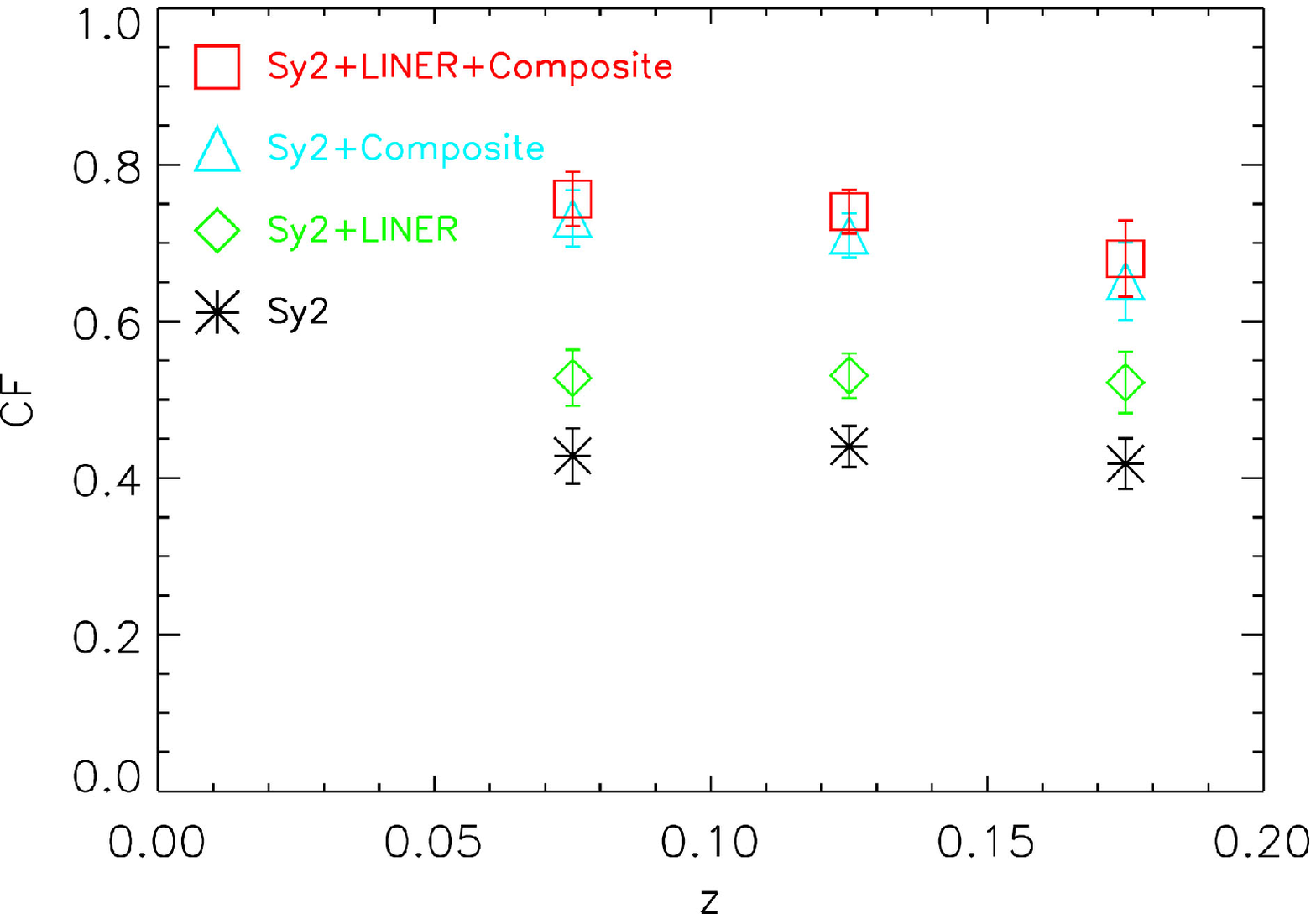}
        \caption{Dependence of the normalization of the CF at $\nu L_{\nu}$(22 $\micron$) = 10.6 L$_{\odot}$ with redshift without considering the AGN wedge.}
    \label{CF_multi_z_22_slope_nonAGN}
    \end{figure}

Figures \ref{CF_multi_z_22_lin_nonAGN} and
\ref{CF_multi_z_22_slope_nonAGN} show the dependence of our measured
CF on redshift, which corresponds to Figures \ref{CF_multi_z_22_lin}
and \ref{CF_multi_z_22_slope} in Section \ref{CF_vs_L},
respectively. We also see that in this case the CF does not appear to
depend on redshift. 


\section{Number Density Ratios for Each Galaxy Type}
In Section \ref{LF}, we constructed the 12- and 22-$\mu$m LFs for each galaxy type in each redshift bin, and these data are listed in Tables \ref{LF_table_12_z} and \ref{LF_table_22_z}.
We also calculated the number density ratios for each galaxy type by integrating the 22-$\mu$m LFs, i.e.,
\begin{equation}
\Phi = \int_L \phi(L) \mathrm{d}L \sim \sum_i \phi_i(L)  \Delta L.
\end{equation}
Table \ref{ND} summarizes the number density ratios normalized by the number density of the type 1 AGNs in each redshift bin.

\begin{deluxetable}{rrrrrrrr}
\tablecolumns{8}
\tablewidth{0pc}
\setlength{\tabcolsep}{0.03in} 
\tabletypesize{\scriptsize}
\rotate
\tablecaption{Number density ratios of each galaxy type in each redshift bin. The number density is estimated from the 22-$\mu$m LFs.\label{ND}}
\tablehead{
\colhead{type}  &  	\multicolumn{1}{c}{$0.006 \leq z \leq 0.05$}	& 
					\multicolumn{1}{c}{$0.05 \leq z \leq 0.1$} 		& 
					\multicolumn{1}{c}{$0.1 \leq z \leq 0.15$} 		& 
					\multicolumn{1}{c}{$0.15 \leq z \leq 0.2$} 		& 
					\multicolumn{1}{c}{$0.2 \leq z \leq 0.25$} 		& 
					\multicolumn{1}{c}{$0.25 \leq z \leq 0.3$} 		& 					
					\multicolumn{1}{c}{$0.006 \leq z \leq 0.3$}		\\
\colhead{(integral range)}  &  	\multicolumn{1}{c}{($8.0 \leq \log \nu L_{\nu} \leq 10.8$)}	& 
								\multicolumn{1}{c}{($9.2 \leq \log \nu L_{\nu} \leq 11.2$)}	& 	
								\multicolumn{1}{c}{($9.6 \leq \log \nu L_{\nu} \leq 11.6$)}	&
								\multicolumn{1}{c}{($10.0 \leq \log \nu L_{\nu} \leq 11.6$)}	&
								\multicolumn{1}{c}{($10.4 \leq \log \nu L_{\nu} \leq 11.6$)}	&
								\multicolumn{1}{c}{($10.8 \leq \log \nu L_{\nu} \leq 11.6$)}	&
								\multicolumn{1}{c}{($8.0 \leq \log \nu L_{\nu} \leq 12.0$)}
}	 
\startdata
type 1 & \multicolumn{1}{c}{1.00} &  \multicolumn{1}{c}{1.00} & \multicolumn{1}{c}{1.00} &  \multicolumn{1}{c}{1.00} &  \multicolumn{1}{c}{1.00} &  \multicolumn{1}{c}{1.00} &  \multicolumn{1}{c}{1.00} \\
type 2 &  4.80 $\pm$ 1.17 &  1.80 $\pm$ 0.14 &  0.96 $\pm$ 0.07 &  0.79 $\pm$ 0.13 &  0.09 $\pm$ 0.02 &  0.06 $\pm$ 0.04 &  4.08 $\pm$ 0.85 \\
LINER &  9.08 $\pm$ 2.00 &  1.37 $\pm$ 0.13 &  0.59 $\pm$ 0.10 &  0.27 $\pm$ 0.04 &  0.03 $\pm$ 0.01 &  0.43 $\pm$ 0.42 &  7.42 $\pm$ 1.37 \\
Composte & 20.76 $\pm$ 4.39 &  8.41 $\pm$ 0.61 &  3.15 $\pm$ 0.20 &  1.13 $\pm$ 0.11 &  0.35 $\pm$ 0.07 &  0.11 $\pm$ 0.05 & 17.38 $\pm$ 3.00 \\
SF & 93.87 $\pm$ 4.34 & 19.68 $\pm$ 0.53 &  4.99 $\pm$ 0.20 &  1.34 $\pm$ 0.13 &  0.16 $\pm$ 0.07 & 0.01 $\pm$ 0.08 & 79.19 $\pm$ 2.96 \\
\enddata
\end{deluxetable}

Note that the luminosity range defines the integral range, so that the integral range is different for each redshift bin.
As shown in Table \ref{ND}, the number density ratio is sensitive to the luminosity range and the redshift. To compare the number density ratio of the type 2 to type 1 AGNs with that of \cite{Toba}, which were based on AKARI 18-$\mu$m LFs, we choose the nearest redshift bin ($0.006 \leq z \leq 0.05$) and an integral range of $\log (\nu L_{\nu})$ $>$ $10^{10}$ $L_{\odot}$ for both AGN types.
We obtained a number ratio of 1.53 $\pm$ 0.50, which is consistent within error with the value of 1.73 $\pm$ 0.36 obtained by \cite{Toba}.

\end{document}